\RequirePackage{fix-cm}
\documentclass{svjour3}                     
\smartqed  

\usepackage[usenames, dvipsnames]{color}
\definecolor{rred}{rgb}{0.7,0,0.1}
\newcommand{\mk}{\color{black}}
\newcommand{\mkr}{\color{black}}
\newcommand{\mkrr}{\color{black}}
\definecolor{greenrb}{rgb}{0.2,0.6,0.2}
 
\usepackage{amssymb}
\usepackage[breaklinks=true,colorlinks=true,linkcolor=blue,urlcolor=blue,citecolor=blue]{hyperref}
\usepackage{subcaption}
\usepackage{bbm}
\usepackage{amsmath}
\usepackage{empheq}
\usepackage{fancyhdr}
\usepackage{tikz}
\usetikzlibrary{arrows,positioning} 
\newtheorem{prop}{Proposition}
\newtheorem{coro}{Corollary}
\newtheorem{thm}{Theorem}

\newtheorem{rmk}{Remark}

\newtheorem{defi}{Definition}
\newcommand\at[2]{\left.#1\right|_{#2}}

\newcommand{\Span}{\mathrm{span~}}
\newcommand{\lr}[1]{\left(#1\right)}
\newcommand{\lrs}[1]{\left[#1\right]}
\newcommand{\smalleq}{\lr{\frac{\epsilon/\sqrt{\delta}}{\sqrt{\delta/\kappa}}}}
\newcommand{\smallin}{\epsilon \sqrt{\kappa} / \delta}

\def\be{\begin{equation}}
\def\ee{\end{equation}}
\def\bes{\begin{equation*}}
\def\ees{\end{equation*}}
\def\bea{\begin{equation} \begin{aligned}}
\def\eea{\end{aligned} \end{equation}}
\def\beas{\begin{equation*} \begin{aligned}}
\def\eeas{\end{aligned} \end{equation*}}
\def\d{\, \mathrm{d}}

\usepackage{lineno}

\numberwithin{equation}{section} 
\begin{document}

\title{Ruelle-Pollicott Resonances of Stochastic Systems in Reduced State Space. Part II: Stochastic Hopf Bifurcation}


\author{Alexis Tantet \and Micka\"el D. Chekroun \and Henk A. Dijkstra \and J. David Neelin}
\institute{A. Tantet \at LMD/IPSL, Ecole polytechnique, Sorbonne Universite, ENS, PSL University, CNRS, Palaiseau, France\\
  \email{alexis.tantet@lmd.polytechnique.fr}
  \and
  M.D. Chekroun \at
   Department of Earth and Planetary Sciences, Weizmann Institute, Rehovot 76100, Israel;
  Department of Atmospheric and Oceanic Sciences, University of California, Los Angeles, CA 90095-1565, USA
 \and 
 J.D. Neelin \at
  Department of Atmospheric and Oceanic Sciences and Institute of Geophysics and Planetary Physics, University of California, Los Angeles, CA 90095-1565, USA
  \and
  H.A. Dijkstra 
  \at Institute of Marine and Atmospheric Research, Department of Physics and Astronomy,  University of Utrecht, Utrecht, The Netherlands}
\date{\today}

\maketitle

\begin{abstract}

The spectrum of the {\mkr generator (Kolmogorov operator)} of a diffusion process, referred to as the Ruelle-Pollicott (RP) spectrum, provides a detailed characterization of {\mkr correlation functions and power spectra of stochastic systems via decomposition formulas in terms of RP resonances; see Part I of this contribution \cite{Chekroun_al_RP2}.} 
Stochastic analysis techniques {\mkr relying on  the theory of Markov semigroups} for the study of the RP spectrum and a rigorous reduction method is presented in Part I \cite{Chekroun_al_RP2}. This framework is here applied to study a stochastic Hopf bifurcation {\mkr in view of characterizing} the statistical properties of nonlinear oscillators perturbed by noise, depending on their stability.

In light of the H\"ormander theorem,
it is first shown that the geometry of the unperturbed limit cycle,
in particular its isochrons, i.e., the leaves of the stable manifold of the limit cycle generalizing the notion of phase, is essential to understand the effect of the noise and the phenomenon of phase diffusion.
In addition, it is shown that the RP spectrum has a spectral gap,
even at the bifurcation point, and that correlations decay exponentially fast.

Explicit small-noise expansions of the RP eigenvalues and eigenfunctions are then obtained, away from the bifurcation point, based on the knowledge of the linearized deterministic dynamics and the characteristics of the noise.
These formulas allow one to understand how the interaction of the noise with the deterministic dynamics affect the decay of correlations. Numerical results complement the study of the RP spectrum at the bifurcation point, revealing {\mkr useful} scaling laws.

The analysis of the Markov semigroup for stochastic bifurcations is thus promising
in providing a complementary approach to the more geometric random dynamical {\mkr system (RDS) approach}.
This approach is not limited to low-dimensional systems and the reduction method presented in~\cite{Chekroun_al_RP2} {\mkr is} applied to a stochastic {\mk model} relevant to climate dynamics in the third part of this contribution~\cite{tantet_ruellepollicott_2019}.

\end{abstract}


\tableofcontents
\section{Introduction}\label{sec:Introduction}

Complex and unpredictable behavior of trajectories is observed in many physical systems.
{\mkr A possible source of such an unpredictable behavior is tied to the} interactions between a large number of degrees of freedom,
 which may be {\mkr either} modeled by the addition of a stochastic forcing,
or {\mkr by} nonlinear coupling {\mkr terms} resulting {\mkr into} chaotic trajectories.
As a result, prediction beyond a certain horizon is hopeless
and one focuses instead on a statistical {\mkr description of the system's evolution}.
The theory presented in the first part of this contribution~\cite{Chekroun_al_RP2} is concerned with the characterization of {\mkr statistical features such as the return to (a statistical) equilibrium, or the description of correlation functions and power spectra}---in both, the reduced and original state spaces---for nonlinear systems {\mkr subject to noise disturbances}, extending thus the approach of~\cite{Chek_al14_RP} to the stochastic framework.
This second part of our three-part article is focused on stochastic perturbations of dynamical systems undergoing a Hopf bifurcation in which a stable {\mk steady state} loses its stability  to give rise to a limit cycle.
It relies on the {\mkr elements of stochastic analysis and the spectral decomposition of Markov semigroups such as framed in Part I \cite{Chekroun_al_RP2}}, {\mkr while the results regarding the notion of reduced RP resonances from Part I are applied to} the third part~\cite{tantet_ruellepollicott_2019} of this contribution.
The latter {\mkr part relies thus} on the preceding two parts to analyze the response to noise of a low-frequency mode of climate variability, El Ni\~no-Southern Oscillation.

Following~\cite{Chekroun_al_RP2}, the evolution of trajectories is modeled by an
 {\mk It\^o} Stochastic Differential Equation (SDE),
\begin{align}
	\d x = F(x) \d t + {\mkr D}(x) \d W_t, \label{eq:SDEGeneral}
\end{align}
on the $N$-dimensional Euclidean space $\mathcal{H} = \mathbb{R}^N$
with $W_t = (W_t^1, \dots, W_t^M)$ an $\mathbb{R}^M$-valued
Wiener process with measure $\mathbb{P}$ and realizations $\omega$ in $\Omega$.
In what follows we assume that the vector field $F$
and the matrix-valued function {\mkr $D : \mathcal{H} \to \mbox{Mat}_{\mathbb{R}}(N\times M)$},
satisfy regularity conditions that guarantee the
existence and the uniqueness of mild solutions, as well as the continuity of the trajectories;
e.g.~\cite{Cerrai2001,Flandoli2010} for such conditions in the case of locally Lipschitz coefficients.
The process $X(t, \omega)$ generated by the SDE (\ref{eq:SDEGeneral}) is thus
a continuous Markov process.

As discussed in~\cite[Appendix A.1]{Chekroun_al_RP2}, while the sample paths generated by the SDE \eqref{eq:SDEGeneral} may be complicated, the evolution of observables, averaged over the noise realizations, may be more regular and more amenable to analysis~\cite{Pavliotis2014}.
The evolution of an observable $u$ in $C_b(\mathcal{H})$, the space of bounded continuous functions, is governed by the \emph{Markov semigroup} $P_t, t \ge 0$, according to
\begin{align*}
	P_t u(x) = \mathbb{E}\left[u(S(t, \cdot) x)\right] = \int_\Omega u(S(t, \omega) x) \d \mathbb{P}(\omega),
\end{align*}
where $S(t, \omega): \mathcal{H} \to \mathcal{H}$ is the stochastic flow
giving the solution at any time $t \ge 0$ to the SDE (\ref{eq:SDEGeneral})
for an initial condition in $\mathcal{H}$ and any noise realization $\omega$ in $\Omega$.
This Markov semigroup can be extended to a strongly continuous semigroup on $L^2_\mu(\mathcal{H})$, the space of square-integrable functions with respect to an invariant measure $\mu$ of the system; e.g.~\cite[Theorem 4]{Chekroun_al_RP2}.
In the remaining, we always work in $L^2_\mu(\mathbb{R}^2)$, for some invariant measure $\mu$ which happens to be unique for the particular two-dimensional system studied here, as shown in Section~\ref{sec:stochastic_hopf}.
In some cases, the generator $K$ of this Markov semigroup
can be identified with the second-order differential operator $\mathcal{K}$
of the (backward) Kolmogorov equation~\cite[Remark~1.(iii)]{Chekroun_al_RP2},
\begin{align}
	\partial_t u &= \sum_{i = 1}^2 F_i(x) \partial_i u
	+ \frac{1}{2} \sum_{i, j = 1}^2 {\mkr \mathbf{\Sigma}_{ij}(x) } \partial_{ij} u, \label{eq:BKEGeneral} \\
	&= \mathcal{K} u,
\end{align}
where {\mkr $\mathbf{\Sigma}_{ij}(x) = \sum_{k = 1}^M  D_{ik}(x) D_{jk}(x)$ is the diffusion tensor}.
In turn, the Kolmogorov equation is dual to the Fokker-Planck equation governing the evolution of probability densities;  {\mk see e.g.~\cite[Sect.~2]{Chekroun_al_RP2} and}~\cite{Risken1989}.

One possible manifestation of unpredictability in chaotic or stochastic systems
is the loss of memory of ensembles on their initial state
as they converge to the statistical {\mk equilibrium} of the system.
In other words, \emph{mixing}~\cite[Chap.~4]{Lasota1994} occurs when densities propagated by the transfer semigroup dual to the Markov semigroup converge to a {\mk unique} statistical {\mk equilibrium}, or invariant measure, in e.g.~the total variation norm.
Conditions ensuring a Markov process to be mixing {\mk in the total variation norm} have been {\mk recalled in~\cite[Theorem 4]{Chekroun_al_RP2}} and rely on the strong Feller and irreducibility properties of the Markov semigroup.
Moreover, as discussed in e.g.~\cite[Remark 1-(i)]{Chekroun_al_RP2}, {\mk a Markov process that is mixing with respect to an invariant measure $\mu$ has its} \emph{correlation function}
\begin{align}\label{Eq_correlation_def}
	C_{f,g}(t) = \int f \cdot P_t g \d \mu - \int f \d\mu \int g \d \mu, \quad t \ge 0,
\end{align}
{\mk that decays asymptotically to zero in time}, for any observables $f$ and $g$ {\mk lying} in $L^2_\mu(\mathbb{R}^2)$.
Together with their Fourier transform, the power spectra, sample estimates of correlation functions are often used in physics, to study the variability of the system.
It is thus important to relate such evolution of the statistics to the dynamics of the system.

The essential point here is that, as {\mkr shown} in~\cite[Theorem~1]{Chekroun_al_RP2}, the spectrum of the generator $K$ of the Markov semigroup {\mk $P_t$ in $L^2_\mu(\mathbb{R}^2)$ associated with~\eqref{eq:BKEGeneral}}, referred to as the \emph{Ruelle-Pollicott (RP) spectrum}, gives a complete characterization of the evolution of observables.
In particular, it allows one to decompose the correlation functions into
several components with different decay rates {\mkr directly related to the RP eigenvalues}; {\mk see~\cite[Corollary 1]{Chekroun_al_RP2}}.
For example, if the spectrum of the generator is only composed of eigenvalues $\lambda_j, j \ge 0$ (as in the case of the stochastic Hopf, Sect.~\ref{sec:stochastic_hopf}) and if these eigenvalues are simple, then the correlation function can be decomposed into the weighted sum of  complex exponentials~\cite[Eq.~(2.13)]{Chekroun_al_RP2}
\begin{align}
	C_{f,g}(t) = \sum_{j = 1}^\infty e^{\lambda_j t} w_j(f, g), \quad f, g \in L^2_\mu(\mathbb{R}^2),
	\label{eq:spectralCorr}
\end{align}
with weights given by
\begin{align}
	w_j(f, g) = \left< f, \psi_j \right>_\mu \left<\psi_j^*, g\right>_\mu,
	\label{eq:spectralWeights}
\end{align}
where $\psi_j$ denotes the $L^2_\mu$-eigenfunction
associated with the eigenvalue $\lambda_j$ of the {\mk $L^2_\mu$-generator $K$ of $P_t$}
and $\psi_j^*$ is the eigenfunction of the adjoint operator $K^*$ of $K$.
A similar decomposition in terms of Lorentzian functions
also holds for the power spectrum $S_{f,g}$~\cite[Sec.~2.3]{Chekroun_al_RP2}:
\begin{align}
	S_{f,g}(z) = - \frac{1}{\pi} \sum_{j = 1}^\infty w_j(f, g) \frac{\Re(\lambda_j)}{(z - \Im(\lambda_j))^2 + \Re(\lambda_j)^2}.
	\label{eq:spectralPower}
\end{align}
The decompositions~\eqref{eq:spectralCorr} and~\eqref{eq:spectralPower} may then be used to reconstruct any correlation function or power spectrum from the RP spectrum.
In addition,~\cite[Theorems~5 and~6]{Chekroun_al_RP2} allow one to analyze the {\mkr the rate of return to the equilibrium $\mu$ (mixing) and the spectral gap in the RP spectrum} from properties of Lyapunov functions and ultimate bounds.
In the following, we refer to the eigenvalues in the RP spectrum, whether simple or not, as the \emph{RP resonances}.

The RP spectrum thus provides a particularly useful description of nonlinear dynamics {\mkr subject to noise disturbances}.
In light of these results, the overarching goal of the present article is to {\mk illustrate the usefulness of}  stochastic analysis techniques {\mk discussed} in~\cite{Chekroun_al_RP2}, 
{\mk for the} study of stochastic bifurcations.
Indeed, while the bifurcation theory of deterministic systems is fairly complete
\cite{Guckenheimer1983,Ruelle1989a,Strogatz1994,Kuznetsov1998},
nonautonomous~\cite{Rasmussen2007,Rasmussen2007a,Zmarrou2007,Potzsche2011} and
stochastic~\cite[Chap.~9]{Arnold2003} bifurcation theory is much less mature.
In particular, the derivation of normal forms, i.e.~finding an equivalent
representation of a system ``as simple as possible,''  can be very tedious in the stochastic case {\mk within the framework of random dynamical systems (RDSs)~\cite[Chap.~8]{Arnold2003}} (see~\cite{Coullet1985,SriNamachchivaya1990a}, for the normal form of the stochastic Hopf bifurcation in particular) and may require the introduction of anticipative terms.
{\mk Such anticipative terms may be avoided in certain cases, by the appropriate use of approximation techniques of local stochastic invariant manifolds~\cite{CLW15_vol1}, or the use of parameterizing manifolds~\cite{CLW15_vol2}, in more general situations.}

It is however important to mention that the {\mkr RDS theory}
has allowed to give {\mkr useful} insights regarding another manifestation of the unpredictability stochastic systems,
namely the divergence of stochastic trajectories characterized by Lyapunov exponents
\cite{Oseledets1968,Arnold1984a,Arnold2003,Cong1997}.
In particular, the Lyapunov exponents have been used to provide a dynamical characterization of stochastic pitchfork~\cite{Crauel1998}, transcritical~\cite{Crauel1999} and Hopf bifurcations \cite{Schimansky-Geier1993,Baxendale1994,Schenk-Hoppe1996,Arnold1996,Arnold2004,Baxendale2004,Deville2011,Anagnostopoulou2015,Engel2016}.
Recently, another approach based on the dichotomy spectrum for random dynamical systems~\cite{Sacker1978,Aulbach2001,Siegmund2002a,Rasmussen2009,Rasmussen2010,Kloeden2011} has been proposed to characterize stochastic bifurcations~\cite{Callaway2013}.


In this study, the focus is on the {\mk description of the change of statistical properties arising at a bifurcation in the presence of noise and as associated with the Markov semigroup}, in the spirit of~\cite{Graham1982a,VandenBroeck1982d}, rather than {\mk those occurring by adopting a pullback approach~\cite{Arnold2003,Rasmussen2007,CSG11,Callaway2013}}.
It is thus ``phenomenological" rather than ``dynamical" {\mkr in the terminology of L.~Arnold \cite[Chap.~9]{Arnold2003}}, {\mkr although new ingredients are brought to describe the phenomenological picture, namely the RP resonances.} 
To be more specific, our study is concerned with the changes occurring in the RP spectrum as the control parameter varies, in the case of a Hopf bifurcation subject to noise.
An example of a three-dimensional {\mkr stochastic slow-fast system} undergoing a Hopf bifurcation perturbed by noise and arising in fluid mechanics is given in~\cite[Sec.~4]{Chekroun_al_RP2},
while the third part {\mkr (\cite{tantet_ruellepollicott_2019})} relies on both the reduction method from~\cite{Chekroun_al_RP2} and the results of this paper to analyze the response to noise of a low-frequency mode of climate variability, El Ni\~no-Southern Oscillation, in a high-dimensional geophysical model of intermediate complexity.
In this article, particular attention is paid to the identification of the key properties of the underlying deterministic dynamical system {\mk that determine} the response to stochastic perturbations.
Our main conclusion is that the geometry of the underlying deterministic {\mk dynamics} is essential to understand the mixing properties of the stochastic {\mk Hopf} bifurcation system. 

{\mk The system considered here} consists of the normal form of the Hopf bifurcation to which white noise is added to the Cartesian coordinates; see Eq.~\eqref{eq:HopfSDECart}.
{\mkrr In Section~\ref{sec:normalForm}, we report on geometric features of the underlying deterministic dynamics that play a key role in the response to stochastic perturbations here: the isochrons that generalize a notion of phase.}
{\mkrr In Section~\ref{sec:StocAnaHopf}, the interaction of the stochastic forcing with the drift term
are assessed in terms of Lie brackets as arising in the so-called H\"ormander's condition pertaining to the elements of stochastic analysis briefly surveyed in~\cite[Appendix~A]{Chekroun_al_RP2} for the unfamiliar reader.}
{\mkrr As a  result for two-dimensional deterministic systems exhibiting} a hyperbolic limit cycle, {\mkrr when perturbed by noise, new geometric insights describing the interactions with nonlinear effects, are provided.  In that respect,} Theorem~\ref{thm:isoHypo} {\mkrr shows that} phase diffusion, responsible for mixing on the limit cycle, occurs when the forcing is transverse to the isochrons.
In addition, we prove with Proposition~\ref{prop:spectral_gap} that the RP spectrum has a spectral gap and that correlations decay exponentially fast, even at the bifurcation point where the deterministic Hopf bifurcation {\mkrr occurs and at which correlations decay at a slower, algebraic rate (i.e.~as a polynomial of time or as a fractional power of time), in absence of noise}.
{\mkrr In Section~\ref{sec:smallNoiseExpansion}, we provide analytic elements describing the RP spectrum}. {\mkrr More precisely, by relying on small-noise expansions, we derive analytic} formulas of  the eigenvalues and eigenfunctions of the {\mk Kolmogorov operator associated with the} Markov semigroup before (Proposition~\ref{prop:eigen_below}) and after (Proposition~\ref{prop:eigen_above}) the bifurcation point when the noise is {\mk relatively} weak. {\mkrr As a byproduct,  these formulas show for the stochastic Hopf equation considered here, that when the noise is sufficiently small, the isochrons still coincide with the isoline of phase of the eigenfunctions of the Kolmogorov operator; see Fig.~\ref{fig:artMixingEigVal} below for a schematic. This is confirmed numerically in Section~\ref{sec:numericalHopf} when the RP spectrum is approximated from  finite-difference approximation of the Kolmogorov operator; see e.g.~Fig.~\ref{fig:numFPeps1beta} below.} {\mkrr Thus, the numerical results of Section~\ref{sec:numericalHopf} allow us to analyze in greater details, the transition from a noisy steady state to a noisy limit cycle. In particular, a transition from a triangular structure of RP eigenvalues to a parabolic one in the left half complex plane, is observed as the bifurcation point is crossed. Finally, Section~\ref{sec:Conclusion} summarizes the insights gained from these numerical results and this article in its whole.}
We build on these {\mkrr insights and  results} in the third part of this contribution to analyze the emergence of noise-induced oscillations in the Cane-Zebiak model of El Ni\~no-Southern Oscillation~\cite{tantet_ruellepollicott_2019}.
The programs used for this analysis are available as an open-source C++ library at \url{https://github.com/atantet/ergoPack/} together with a link to its documentation.

\section{A stochastically perturbed nonlinear oscillator}\label{sec:normalForm}

Nonlinear oscillators are found in many different applications of physics and engineering.
Particularly important is to understand the statistical properties of such systems
in response to noise.
For example, Hopf bifurcations resulting in the emergence of a stable limit cycle 
are found in several climate models, such as in quasi-geostrophic models of the midlatitude ocean circulation~\cite{Dijkstra2005b,Simonnet2009}, while fast atmospheric {\mk processes} forcing the ocean are sometimes modeled by a stochastic {\mk process~\cite{Roulston2000,sura2001regime,schmeits2001bimodal,Ghil2008,sura2002sensitivity,Dijkstra2013}.}

Thus, as a first step towards understanding {\mk more} complex stochastically perturbed nonlinear oscillators,
the RP spectrum of a simple form of stochastic Hopf bifurcation is analyzed.
In this section, we recall some known results regarding the Hopf bifurcation
and its stochastic counterpart.
In particular, we stress the role played in the phenomenon of phase diffusion by relying on the concept of isochrons associated with the underlying deterministic limit cycle.
This approach provides new geometric insights concerning the response of nonlinear oscillators to noise on one hand --- see Section~\ref{sec:StocAnaHopf} --- and concerning the associated RP spectrum, on the other; see Sections~\ref{sec:smallNoiseExpansion}
and~\ref{sec:numericalHopf}.

\subsection{{\mk RP spectrum of the deterministic Hopf normal form}}
Nonlinear systems with a fixed point losing stability to a limit cycle
as a parameter is changed are prominent in physics and engineering; {\mk see e.g.~\cite{Strogatz1994}}.
For instance, this kind of bifurcation, namely the Hopf bifurcation, 
is found in the climate models of El Ni\~no-Southern Oscillation analyzed in the third part of this contribution~\cite{tantet_ruellepollicott_2019}.
The genericity of the {\mk reduced} dynamics close to a Hopf bifurcation is captured by the following normal form~\cite{Guckenheimer1983,Arnold2012geometrical}, in polar coordinates $(r, \theta)$,
\bea\label{eq:normalHopf} 
	\d r		&= \lr{\delta r - \kappa r^3} \d t \\
	\d \theta	&=  \lr{\gamma - \beta r^2} \d t,
\eea
where we assume that $\kappa > 0$, so that only the case of the supercritical Hopf bifurcation is considered here.
The parameter $\delta$ controls the stability of the fixed point $x_*$ (for $\delta < 0$) or of the limit cycle $\Gamma$ (for $\delta > 0$).
The parameter $\gamma$ controls the period of the oscillations, while $\beta$ regulates their dependence on the radius.
Such a dependance may for example arise in systems conserving angular momentum~\cite{Arnold2012geometrical}.
As a result, the limit cycle $\Gamma$ has a radius
\begin{align*}
  R(\delta, \kappa) = \sqrt{\delta / \kappa},
\end{align*}
and a period $T(\gamma, \beta, \delta, \kappa) = 2\pi/\omega_f(\gamma, \beta, \delta, \kappa)$, where $\omega_f$ is the angular frequency
\begin{align*}
  \omega_f(\gamma, \beta, \delta, \kappa)
  = \gamma - \beta \delta / \kappa,
\end{align*}
simply noted $R$, $T$ and $\omega_f$, respectively, in the following.
Denoting by {\mk $(S_t)_{t\in \mathbb{R}}$} the deterministic flow generated by~\eqref{eq:normalHopf},
one has that $S_T p = p$ for any point $p$ on the limit cycle $\Gamma$.
For reasons that become apparent below, we refer to the adimensional parameter $\beta / \kappa$, noted $\tilde \beta(\beta, \kappa)$ or simply $\tilde \beta$, as the \emph{twist factor}.
In section~\ref{sec:smallNoiseExpansion} we show that while the nature and stability of the solutions is controlled by $\delta$, the {\mk geometry of both the locations of the RP eigenvalues and their corresponding eigenfunctions, is strongly dependent on $\tilde \beta$.}

The RP resonances of {\mk system~\eqref{eq:normalHopf} obtained as the eigenvalues in of the corresponding (backward) Liouville eigenvalue problem 
\be
(\delta r -r^3) \partial_r \Psi(r,\theta)+(\gamma -\beta r^2) \partial_\theta \Psi(r,\theta)=\lambda \Psi(r,\theta),
\ee
have} been calculated analytically in~\cite{Gaspard2001a} using trace formulas; for comparison, we give a brief summary of the results in~\cite{Gaspard2001a} below.
To do so, care is given to the functional setting, since, due to the deterministic dissipative dynamics and in order to capture the decay of correlations, eigenfunctions should be sought as distributions acting on observables given by smooth enough test functions (see~\cite[Sect.~B]{Gaspard2001a} and~\cite{Gaspard1995}).
The {\mk authors of~\cite{Gaspard2001a}} found that below the bifurcation point, i.e., for $\delta$ smaller than its critical value $0$, the RP resonances $\lambda_k, k \ge 0$ are given by integer linear combinations of the complex pair of eigenvalues $\lambda^{\pm} = \delta \pm i \gamma$ of the tangent map of the vector field at the fixed point.
As a result, the RP resonances are organized in a triangular array of eigenvalues {\mk~\cite[Eq.~(43)]{Gaspard2001a}}
\begin{align}
	\lambda_{ln} = (l + n) \delta + i (n - l) \gamma, \quad l, n \in \mathbb{N}.
\end{align}
Above the bifurcation point, i.e., for $\delta > 0$, the RP resonances are composed of two families of eigenvalues associated with the limit cycle and the unstable fixed point respectively.
The family associated with the limit cycle is organized in an array
of equally spaced eigenvalues~\cite[Eq.~(44)]{Gaspard2001a}
\begin{align}\label{Eq_mixing_spec_det}
  \lambda_{ln} = - 2 l \delta + i n~\omega_f,
  \quad l \in \mathbb{N}, n \in \mathbb{Z}.
\end{align}
{\mk These eigenvalues have their real parts spaced by a gap} corresponding to the characteristic exponent
of a linearized {\mk Poincar\'e map for the limit cycle $\Gamma$}
and {\mk their imaginary parts spaced by} a gap given by the angular frequency $\omega_f$.
Each multiple of the angular frequency corresponds to a harmonic which may be excited for {\mk certain} nonlinear observables. {\mk We refer to Sect.~\ref{sec:numericalHopf}  below for such nonlinear observables.}

{\mk The spectrum given by  \eqref{Eq_mixing_spec_det} contains pure imaginary eigenvalues $\lambda_{0n}, n$  in $\mathbb{Z}$, showing in particular that the  deterministic system \eqref{eq:normalHopf} is not mixing.} {\mk This can be intuitively understood by the neutral dynamics} along $\Gamma$, i.e the dynamics is neither contracting nor expanding.  {\mk Indeed due to this dynamics,} a density with support {\mk contained in} $\Gamma$ is simply rotated without {\mk mixing along} $\Gamma$. On the other hand, there is also a family of eigenvalues~\cite[Eq.~(44)]{Gaspard2001a} {\mk forming a triangular array}
\begin{align}
	\lambda_{ln} = -(l + n + 2) \delta - i (l - n) \gamma, \quad l, n \in \mathbb{N},
\end{align}
associated with the unstable fixed point.
All these eigenvalues are {\mk located} to the left of the imaginary axis, in agreement with the fact that the unstable fixed point
is a repeller. To this repeller can then be associated an escape rate of densities given by
the real part $|\Re(\lambda_{00})| = 2 \delta$ of the leading eigenvalue. Finally, exactly at the critical value $0$, the spectrum is continuous, resulting in an algebraic decay of correlations, at a rate $t^{-1/2}$ (\cite{Gaspard2001a}, Eq.~(82)), known as critical slowing down. 

{\mk As shown hereafter, when subject to the appropriate noise perturbations, the critical slowing down disappears and the system becomes mixing at the criticality and after; see Sect.~\ref{sec:discreteSpectrum}.}

\subsection{Stochastic Hopf {\mk equation}}\label{sec:stochastic_hopf}

%
As a minimal model of nonlinear oscillator perturbed by noise,
we are thus led {\mk naturally} to analyze the Hopf normal form (\ref{eq:normalHopf})
{\mk subject to white noise disturbances} added to its Cartesian coordinates, as in~\cite{Deville2011}.
This stochastic process is thus governed by the SDE
\begin{equation}
  \begin{aligned}
    \d x
    &= \underbrace{
      \lrs{\lr{\delta - \kappa \lr{x^2 + y^2}} x
        - \lr{\gamma -\beta \lr{x^2 + y^2}} y}
    }_{F_x(x, y)} \d t + \epsilon \d W_x\\
    \d y
    &= \underbrace{
      \lrs{\lr{\gamma - \beta \lr{x^2 + y^2}} x
        + \lr{\delta - \kappa \lr{x^2 + y^2}} y}
    }_{F_y(x, y)} \d t + \epsilon \d W_y,
\end{aligned}\label{eq:HopfSDECart}
\end{equation}
where $W_x$ and $W_y$ are two independent Wiener processes with differentials interpreted in the It\^o sense~\cite{Ikeda1989} and $\epsilon$ is a parameter controlling the level of noise.
In the following, {\mk Eq.~\eqref{eq:HopfSDECart}} will be referred to as the
{\it Stochastic Hopf Equation (SHE)} in Cartesian coordinates.
The Kolmogorov equation corresponding to (\ref{eq:HopfSDECart}) is then given by
\begin{align}
	\partial_t u =
	& F_x \partial_x u + F_y \partial_y u
	+ \frac{\epsilon^2}{2} \partial^2_{xx} u + \frac{\epsilon^2}{2} \partial^2_{yy} u.
	\label{eq:HopfBKECart}
\end{align}
As {\mkr recalled in Introduction}, the solutions to the Kolmogorov equation {\mk have a natural probabilistic interpretation in terms of  expectation, i.e.}~$u(t, (x, y)) = \mathbb{E}\left[u(S(t, \omega) (x, y))\right]$, where, {\mk loosely speaking}, the stochastic flow $S(t, \omega)$ applied to $(x, y)$ yields {\mkr to the solution at time $t$ to  Eq.~\eqref{eq:HopfSDECart} that emanates from $(x, y)$, when driven by the noise realization $\omega$}. {\mkr Here and below,} $\mathbb{E}[\cdot] = \int_\Omega \cdot \d \mathbb{P}$ {\mk denotes the expected value} with respect to {\mk such noise realizations}.
The diffusion in~\eqref{eq:HopfBKECart} is elliptic by construction, a condition that is relaxed in Section~\ref{sec:StocAnaHopf}.

The Kolmogorov equation~\eqref{eq:HopfBKECart} in Cartesian coordinates will be useful to perform expansions {\mk about} the stable fixed point for $\delta < 0$ in Section~\ref{sec:ExpansionSub}.
For $\delta > 0$, however, deterministic solutions converge {\mk (i.e.~when $\epsilon=0$)} to the limit cycle $\Gamma$ with radius $R$ so that it is sometimes more convenient to work in polar coordinates $(r, \theta)$ with $x = r \cos \theta$ and $y = r \sin \theta$.
Applying It\^o's formula~\cite[Theorem~5.1]{Ikeda1989}, the SHE~\eqref{eq:HopfSDECart} transforms to polar coordinates as follows,
\bea
\d r &=
\lr{\delta r - \kappa r^3 + \frac{\epsilon^2}{2 r}} \d t
+ \epsilon \d W_r \\
\d \theta &=
\lr{\gamma - \beta r^2} \d t
+ \frac{\epsilon}{r} \d W_\theta, \label{eq:HopfSDEPolar}
\eea
where $W_r$ and $W_\theta$ are two {\mk Wiener processes satisfying the SDE system}
\beas
\d W_r &= \cos \theta \d W_x + \sin \theta \d W_y,\\
\d W_\theta &= -\sin \theta \d W_x + \cos \theta \d W_y,
\eeas
and with $r > 0$ and $\theta$ in $[-\pi, \pi]$, i.e. the largest domain on which the change of variables to polar coordinates is defined (and twice continuously differentiable).
The Kolmogorov equation in polar coordinates corresponding to the SDE~\eqref{eq:HopfSDEPolar} has a diffusion matrix
\begin{align*}
	D = \epsilon^2 \begin{pmatrix} 1 & 0 \\ 0 & 1 / r^2 \end{pmatrix},
\end{align*}
and is thus given by
\bea
\partial_t u
&= \left(\delta r - \kappa r^3
  + \frac{\epsilon^2}{2 r} \right) \partial_r u
+ \lr{\gamma - \beta r^2} \partial_\theta  u
+ \frac{\epsilon^2}{2} \partial^2_{rr} u
+ \frac{\epsilon^2}{2 r^2} \partial^2_{\theta\theta} u\\
&= \mathcal{K} u. \label{eq:HopfBKEPolar}
\eea
We refer to the second-order differential operator $\mathcal{K}$ {\mk of the} right-hand side of (\ref{eq:HopfBKEPolar}) as the {\it Kolmogorov operator} of the SHE; {\mk see also~\cite[{\mkr Eq.~(2.16)}]{Chekroun_al_RP2}.}
One {\mk observes} that the nonlinear drift term $\gamma - \beta r^2$ in {\mk the $\theta$-direction}
hinders the separation of the Kolmogorov equation~\eqref{eq:HopfBKEPolar} in $r$ and $\theta$.
However, this difficulty is partially overcome in the following Section~\ref{sec:isochrons} by the introduction of coordinates adapted to the geometry of the deterministic flow about the limit cycle.

Due to the rotational symmetry of the SHE~(\ref{eq:HopfSDEPolar}), a stationary density $\rho_\infty$\footnote{
  Recall  that a density $\rho$ is a stationary solution if $\mathcal{K}^* \rho = 0$, where $\mathcal{K}^*$ {\mk denotes the (formal)} adjoint of the Kolmogorov operator $\mathcal{K}$; {\mk see e.g.~\cite[Sect.~2]{Chekroun_al_RP2}.}
}
for the Fokker-Planck equation dual to the Kolmogorov equation~\eqref{eq:HopfBKEPolar} has to be independent of $\theta$.
On the other hand, the radial component of the drift is gradient with potential
\begin{align}
  U(r) = -\delta r^2 /2 + \kappa r^4 / 4 - \epsilon^2 \log{r}/2, \quad r > 0.
  \label{eq:potential_polar}
\end{align}
From the classical results relating the stationary density $\rho_\infty$ of a gradient SDE to its potential (see e.g.~\cite[Chap.~2.4]{Pavliotis2014}) one has, for $\epsilon > 0$ and any $\delta, \kappa, \gamma$ and $\beta$ in $\mathbb{R}$, that
\begin{align}
  \rho_\infty(r)
  &= \frac{N}{2\pi} e^{- \frac{2 U(r)}{\epsilon^2}}
    = \frac{N}{2 \pi} r e^{\frac{\delta}{\epsilon^2} r^2
    - \frac{\kappa}{2 \epsilon^2} r^4},
    \label{eq:statDist}
\end{align}
with $N = (\int_0^\infty e^{- \frac{2 U(r)}{\epsilon^2}} dr)^{-1}$ a normalization constant.
This density does not depend on the parameters $\gamma$ and $\beta$ defining the azimuthal component of the deterministic vector field.
As expected from the rotational symmetry, equal weights are given to any set of points on a circle when calculating long-term averages.

The following Proposition~\ref{prop:strong_mixing} ensures the existence of a unique invariant measure $\mu$ and the discreetness of the RP spectrum.
Its proof, below, is a straightforward application of e.g.~Theorem 4 recalled in~\cite{Chekroun_al_RP2}.
In the remaining, this invariant measure $\mu$ defines the space $L^2_\mu(\mathbb{R}^2)$ from which all functional-analytic results from the present work are derived.
\begin{prop}\label{prop:strong_mixing}
  For any $\delta, \kappa, \gamma$ and $\beta$ in $\mathbb{R}$, and $\epsilon > 0$, the Markov semigroup associated with the Kolmogorov equation~\eqref{eq:HopfBKECart}
  \begin{enumerate}
  \item has a unique invariant measure $\mu$ and it is strongly mixing,
  \item is compact on $L^2_\mu(\mathbb{R}^2)$,
  \item has a discrete (RP) spectrum of finite multiplicity on $L^2_\mu(\mathbb{R}^2)$.
  \end{enumerate}
\end{prop}
\begin{rmk}
  From Proposition~\eqref{prop:strong_mixing}, it follows that the unique invariant measure $\mu$ is necessarily associated with the stationary density $\rho_\infty$ given by~\eqref{eq:statDist}.
  This formula is given for completeness, but none of the results of this article rely on the knowledge of $\rho_\infty$.
\end{rmk}
\begin{rmk}
  The additional drift term $\epsilon^2 / (2 r)$ in~\eqref{eq:HopfSDEPolar} can be understood by visualizing a circle of radius $r$ centered at the origin in the $Oxy$ plane and figuring the impact of the noise on a state lying on this circle.
  On average, tangential perturbations will push the state away from the centre, with an intensity increasing with the noise level $\epsilon$ and with the curvature $1 / r^2$ of the circle.
\end{rmk}
\begin{rmk}
  One {\mkr observes} in the Kolmogorov equation~\eqref{eq:HopfBKEPolar} {\mkr written} in polar coordinates that the diffusion in the azimuthal direction is inversely proportional to the square of the radius $r$.
  Indeed, for larger $r$, the effect on the angle $\theta$ of a noisy perturbation on the Cartesian coordinates will be weaker.
\end{rmk}
\begin{proof}
  We first show that the conditions of Theorem 4 recalled in~\cite{Chekroun_al_RP2} are verified, that is, that the Markov semigroup associated with~\eqref{eq:HopfBKECart} is strong Feller and irreducible.
  To do so, we follow the approach recalled in~\cite[Appendix~A.2]{Chekroun_al_RP2}.
  
  The diffusion operator,
\begin{align*}
  D = \epsilon^2 (\partial^2_{xx} + \partial^2_{yy}),
\end{align*}
in Cartesian coordinates in the right-hand side of~\eqref{eq:HopfBKECart}, is \emph{uniformly elliptic}.
In other words, there exists $\alpha > 0$ such that,
\begin{align*}
  \left< \xi, D \xi \right> \ge \alpha \| \xi \|^2, \quad \forall \xi \in \mathbb{R}^2,
\end{align*}
so that the noise is nondegenerate and the strong Feller property holds by Weyl's lemma~\cite[Chap.~4]{Pavliotis2014}.

Moreover, $D$ is constant and the deterministic vector field $F$ given by \eqref{eq:HopfSDECart} in Cartesian coordinates is polynomial of degree 3.
The result by~\cite{Jurdjevic1985} then ensures that the associated control system
\begin{align}
	\begin{cases}
		\dot{x}(t) &= F_x(x, y) + \epsilon u_1(t) \\
		\dot{y}(t) &= F_y(x, y) + \epsilon u_2(t),
	\end{cases}
	\label{eq:HopfControl}
\end{align}
is controllable and the irreducibility of the Markov semigroup follows from the result by~\cite{Stroock1972a}.

{\mkr Thus, the invariant measure $\mu$ is unique and strongly mixing; see e.g.~\cite[Theorem 4]{Chekroun_al_RP2}.}

To prove the compactness of the Markov semigroup on $L^2_\mu(\mathbb{R}^2)$, we note that the potential $U$ in~\eqref{eq:potential_polar} can be written in Cartesian coordinates as a fourth-order polynomial with the appropriate growth conditions to apply Theorem~8.5.3 in~\cite{lorenzi2006analytical} and recalled in~\cite[Remark 2-(ii)]{Chekroun_al_RP2}.
In particular, $\lim_{|x| \to +\infty} |U'(x)| = +\infty$.
The Markov semigroup on $L^2_\mu(\mathbb{R}^2)$ generated by the generator associated with the Kolmogorov equation~\eqref{eq:HopfBKECart} is thus compact as long as $\epsilon > 0$.
The RP spectrum for~\eqref{eq:HopfBKECart} is then discrete and of finite multiplicity~\cite[Theorem~III.6.26]{Kato1995} for any values of the parameters $\delta, \kappa, \gamma$ and $\beta$ and for $\epsilon > 0$.
\end{proof}

\subsection{Isochrons and phase diffusion, for $\delta > 0$}
\label{sec:isochrons}

When the twist factor $\tilde \beta = \beta / \kappa$ in the SHE~\eqref{eq:HopfSDEPolar} is nonzero the evolution of the radial and azimuthal coordinates is coupled, resulting in a non-trivial response to perturbation of the system in the azimuthal direction.
We now identify the geometric structures, the isochrons, that help better understand this response when the underlying deterministic dynamics are hyperbolic about a limit cycle, i.e.~for $\delta > 0$ in our case.
While the relationship between isochrons and the phenomenon of phase diffusion {\mk has been discussed in previous works}~\cite{Winfree1974,Kuramoto1984,Djurhuus2009,Bonnin2014a},
the {\mk main contribution of} our approach is to relate these structures to the regularity of the Markov semigroup in Section~\ref{sec:StocAnaHopf}
and to the RP spectrum in Section~\ref{sec:smallNoiseExpansion},
thus giving a more detailed geometric understanding
of {\mk the stochastic Hopf bifurcation}.

{\mk The isochrons have been used in~\cite{Winfree1974} to study} chemical mixing in
perturbed periodic biochemical systems {\mk and} a new coordinate {\mk system} generalizing the notion of phase {\mk was introduced for that purpose namely} the \emph{asymptotic phase},
whose evolution by the autonomous flow is independent of the distance to the limit cycle.
{\mk This approach has been used also in}~\cite[Chap.~3-4]{Kuramoto1984}
to study the interaction of nonlinear oscillators and has recently been introduced to the engineering literature
by~\cite{Djurhuus2009} to study the response of nonlinear oscillators to forcing and the phenomenon of phase diffusion. Moreover, the important role of the twist of the isochrons regarding the stability of trajectories
measured by the Lyapunov exponents in periodically kicked limit cycles has been
shown in~\cite{Wang2003c,Lin2010} and corresponding results have been obtained
by~\cite{Engel2016} for stochastically driven limit cycles
(see also~\cite{Wieczorek2009}, for numerical results on coupled stochastic oscillators).
Perhaps the most relevant results to our study are, however, those of~\cite{Mauroy2012,Mauroy2013}
where it was shown that, in the deterministic autonomous case, the isochrons
coincide with isolines of phase of the RP eigenfunctions associated with purely imaginary eigenvalues.

The following definition and proposition from~\cite{Guckenheimer1975} and adapted to~\eqref{eq:HopfSDECart} are used {\mk in Section~\ref{sec:StocAnaHopf}} within the framework of stochastic analysis to show how and when the interaction of a stochastic forcing with the autonomous dynamics of the {\mk Hopf normal form results in mixing.}
Corresponding analytical formulas for the equation of phase derived in the next subsection~\ref{Sec_phase_BKE} are then applied in Section~\ref{sec:smallNoiseExpansion} to give a detailed description of mixing via {\mk small-noise} expansions of the RP spectrum.

\begin{defi}[Isochron~\cite{Guckenheimer1975}]
  With $\delta > 0$ and $\epsilon = 0$, let $\Gamma \subset \mathbb{R}^2$ be the hyperbolic limit cycle of the smooth flow {\mk $S_t$} generated by~\eqref{eq:HopfSDECart} on $\mathbb{R}^2$.
  The {\mk collection of sets} $I(p)$ such that
\begin{align}
	I(p) := \{ q \in \mathbb{R}^2: \lim_{t \to \infty} \|S_t q - S_t p\| = 0 \}, \label{eq:defIsochrons}
\end{align}
{\mk as $p$ varies on $\Gamma$}, are the isochrons of $\Gamma$.
\end{defi}
In other words, {\mk given a point $p$ on the limit cycle $\Gamma$, the set of points $I(p)$ of $\mathbb{R}^2$ is identified with all points that share the same phase asymptotically on the limit cycle.
{\mk More specifically,} each isochron $I(p)$ is by definition a leaf $W_{ss}(p)$ of the stable manifold $W_s(\Gamma) = \mathbb{R}^2 \setminus \{0\}$ of $\Gamma$.
As a matter of fact, the following result follows from the Invariant Manifold Theorem; see e.g.~\cite[Chap.~4]{Hirsch1977}.
\begin{prop}[\cite{Guckenheimer1975}]
\label{prop:isochron}
With $\delta > 0$ and $\epsilon = 0$, let $\Gamma$ be the hyperbolic limit cycle of period $T$ for the smooth flow $S_t$ generated by~\eqref{eq:HopfSDECart} on $\mathbb{R}^2$.
Then:
\begin{enumerate}
\item[{\mk (i)}] For each $p \in \Gamma$, there exists an one-dimensional isochron $I(p) = W_{ss}(p)$ transverse to $\Gamma$ at $p$
  (and \eqref{eq:defIsochrons} holds a fortiori).
\item[(ii)] The isochrons commute with the flow, i.e. $S_t I(p) = I(S_t p), p \in \Gamma, t \ge 0$.
\item[(iii)] The tangent map $D S_{T}$ at $p \in \Gamma$ leaves invariant
  the subspace tangent to the isochron $I(p)$.
\end{enumerate}
\end{prop}
These properties are key to understand the role played by the isochrons
{\mk to analyze the response of the dynamics to stochastic perturbations}.
The isochrons for the Hopf normal form are illustrated in Fig.~\ref{fig:artIsochrons} based on the calculations of the following Section~\ref{sec:asymptotic_phase}.
There,  the first property is used to associate a new coordinate $\phi$ playing the role of phase to all points on the same isochron.
The second property guaranties that the points of a same isochron are all mapped by the flow to another single isochron.
In particular, after one period $T$ of the limit cycle, all these points return to the same isochron, i.e. $S_T I(p) \subset I(p)$.
As a result, the evolution of their common phase $\phi$ by the flow
is independent of the transverse coordinates.
The last property relates the isochrons to the tangent map of the Poincar\'e map, as discussed in the following Section~\ref{sec:twistPerturbation}.

\subsubsection{Twist factor $\tilde \beta$ and response to perturbation}
\label{sec:twistPerturbation}
Applying the Floquet theory to~\eqref{eq:HopfSDECart} for $\delta > 0$ and $\epsilon = 0$, we first show how the twist of the isochrons in the neighborhood of the limit cycle $\Gamma$ is controlled by the twist factor $\tilde \beta$ and relates to the nonorthogonality of the eigenspaces {\mk associated with} the tangent map {\mk to} the flow of the Hopf normal form.

One {\mkr observes} from \eqref{eq:normalHopf} and \eqref{eq:HopfSDEPolar} that the evolution of the angular position $\theta$ is dependent on the radial position $r$, when the twist factor $\tilde \beta$ is nonzero.
This dependence impacts the response of the autonomous system \eqref{eq:normalHopf} to perturbations.
This can be understood from the Floquet representation of the fundamental matrix associated with the tangent map of the limit cycle of the deterministic vector field
\begin{align}
	F(p) =
	\begin{pmatrix}
		\delta r - \kappa r^3 \\
		\gamma - \beta r^2
	\end{pmatrix}.
	\label{eq:HopfField}
\end{align}
Here, care is taken not to include the drift term $\epsilon / (2r)$ as we focus on the deterministic dynamics.
The application of Floquet theory to the Hopf normal form (\ref{eq:normalHopf}) is reviewed in~\ref{sec:Floquet}.
In this case, the Floquet vectors coincide with the eigenvectors of the 
Jacobian matrix $J_\Gamma$ in polar coordinates
\begin{align}
	J_\Gamma(p) = 
	\begin{pmatrix}
		-2\delta & 0 \\
		-2 \beta R & 0
	\end{pmatrix},
	\label{eq:JacobianCycle}
\end{align}
for some point $p$ on $\Gamma$ and are rotated along the limit cycle.
The Jacobian matrix is diagonalizable with right eigenvectors
\begin{align}
	\vec{e}_1(p) =
	\begin{pmatrix}
		1 \\
		\frac{\tilde \beta}{R}
	\end{pmatrix}
	\quad \mathrm{and} \quad
	\vec{e}_2(p) =
	\begin{pmatrix}
		0 \\
		1
	\end{pmatrix},
	\label{eq:eigVecJacobian}
\end{align}
and left eigenvectors
\begin{align}
	\vec{f}_1(p) =
	\begin{pmatrix}
		1 \\
		0
	\end{pmatrix}
		\quad \mathrm{and} \quad
	\vec{f}_2(p) =
	\begin{pmatrix}
		-\frac{\tilde \beta}{R} \\
		1
	\end{pmatrix},
	\label{eq:eigVecAdjointJacobian}
\end{align}
respectively associated with the Floquet values
\begin{align}
	\alpha_1 =  -2 \delta \quad \mathrm{and} \quad \alpha_2 = 0.
	\label{eq:eigValJacobian}
\end{align}
For all $p$ in $\Gamma$, the Floquet vector $\vec{e}_2(p)$ is tangent to $\Gamma$, in the direction of the flow, while $\vec{e}_1(p)$ is transverse to it.
The latter is tangent to the isochron $I(p)$ and is associated with the stability of $\Gamma$ to small perturbations; see Sect.~\ref{sec:asymptotic_phase} and Appendix~\ref{sec:Floquet}.
It follows that
\begin{align*}
  \frac{\left< \vec{e}_1(p), \vec{e}_2(p) \right>}{\|\vec{e}_1(p)\| \|\vec{e}_2(p)\|}
  &= \frac{\tilde \beta}{\sqrt{\tilde \beta^2 + R^2}}
    \ne 0 \enskip \mathrm{if} \enskip \beta \ne 0,
\end{align*}
where $\left< v, w \right>$ is the inner product in $\mathbb{R}^2$ and $\|v\|$ the induced norm.

As a result, when the twist factor is nonzero, the eigenvectors of $J_\Gamma$ are not orthogonal under the inner product $\left<\cdot, \cdot \right>$ and the Jacobian matrix is nonnormal by definition; {\mk see e.g.~\cite[Chap.~I.2]{Trefethen2005}.}
It is known that the nonnormality of a linear evolution operator is associated with a nontrivial response of the system to forcing.
In the particular situation considered here, the stochastic forcing is responsible for perturbing trajectories away from $\Gamma$, making these trajectories vulnerable to the effect of the twist factor $\tilde \beta$.
It is then crucial to take into account the dependence of the angular frequency on the radius, as controlled by $\tilde \beta$.

\subsubsection{Asymptotic phase and isochrons}\label{sec:asymptotic_phase}

In the case of the Hopf normal form \eqref{eq:normalHopf} considered here, with $\delta > 0$, explicit formulas for the isochrons can be obtained, allowing us to analyse the twist of the isochrons away from the limit cycle and to derive a phase diffusion equation in the stochastic case.

From Proposition~\ref{prop:isochron} and the definition of a foliation~\cite[Chap.~6]{Spivak1999b}, there exists a coordinate system $(\nu, \phi)$ on $\mathbb{R}^2 \setminus \{0\}$ such that
\begin{align*}
	\phi(q) = \text{constant} = \phi(p) \quad \text{for any point} \enskip q \in I(p), p \in \Gamma.
\end{align*}
In other words, the coordinate $\phi$ is the same for all points of a same isochron.
On the other hand, thanks to the transversality of the isochrons to $\Gamma$ ({\mk Proposition~\ref{prop:isochron}-(i)}),
one can choose $\phi$ such that for all $q \in I(p)$, $\phi(q) = \theta(p)$, where $\theta$ is the angle coordinate of the unique point $p$ at the intersection of $I(p)$ with $\Gamma$.
In addition, the second coordinate $\nu$ in the direction transverse to $\Gamma$ can simply be chosen {\mk as} the radius $r$.
Last, from the invariance of the stable foliation by the flow (proposition (\ref{prop:isochron}).2), one has that
\begin{align*}
	\frac{d \phi}{dt}(p) = \omega_f, \quad p \in \mathbb{R}^2 \setminus \{0\}.
\end{align*}
In other words, as opposed to $\theta$ in (\ref{eq:normalHopf}),
the evolution of the coordinate $\phi$ by the autonomous flow does not depend on the radius.
These properties thus make $\phi$ {\mk a} perfect candidate for playing
the role of \emph{phase} for points $q$ not necessarily on $\Gamma$.

To define the change of coordinates from $(r, \theta) \to (r, \phi)$ for any point $q \in \mathbb{R}^2 \setminus \{0\}$ explicitly,
one can use the rotational symmetry of the vector field (\ref{eq:HopfField}),
\begin{align*}
	F(r, \theta_1) = F(r, \theta_2), \quad \text{for any} \quad \theta_1, \theta_2 \in [0, 2\pi], \enskip r > 0,
\end{align*}
to look for a constraint~\cite[Chap.~1.5]{Fecko2006} of the type
\begin{align*}
	\phi(q) = \theta + f(r).
\end{align*}
Differentiating with respect to time, one finds that
\begin{align*}
  \frac{\d \phi}{\d t}
  = \omega_f
  = \frac{\d \theta}{\d t} + \frac{\d f}{\d r} \frac{\d r}{\d t},
\end{align*}
Considering autonomous trajectories governed by the normal form (\ref{eq:normalHopf})
with vector field $F$ given in (\ref{eq:HopfField}), one finds that
\begin{align}
	\frac{\d f}{\d r} = -\frac{\tilde \beta}{r},
	\label{eq:dphidr}
\end{align}
which does not depend on $\delta$.
Finally, integrating with the condition that the phase coincides with the angle on the limit cycle,
i.e.~that $\phi(p) = \theta(p)$ for $p \in \Gamma$, gives
\begin{align}
  \phi
  &= \theta - \tilde \beta \log{\frac{r}{R}}
    \quad \mathrm{for~} \delta > 0. \label{eq:phaseIsochron}
\end{align}
We have hence defined a new coordinate system $(r, \phi)$, such that all points on the same isochron
have the same asymptotic phase $\phi$.
One {\mkr notes} from the constraint \eqref{eq:phaseIsochron}, implicitly defining the isochrons, that, while the latter are rectilinear for vanishing $\tilde \beta$,
they undergo a nonlinear twist when $\tilde \beta \ne 0$.
Moreover, in agreement with {\mk Proposition~\ref{prop:isochron}-(iii)},
one can verify that the eigenvector $\vec{e}_1 = (1, \tilde \beta / R)$ of the polar Jacobian $J_\Gamma$ is tangent to the isochron $I(S_t p)$.
Thus, the nonorthogonality of the eigenvectors of the polar Jacobian $J_\Gamma$ is
directly associated with the twist of the isochrons.

These results are illustrated in Fig.~\ref{fig:artIsochrons},
for the particular case of the Hopf normal form (\ref{eq:normalHopf})
considered here, with $\delta > 0$ and $\beta > 0$.
The limit cycle $\Gamma$ is the circle of radius $R$ represented by a thin black line.
Three different isochrons $I^\beta$ are represented in red.
The first one is transverse to $\Gamma$ at $p$, while the other two are transverse to $\Gamma$
at the images of $p$ by the flow at times $t_1  = T / 3$ and $t_2 = 2 T / 3$.
{\mk Each of these points is} marked by a black dot and, from the invariance of $\Gamma$, are also on $\Gamma$.
In addition, two trajectories starting from distinct points on the isochron $I^\beta(p)$ are represented
by a dashed line. Their states at $t_1$ and $t_2$ are also marked by black dots
and, since the isochrons commute with the flow {\mk Proposition~\ref{prop:isochron}-(ii)},
they also belong to the isochrons $I(S_{t_1} p)$ and $I(S_{t_2} p)$, respectively,
and share the same asymptotic phase $\phi$ given by (\ref{eq:phaseIsochron}).
Moreover, from the stability of the foliation,
the distance between the trajectories vanishes as time approaches infinity.
To see the effect of the twist factor $\tilde \beta$ on the isochrons, the isochron $I^0(p)$, for $\beta = 0$, is represented in blue.
In agreement with (\ref{eq:phaseIsochron}), $I^0(p)$ is rectilinear, while $I^\beta(p)$ is twisted due to the shear in the angular velocities when $\beta \ne 0$.
It follows that the eigenvector $\vec{e}^\beta_1$ of the polar Jacobian at $p$, tangent to the isochron $I^\beta(p)$, is not orthogonal to the eigenvector $\vec{e}_2$ tangent to $\Gamma$, when $\beta \ne 0$.
\begin{figure}
	\centering
	\includegraphics[width=.6\textwidth,height=.6\textwidth]{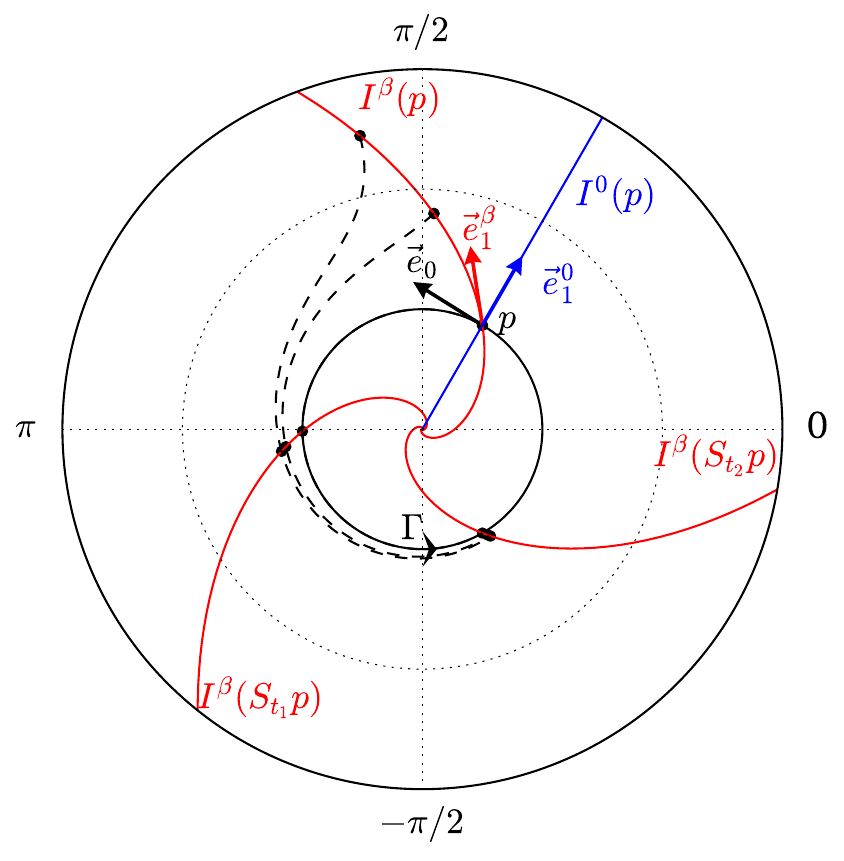}
	\caption{
          \footnotesize Limit cycle $\Gamma$ (thin black line) and its isochrons (thin red lines) for $\delta > 0$, $\kappa > 0$ and $\beta > 0$ at $p$ and at the image of $p$ by the flow at times $t_1 = T / 3$ and $t_2 = 2 T / 3$.
          Trajectories starting from three points on $I^\beta(p)$, are represented by a dashed line.
          One of them belongs to $\Gamma$ and overlaps the thick black line representing it.
          The states of the trajectories at times $0$, $t_1$ and $t_2$ are represented by black dots.
          The isochron at $p$ for vanishing $\beta$ is also represented in blue.
          Finally, the vectors $\vec{e}_1$ associated with the characteristic exponent $-2 \delta$ and tangent to the isochrons at $p$ are also plotted for $\beta \ne 0$ and $\beta = 0$.}\label{fig:artIsochrons}
\end{figure}
\subsubsection{Phase diffusion equation}\label{Sec_phase_BKE}
After introducing the change of variable $(r, \theta) \to (r, \phi)$ according to (\ref{eq:phaseIsochron})
and such that $d \phi / dt = \omega_f$ for a deterministic trajectory of the normal form (\ref{eq:normalHopf}),
one can now apply It\^o's formula~\cite[Theorem~5.1]{Ikeda1989}
to derive the SDE corresponding to \eqref{eq:HopfSDEPolar} in coordinates $(r, \phi)$.
Hence, one finds the following {\mk phase diffusion equation}
\bea
\d r &= (\delta r - \kappa r^3 + \frac{\epsilon^2}{2 r}) \d t + \epsilon \d W_r  \\
\d \phi	&= \omega_f \d t + \epsilon\frac{\d W_\theta}{r} - \tilde \beta \epsilon \frac{\d W_r}{r}. \label{eq:SDEIsochron}
\eea
As expected, the $\phi$-component of the drift is now independent of the radius, as in the deterministic case.
This is, however, to the expense of the statistical dependence of the noise terms acting on $r$ and $\phi$.
The phase $\phi$ thus experiences advection at constant angular velocity $\omega_f$
together with nonuniform diffusion.
As a consequence, the Kolmogorov equation~\eqref{eq:HopfBKEPolar} with $u'(r, \phi, t) = u(r, \theta, t)$ becomes
\be
	\partial_t u'
= \left(\delta r - r^3 + \frac{\epsilon^2}{2 r} \right) \partial_r u' + \omega_f \partial_\phi u'
	+ \frac{\epsilon^2}{2} \partial^2_{rr} u' - \frac{\tilde \beta \epsilon^2 }{r} \partial^2_{r \phi} u' + \frac{\epsilon^2 (1 + \tilde \beta^2)}{2 r^2} \partial^2_{\phi \phi} u'.
	\label{eq:BKEIsochron}
\ee
{\mk Compared to~\eqref{eq:HopfBKEPolar},} the coefficients
in front of the first-order differential operators associated with the drift in~\eqref{eq:BKEIsochron} 
are now separated {\mk in their arguments, here in their $r$- and $\phi$-dependences}.
This {\mk feature} is key to the derivation of small-noise expansion of the RP spectrum in Section~\ref{sec:smallNoiseExpansion}.
As a by-product, however, the dependence of the angular frequency on the radius for $\beta \ne 0$ is responsible for an effective increase of the phase diffusion by a factor $1 + \tilde \beta^2$; {\mk cf.~the coefficient in front of $\partial^2_{\theta\theta}$ in~\eqref{eq:HopfBKEPolar}}.
This effect could have been anticipated from the nonnormality of the
polar Jacobian $J_\Gamma$
and is explained in greater detail in light of the H\"ormander theorem in Section~\ref{sec:StocAnaHopf} below.
\begin{rmk}
Equations of the type~\eqref{eq:SDEIsochron} and~\eqref{eq:BKEIsochron}
for more general systems with an adiabatic phase reduction have recently received much attention for the study 
of the impact of noise on nonlinear oscillators in physics and engineering; see e.g.~\cite{Demir2000,Djurhuus2009,Bonnin2013,Bonnin2014a}.
\end{rmk}

\tikzset{
    >=stealth',
    punkt/.style={
           rectangle,
           rounded corners,
           draw=black, thick,
           text width=0.35\textwidth,
           minimum height=2em,
           text centered},
    pil/.style={
           ->,
           thick,
           shorten <=2pt,
           shorten >=2pt,}
}
\section{Analysis of the stochastic Hopf bifurcation}\label{sec:StocAnaHopf}
 
In this section, we apply the stochastic analysis approach {\mkr as (briefly) surveyed in}~\cite[Appendix~A]{Chekroun_al_RP2}, to study the general properties of the Markov semigroup of the SHE~\eqref{eq:HopfSDECart} and its spectrum for any values of the parameters $\delta, \gamma$ and $\beta$, and for $\kappa > 0$.
{\mkr The material presented in this section is mostly known by the expert working on the ergodic theory of stochastic systems but contains also useful insights about} the role played by the {\mkr the geometric structures organizing the underlying dynamics and their interactions with the noise}.
{\mkr In that respect, Theorem~\ref{thm:isoHypo} provides interesting relationships between the isochrons of Sec.~\ref{sec:isochrons} and the violation of the H\"ormander condition, positioning thus the material exposed hereafter to be also useful for the expert, while having in mind a wider audience in the geosciences and macroscopic physics.}

We start by showing in Sec~\ref{sec:FellerIrreducibility} {\mkr below  how the existence of a unique ergodic and smooth invariant measure to SHE~\eqref{eq:HopfSDECart} as well as its mixing properties, 
relate directly to} the configuration of the stochastic forcing with respect to the isochrons.
The existence of a spectral gap at the bifurcation and the exponential decay of correlations is then {\mk proved} in Section~\ref{sec:discreteSpectrum}.

\subsection{Smoothing and mixing by the noise: a geometric perspective}
\label{sec:FellerIrreducibility}

We have {\mkr discussed} in Section~\ref{sec:normalForm} {\mkr how} the tilt of the isochrons, as measured by the twist factor $\tilde \beta$, is associated with an increase of the diffusion coefficient in the phase in the Kolmogorov equation~\eqref{eq:BKEIsochron} by a factor $1 + \tilde \beta^2$.
This simple result shows the importance of the underlying geometry of the drift and diffusion operators in the study of the ergodic properties of continuous Markov processes.
The novel approach which is followed in this section is to place the isochrons in the context of stochastic analysis and to show in Theorem~\ref{thm:isoHypo} that, for fairly general nonlinear oscillators with diffusion, the smoothing and mixing effects of the noise may critically depend on the interaction of the stochastic forcing fields with the isochrons.

Recall that, according to Doob's theorem~\cite{Doob1948}, the existence of {\mk at most one ergodic} invariant measure with a smooth Lebesgue density for a continuous Markov process {\mkr is a consequence of} the \emph{regularity} of the Markov semigroup $P_t$; see e.g.~\cite[Chap.~4]{DaPrato1996}.
A result, due to~\cite{KhasMinskii1960}, shows that the regularity of the Markov semigroup {\mkr is in turn ensured} from the \emph{irreducibility} and the \emph{strong Feller} property of the Markov semigroup. 
The {\mkr irreducibility and strong Feller properties follow from} the \emph{controllability} of the corresponding control system~\cite{Stroock1972a} the (hypo-)ellipticity of the operators, respectively.

This well-known approach is used in Proposition~\ref{prop:eigen_below} to show that the measure $\mu$ with density $\rho_\infty$ given by~\eqref{eq:statDist} is the unique invariant measure of the SHE~\eqref{eq:HopfSDECart}.
It is {\mkr recalled} in~\cite[Appendix~A.2]{Chekroun_al_RP2} {\mkr for the unfamiliar reader that along with \cite[Theorem 4]{Chekroun_al_RP2}}  relating the smoothness and the strong mixing property of the invariant measure to the strong Feller and irreducibility properties of Markov semigroup.
This approach is summarized {\mk here} by the diagram {\mk shown in} Fig.~\ref{fig:FellerIrreducibility}.
\begin{figure}
\begin{tikzpicture}[scale=.85, transform shape]
    \node[punkt] (mu){{\mk Existence and smoothness of at most one ergodic measure} $\mu$} ;
    \node[punkt, below=of mu] (regularity) {Regularity of $P_t$: Doob theorem}
       edge[pil] (mu.south);
    \node[below=of regularity] (dummy) {};
    \node[punkt, left=of dummy] (irreducibility) {Irreducibility of $P_t$}
       edge[pil] (regularity.south);
    \node[punkt, below=of irreducibility] (control) {Controllability: Stroock-Varadhan support theorem}
       edge[pil] (irreducibility.south);
    \node[below=of control] (dummy2) {};
    \node[punkt, left=of dummy2] (poly) {Control of polynomial systems}
       edge[pil] (control.south);
    \node[punkt, right=of dummy2, text width=0.05\textwidth] (else) {...}
       edge[pil] (control.south);
    \node[punkt, right=of dummy] (feller) {$P_t$ is Strong Feller}
       edge[pil] (regularity.south);
    \node[punkt, below=of feller] (ellipticity) {Hypoellipticity of the generator of $P_t$}
       edge[pil] (feller.south);
    \node[punkt, below=of ellipticity] (parabolic) {H\"ormander bracket condition}
       edge[pil] (ellipticity.south);
\end{tikzpicture}
\caption{Schematic of the strong Feller-irreducibility approach to prove the existence and uniqueness
of a smooth invariant measure for a continuous Markov process.}
\label{fig:FellerIrreducibility}
\end{figure}
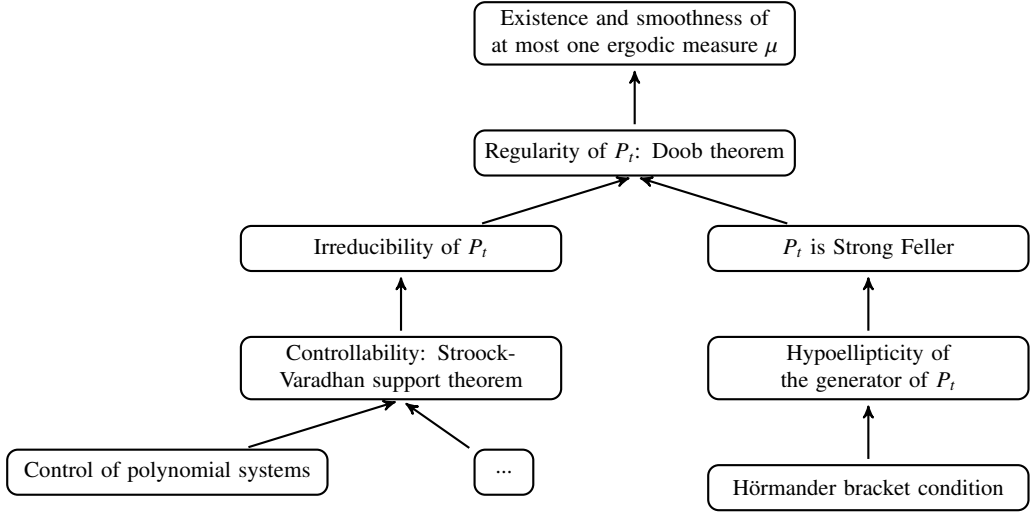

Yet, the ellipticity of the Kolmogorov operator~\eqref{eq:HopfBKECart} stems from the fact that noise is added to both coordinates of the two-dimensional SHE~\eqref{eq:HopfSDECart}.
To reveal the role played by the isochrons from a stochastic analysis perspective, we consider next degenerate cases in which noise is not added to both coordinates and study under which conditions the corresponding  Markov semigroup is still strongly Feller and irreducible, and thus has a smooth density.}

To do so, we rely on the H\"ormander theorem for hypoelliptic operators~\cite{Hormander1968a}.
For further reference, we recall the H\"ormander's bracket condition for an
SDE on $\mathbb{R}^N$ {\mk written in its Stratonovich interpretation for $m$ independent 1D Wiener process $W_i$,}
\begin{align}
	\d x =V_0(x)dt + \sum_{i = 1}^m V_i(x) \circ \d W_i.
	\label{eq:SDEFields}
\end{align}
{\mk One defines the following} collection of vector fields $\mathcal{V}_k$ by 
\begin{align}
	\mathcal{V}_0 = \{V_i : 1\leq  i \le m\}, \quad
	\mathcal{V}_{k+1} = \mathcal{V}_k \cup \{[U,V_j ]: U \in \mathcal{V}_k \enskip \& \enskip 0 \le j \le m \}.
	\label{eq:generatorLie}
\end{align}
The main assumption {\mkr to be checked for application of the H\"ormander theorem is then the following
 \emph{H\"ormander bracket condition}}
\begin{align}
	\cup_{k \ge 1} \Span\{V(q): V \in \mathcal{V}_k\} = \mathbb{R}^N \quad \text{for every} \quad q \in \mathbb{R}^N,
	\label{eq:HCondition}
\end{align}
\subsubsection{The case of the SHE~\eqref{eq:HopfSDECart} with degenerate noise}

{\mk Let us consider the following modification of the SHE~\eqref{eq:HopfSDEPolar} written in Stratonovich form},
\begin{align}
	\d X = V_0(X) \d t + V_1(X) \circ \d W_1.
	\label{eq:fieldSDE}
\end{align}
{\mk In~\eqref{eq:fieldSDE}, $V_0$ denotes the deterministic vector field in~\eqref{eq:HopfSDEPolar}.
  However, whereas the original SHE~\eqref{eq:HopfSDEPolar} is driven by two one-dimensional Wiener processes $W_r$ and $W_\theta$,~\eqref{eq:fieldSDE} is driven by a single one-dimensional Wiener process $W_1$ with an arbitrary smooth vector field $V_1$ of $\mathbb{R}^2$.}


{\mk Using the coordinate-free formalism (see Remark~\ref{coco_free} below), the Kolmogorov operator $\mathcal{K}$ associated with~\eqref{eq:fieldSDE} can be written as}
\begin{align*}
	\mathcal{K}  = V_0 + (V_1)^2.
\end{align*}
Here, as opposed to the original SHE~\eqref{eq:HopfSDECart} we have chosen $V_1$ to be nonconstant {\mk and to be multiplied by a one-dimensional} Wiener process, {\mk only}.
Thus, at each point $q$ in $\mathbb{R}^2$, the vector $V_1(q)$ {\mk alone} cannot span $\mathbb{R}^2$ and the Kolmogorov operator $\mathcal{K}$ is no longer elliptic.
{\mk It may turn out, however, that the operator is hypoelliptic, ensuring, roughly speaking, to have the noise to propagate out in the whole space; see next subsection.}
Our aim is then to check under which condition on $V_1$ the Kolmogorov operator $\mathcal{K}$ is hypoelliptic.
For that purpose we need to verify under which conditions the H\"ormander condition~\eqref{eq:HCondition} holds.

We thus calculate the Lie bracket of $V_1$ with $V_0$.
The vector fields $V_0$ and $V_1$ are given in polar coordinates by
\begin{align*}
	V_0(r, \theta) &= F(r, \theta)
	= (\delta r - \kappa r^3)~\partial_r + (\gamma - \beta r^2)~\partial_\theta \\
	V_1(r, \theta) &= 
	V_1^r(r, \theta)~\partial_r + V_1^\theta(r, \theta)~\partial_\theta,
\end{align*}
where $V_1^r$ and $V_1^\theta$ are the (smooth) components of $V_1$.
%
%
Let us first consider the simple yet instructive case when
\begin{align*}
	V_1^r = \epsilon_r, \quad \mathrm{and} \quad V_1^\theta = 0,
\end{align*}
for some constant $\epsilon_r > 0$.
Then the Lie bracket $[V_0, V_1]$ yields
\begin{align}
	[V_0, V_1] = -\epsilon_r (\delta - 3 \kappa r^2)~\partial_r - 2 \epsilon_r \beta r~\partial_\theta.
	\label{eq:LieBracketRadial}
\end{align}
{\mkr Observe} that $\Span \mathcal{V}_1 = \Span \{V_1, [V_0, V_1]\} = \mathbb{R}^2$
if and only if $\beta$, or equivalently the twist factor $\tilde \beta$, is nonzero.
This is also true when further iterating the Lie brackets.

Thus, even in the case of a purely radial stochastic forcing,
the twist of the isochron controlled by $\tilde \beta$ allows for the noise to be injected in the azimuthal direction and for the Markov semigroup to be strongly Feller.
This also explains the increase {\mk by a factor $1 + \tilde \beta^2$} of the diffusion coefficient in the Kolmogorov equation~\eqref{eq:BKEIsochron} {\mk written in the phase-coordinate, compared to that of~\eqref{eq:HopfBKEPolar} written in polar coordinates. It will have also} important consequences on the RP eigenfunctions in Section~\ref{sec:smallNoiseExpansion}.
On the other hand, if $\beta = 0$ and $V_1$ is radial, the noise is not felt in the azimuthal direction and no phase diffusion may occur.

This result is illustrated in Fig.~\ref{fig:isochronlie} for the SHE~\eqref{eq:HopfSDECart} with $\delta = \kappa = 1$ and $\beta = 0$ (upper panels)
and $\beta = 0.8$ (lower panels).
On the left panels, the Lie bracket $[V_0, V_1]$, for $V_1 = \epsilon_r~\partial_r$ (blue vector), is applied to a point $p$.
There, $S_t$ and $S_t^{V_1}$ are the flows generated by $V_0$ and $V_1$, respectively,
and $\delta t$ is a small time.
The Lie bracket (red vector {\mk in Fig.~\ref{fig:isochronlie}}) is given by the tangent vector to the curve obtained
by successively applying $S_{\delta t}$ and $S_{\delta t}^{V_1}$ forward
and then backward in the limit when $\delta t \to 0$.
On the right panels, samples of simulated time series of the asymptotic phase
$\phi$ given by (\ref{eq:phaseIsochron}) are represented.
One {\mkr observes} that when $\beta = 0$ (upper panels {\mk of Fig.~\ref{fig:isochronlie}}),
the integral curves of the forcing field $V_1$ (dashed blue lines {\mk in Fig.~\ref{fig:isochronlie}})
coincide with the isochrons (red lines {\mk in Fig.~\ref{fig:isochronlie}}) and the resulting Lie bracket
is collinear to $V_1$, in agreement with (\ref{eq:LieBracketRadial}).
As a result, no phase diffusion is observed on the corresponding upper right panel {\mk of Fig.~\ref{fig:isochronlie}}.

On the other hand, when $\beta$ is nonzero (lower panels), the forcing field $V_1$ is not tangent to the isochrons anymore  and the resulting Lie bracket is not collinear to $V_1$.
This allows for the noise to be injected in the azimuthal direction,
as can be seen from the phase diffusion occurring in the lower right panel {\mk of Fig.~\ref{fig:isochronlie}}.
This figure reveals that the dependence of the Lie bracket (\ref{eq:LieBracketRadial}) on the twist factor $\tilde \beta$ is directly related to the orientation of the forcing field $V_1$ with respect to the isochrons.
This observation will now be made rigorous for the more general case of a dynamical system with a hyperbolic limit cycle.

\begin{rmk}\label{coco_free}
	{\mk In the coordinate-free framework of differential geometry, a vector field $V$ defined on the plane $\mathbb{R}^2$ and decomposed in Cartesian coordinates as $V(x, y) = V^1(x, y) \vec{e}_1 + V^2(x, y) \vec{e}_2$,  is identified (by isomorphism) with} the first-order differential operator 
	\bes
	V = V^1(x, y)~\partial_x + V^2(x, y)~\partial_y.
	\ees
	See e.g.~\cite{Fecko2006} for an introduction to differential geometry.
\end{rmk}
\subsubsection{H\"ormander bracket condition for a general hyperbolic limit cycle}

For more generality, let us consider a dynamical system with flow $S_t, t \ge 0$,
generated by the smooth vector field $V_0$
on the $N$-dimensional Euclidean space $\mathbb{R}^N$.
Assume that the flow has a hyperbolic limit cycle $\Gamma$
with basin of attraction $U_\Gamma \subseteq \mathbb{R}^N$,
so that the isochrons $W_{ss}(p)$ at any point $p$ on $\Gamma$ can be defined as the stable foliation of $\Gamma$; see Section~\ref{sec:normalForm}.
Consider then the SDE \eqref{eq:SDEFields} {\mk in which the} deterministic field $V_0$
is perturbed by $m$ smooth vector fields $\{V_i, {\mk 1 \leq i \le m}\}$ {\mk each multiplied  by independent one-dimensional} Wiener processes.
We would like to know when the interaction of this stochastic forcing with the isochrons
allows for the parabolic H\"ormander condition~\eqref{eq:HCondition} in $U_\Gamma$ to be fulfilled.
The following Theorem~\ref{thm:isoHypo} is proved in Appendix~\ref{sec:proofIso}, as a direct consequence of the definition of the Lie derivative in terms of pullback {\mk of a vector field by a diffeomorphism}.
\begin{thm}
\label{thm:isoHypo}
If, for some point $q$ in $U_\Gamma$, the vector fields $\{V_i, {\mk 1 \leq i \le m}\}$ of the stochastic forcing in~\eqref{eq:fieldSDE} are all tangent to the isochron $W_{ss}(p)$ passing through $q$, then the {\mk vector} space 
\bes
\cup_{k \ge 1} \Span\{V(q): V \in \mathcal{V}_k\},
\ees
generated by the vector fields $V_0$ and $\{V_i, {\mk 1 \leq i \le m}\}$ according to~\eqref{eq:generatorLie} is also tangent at $q$ to the isochron $W_{ss}(p)$.
\end{thm}
Keeping the same notations, the contraposition of Theorem~\ref{thm:isoHypo} yields the following corollary.
\begin{coro}\label{coro:isoHypo}
  For the parabolic H\"ormander condition~\eqref{eq:HCondition} to be fulfilled, it is necessary that, for each point $q$ in $U_\Gamma$, at least one of the vector fields in $\{V_i, {\mk 1 \leq i \le m}\}$ is transverse to the isochron passing through this point.
\end{coro}

\begin{figure}
	\centering
	\begin{subfigure}{0.62\textwidth}
		\includegraphics[width=\textwidth]{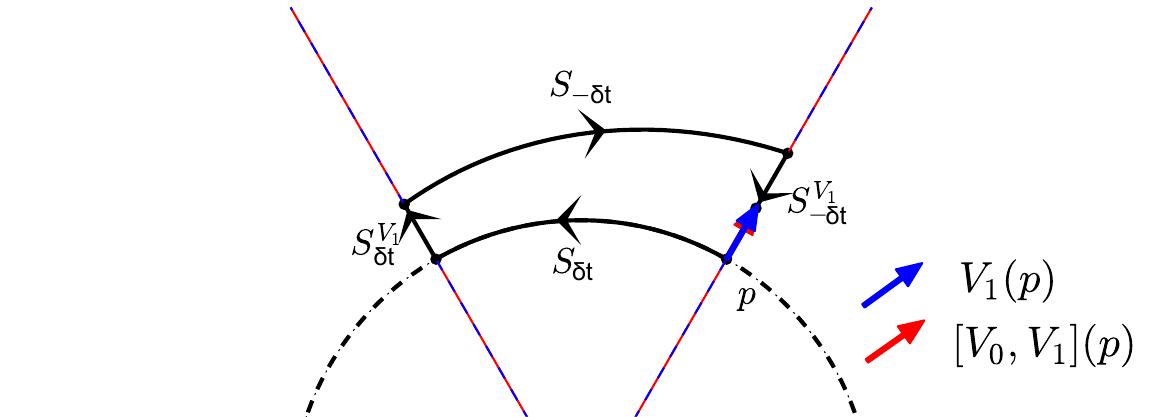}
	\end{subfigure}
	\begin{subfigure}{0.36\textwidth}
		\includegraphics[width=\textwidth]{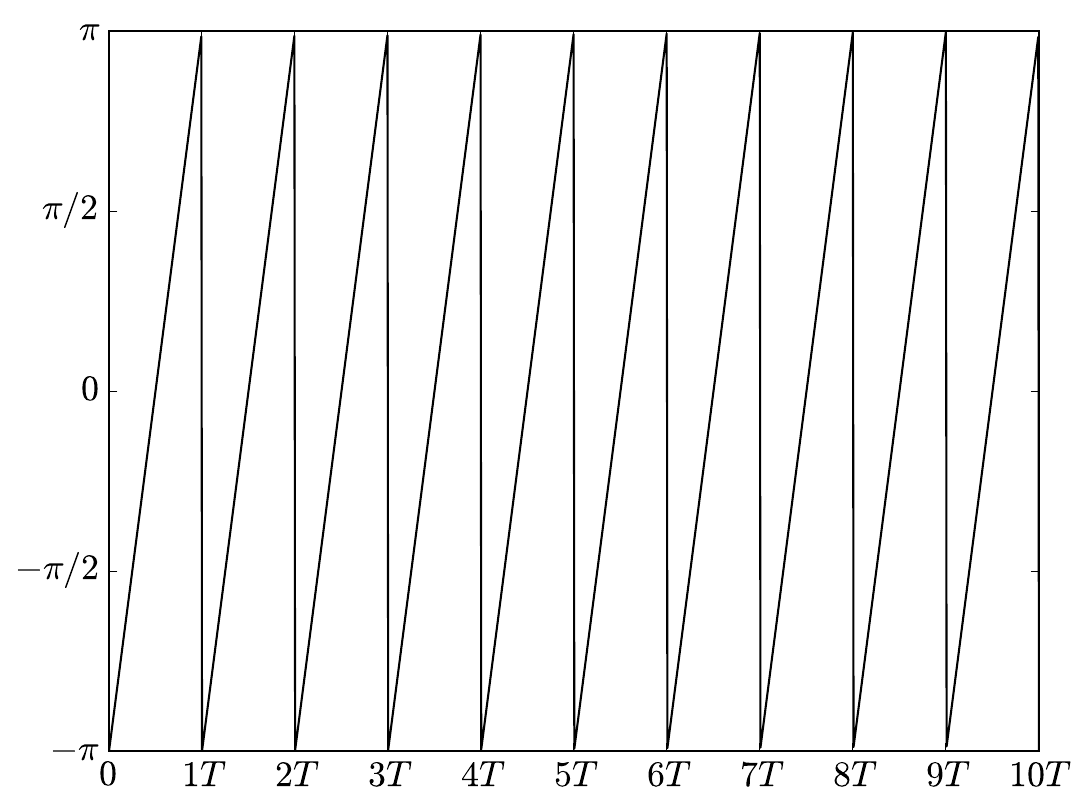}
	\end{subfigure}\\
	\begin{subfigure}{0.62\textwidth}
		\includegraphics[width=\textwidth]{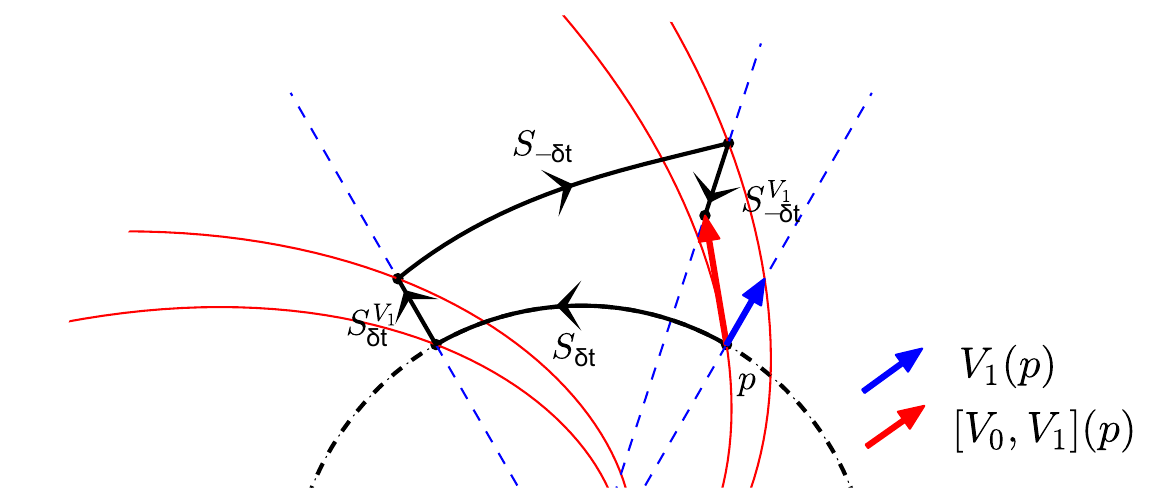}
	\end{subfigure}
	\begin{subfigure}{0.36\textwidth}
		\includegraphics[width=\textwidth]{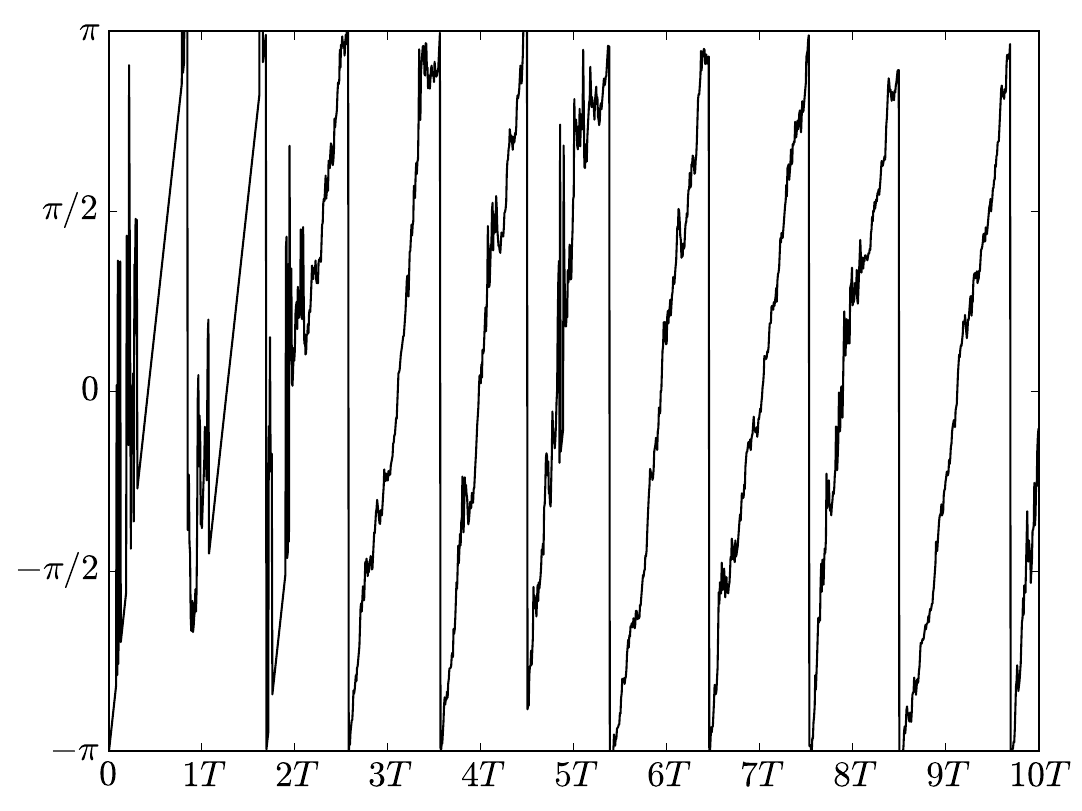}
	\end{subfigure}
	\caption{{\bf Left}: Illustration of the action of the Lie bracket $[V_0, V_1]$ (red arrow) at a point $p$
	between the vector field $V_0$ generating the autonomous flow $S_t, t \ge 0$
	with the forcing field $V_1$ (blue arrow at $p$) generating the flow $S_t^{V_1}, t \ge 0$
	in the radial direction for $\beta = 0$ (upper panels) and $\beta \ne 0$ (lower panels).
	The thick black curve represents the composition of the two flows for a short time ${\delta t}$ and then back.
	The isochrons passing through the different images of the initial point $p$
	by the flow are represented by a thin red line, while
	the integral curves of the forcing field $V_1$ are represented by dashed blue lines.
	{\bf Right}: Sample time series of the phase $\phi$ corresponding to the left panels
	when the forcing field $V_1$ acts on a Wiener process as in the SDE (\ref{eq:fieldSDE}).
	}
	\label{fig:isochronlie}
\end{figure}

The dependence of the Lie bracket $[V_0, V_1]$ on the twist factor $\tilde \beta$ in Fig.~\ref{fig:isochronlie} is now understood thanks to Theorem~\ref{thm:isoHypo} in terms of orientation of the forcing vector field $V_1$ with respect to the isochrons.
There, $V_1$ acts on the radial direction only.
For $\beta = 0$ (upper panels), the isochrons are rectilinear
and coincide with the integral curves of $V_1$.
In agreement with Theorem~\ref{thm:isoHypo},
the Lie bracket $[V_0, V_1]$ is also tangent to the isochrons.
Thus, 
\bes
\cup_{k \ge 1} \Span\{V(q): V \in \mathcal{V}_k\} = TW_{ss}(p) \ne \mathbb{R}^2,
\ees
and the Kolmogorov operator $\mathcal{K}$ is not hypoelliptic,
which explains the absence of phase diffusion on the upper right panel.
For $\beta \ne 0$ (lower panels), however,
the stochastic field $V_1$ is not tangent to the isochrons anymore.
As a consequence and in agreement with (\ref{eq:LieBracketRadial}),
the Lie bracket $[V_0, V_1]$ is able to span the azimuthal direction,
so that the H\"ormander condition (\ref{eq:HCondition}) is fulfilled.
It follows that the Kolmogorov operator $\mathcal{K}$ is hypoelliptic, by H\"ormander's theorem,
which is {\mk manifested} by the occurrence of phase diffusion in the lower right panel {\mk of Fig.~\ref{fig:isochronlie}.}
\subsection{Spectral gap property of the SHE~\eqref{eq:HopfSDECart}}
\label{sec:discreteSpectrum}

We now turn to the spectral properties of the Markov semigroup of the SHE~\eqref{eq:HopfSDECart} and to the nature of the decay of correlations depending on the control parameter $\delta$ and $\kappa$ and for $\epsilon > 0$.
In this case, recall that the diffusion operator,
\begin{align*}
D = \epsilon^2 (\partial^2_{xx} + \partial^2_{yy}),
\end{align*}
in Cartesian coordinates in the right-hand side of~\eqref{eq:HopfBKECart}, is \emph{uniformly elliptic}, in the sense that there exists $\alpha > 0$ such that,
\begin{align*}
	\left< \xi, D \xi \right> \ge \alpha \| \xi \|^2, \quad \forall \xi \in \mathbb{R}^2.
\end{align*}

In addition exponential decay of correlation is expected below the bifurcation point, for $\delta < 0$, since, for the deterministic case, the RP spectrum in spaces of distributions has a spectral gap~\cite{Gaspard2001a}.
However, this is not the case above the bifurcation for which some resonances are on the imaginary axis and prevent mixing, nor is it the case exactly at the bifurcation point where the RP spectrum is continuous and responsible for an algebraic decay of correlations.
The latter is not possible here, since we know that the spectrum of the SHE~\eqref{eq:HopfSDECart} is discrete (see Sect.~\ref{sec:stochastic_hopf}).
We also know from the previous Section~\ref{sec:FellerIrreducibility} that purely imaginary are not to be expected since the invariant measure is strongly mixing.
Yet an accumulation point at 0 in the complex plain could still prevent the existence of a spectral gap.

The following proposition states that, for all values of the control parameter $\delta$, a spectral gap in fact exists in $L^2_\mu(\mathbb{R}^2)$, where $\mu$ is the invariant measure associated with the density $\rho_\infty$ (Eq.~\eqref{eq:statDist}).
The proof is given in Appendix~\ref{sec:proofGap} and relies on the theory of Lyapunov functions and ultimate bounds reviewed in~\cite[{\mkr Appendix A.5}]{Chekroun_al_RP2}.
\begin{prop}\label{prop:spectral_gap}
  For any $\delta$ in $\mathbb{R}$, $\beta$ in $\mathbb{R}$, $\kappa > 0$ and $\epsilon > 0$, the SHE~\eqref{eq:HopfSDECart} has a spectral gap and correlations decay exponentially in $L^2_\mu(\mathbb{R}^2)$.
\end{prop}

{\mkr This result is thus just a consequence of stochastic analysis techniques as reviewed in~\cite{Chekroun_al_RP2}}, without explicit calculations of the RP spectrum.
{\mkr In Section~\ref{sec:smallNoiseExpansion} below,} we provide however a more precise description of the latter by using small-noise expansion techniques; {\mkr see Propositions~\ref{prop:eigen_below} and~\ref{prop:eigen_above}}.

\begin{rmk}\label{Remark_2delta}
	Note {\mk that for $\delta<0$}, the rate $2\delta$ of the exponential bound in \eqref{eq:boundSub} is given by the real part of the second eigenvalue of the Kolmogorov operator of the linearized system {\mk at the origin},
	i.e.~the leading eigenvalue associated with an eigenfunction {\mk on which the projection of $\varphi(r,\theta)=r^2$ is nonzero}; see Section~\ref{sec:ExpansionSub}.
\end{rmk}
\begin{rmk}
Interestingly, for $\delta = 0$ and $\epsilon = 0$, the ultimate bound is, however, not verified.
This is not surprising, since we know from~\cite{Gaspard2001a} that the decay of correlation is in this case only algebraic.
On the other hand, for $\delta \ne 0$ but $\epsilon = 0$, the ultimate bound still holds {\mk but one cannot apply Theorem~6 from~\cite{Chekroun_al_RP2} anymore, since the system is no longer stochastic}.
However, the existence of a spectral gap and the exponential decay of correlations in this deterministic case may be inferred from~\cite{Gaspard2001a}.
\end{rmk}

\section{Small-noise approximation of the RP resonances, $\delta \ne 0$}\label{sec:smallNoiseExpansion}

In this section, we look for expansions of the RP eigenvalues and eigenfunctions for relatively low values of the noise level $\epsilon > 0$ and away from the bifurcation as singular perturbations of the deterministic case.

General small-noise expansion formulas for the RP resonances have been derived by~\cite{Gaspard2002a} using a WKB approximation and his results have been discussed for a form of the SHE~\eqref{eq:HopfSDECart} considered here by~\cite{Bagheri2014}.
However, to learn more about the geometrical properties of the stochastic system
and to be able to calculate power spectra between any pair of observables according to the spectral decomposition (\ref{eq:spectralPower}), we derive {\mk analytic approximations of} the eigenfunctions of the Kolmogorov operator $\mathcal{K}$ {\mk as well as of those of its adjoint, $\mathcal{K}^\ast$}.

To do so, we rely on a rescaling of the coordinates depending both on the noise level $\epsilon$ and on the parameter $\delta$ controlling the stability of the solutions to adimensionalize the SHE~\eqref{eq:HopfSDECart}.
A natural time scale is given by $\delta^{-1}$, while a spatial scale $L_\epsilon(\delta)$ capturing the effect of the noise with respect to the stability of the deterministic solutions is given by $\epsilon / \sqrt{-\delta}$ if $\delta < 0$ or by $\epsilon / \sqrt{\delta}$ if $\delta > 0$.
Applying It\^o's formula, the change of variable $r\to r' = r / L_\epsilon(\delta)$, $\theta \to \theta' = \theta$ or $\phi \to \phi' = \phi$, and $t \to t' = \delta t$ yields for the SHE~\eqref{eq:HopfSDECart},
\bea
\label{eq:SHEScaled}
\d r'
&= r' \lr{\frac{\delta}{|\delta|} - \frac{r'^{2}}{r_\epsilon^2} + \frac{1}{2r'^2}} \d t' + \d W_r \\
\d \theta'
&= \lr{\tilde \gamma - \tilde \beta \frac{r'^2}{r_\epsilon^{2}}} \d t' + \frac{\d W_\theta'}{r'}\\
\mathrm{or~} \d \phi'
&= \lr{\tilde \gamma - \tilde \beta} \d t' - \tilde \beta \frac{\d W_r}{r'} + \frac{\d W_\theta}{r'},
\eea
where $\delta / |\delta| = -1$ if $\delta < 0$, $1$ if $\delta > 0$.
In addition to the twist factor $\tilde \beta = \beta / \kappa$, we have introduced the adimensional parameters $\tilde \gamma(\gamma, \delta) = \gamma / \delta$ and $r_\epsilon(\delta, \kappa) = \delta / (\sqrt{\kappa} \epsilon)$, simply noted $\tilde \gamma$ and $r_\epsilon$, respectively, in the remaining.
Defining $R$ by $\sqrt{-\delta / \kappa}$ for $\delta < 0$, the adimensional parameter is such that $r_\epsilon = R / L_\epsilon(\delta)$.
Thus the effect of the noise on the adimensional dynamics~\eqref{eq:SHEScaled} is bound to that of the parameters $\delta$ and $\kappa$ in a single coefficient $r_\epsilon$.
For a fixed $R$, this effect increases with the noise-level $\epsilon$ and decreases with the square root of $\delta$.
Since all coefficients in~\eqref{eq:SHEScaled} involving the noise level $\epsilon$ enter as $\sigma_\epsilon = 1 / r_\epsilon$, we are led to expand the eigenvalues and eigenfunctions of the Kolmogorov operator as,
\begin{align*}
  \lambda &= \lambda^{(0)} + \sigma_\epsilon \lambda^{(1)} +
              \sigma_\epsilon^2 \lambda^{(2)} + \dots \\
  \psi' &= \psi^{(0)} + \sigma_\epsilon \psi^{(1)} +
            \sigma_\epsilon^2 \psi^{(2)} + \dots .
\end{align*}
Since, the deterministic solutions and the change of variables differ for $\delta < 0$ and $\delta > 0$, each case is treated separately in the next subsections~\ref{sec:ExpansionSub} and~\ref{sec:ExpansionSup}, respectively.
From the definition of the small parameter $\sigma_\epsilon$, whether for $\delta < 0$ or for $\delta > 0$, the small-noise expansions will be more precise when the noise level $\epsilon$ is small with respect to $\delta$, for a fixed $\kappa$.

\subsection{Below the bifurcation ($\delta < 0$)}
\label{sec:ExpansionSub}

All deterministic solutions converge to the {\mk steady state} $x_*$ at the origin.
An example of stochastic trajectory is represented in blue in Fig.~\ref{fig:artMixingEigVal}-(a) for $\delta = -1$, $\kappa = 1, \gamma = 4, \beta = 0.5$ and $\epsilon = 0.4$ on top of the corresponding stationary density given by~\eqref{eq:statDist}.
As expected, the process {\mkr meanders near}, $x_*$, although the maximum in density is slightly away from $x_*$, due to the additional drift term $\epsilon / (2r)$ in~\eqref{eq:HopfSDEPolar}.
The following proposition yields the small-noise expansion of the leading part of the spectrum of the SHE~\eqref{eq:HopfSDECart} for $\delta < 0$.
The proof is given in Appendix~\ref{sec:proof_eigen_below} and relies on known results for the complex Ornstein-Uhlenbeck process~\cite{Metafune2002a,Chen2014}.

\begin{prop}\label{prop:eigen_below}
{\mk For $\delta < 0$ and $\smallin << 1$ the approximation of the leading eigenvalues and eigenfunctions associated with the SHE~\eqref{eq:HopfSDECart} are given by:}

\begin{minipage}{\linewidth}
\begin{itemize}
\item {\mk Eigenvalues associated with} the stable {\mk steady state}:
\begin{equation}
	\lambda_{ln} = (l + n) \delta + i (n - l) \gamma + \mathcal{O}_{\tilde \beta}\lr{\smalleq^2}, \quad l, n \in \mathbb{N}.
	\label{eq:eigValFPSub}
\end{equation}
\item {\mk Eigenfunctions associated with} the stable {\mk steady state}:
\begin{equation}
  \psi_{ln}(r, \theta) \approx
	\begin{cases}
          e^{i(n-l)\theta} \enskip
          \sqrt{\frac{l!}{n!}}
          \lr{\sqrt{-\frac{\delta}{\epsilon^2}} r}^{n-l}
          L_l^{n-l}\lr{-\frac{\delta r^2}{\epsilon^2}}, \quad &n \ge l \\
          e^{i(l-n)\theta} \enskip
          \sqrt{\frac{n!}{l!}}
          \lr{\sqrt{-\frac{\delta}{\epsilon^2}}r}^{l-n}
          L_n^{l-n}\lr{-\frac{\delta r^2}{\epsilon^2}}, \quad &n < l,
	\end{cases}
	\label{eq:complexHermite}
\end{equation}
{\mk where $L^\alpha_l(r) = \frac{r^{-\alpha}}{l!} e^r \frac{d^l}{dr^l}(e^{-r} r^{l + \alpha})$
denotes the Laguerre polynomial of degree $l$~\cite[p.~76]{Lebedev1972} in the radius $r$.}
\item {\mk Adjoint eigenfunctions associated with} the stable {\mk steady state}:
\begin{equation}
	\psi^*_{ln} \approx \psi_{ln}
	\enskip \rho_{x_*}.
\end{equation}
\item Decorrelation time:
\begin{equation}
	\tau = -\frac{1}{\delta} + \mathcal{O}_{\tilde \beta}\lr{\smalleq^2}.
	\label{eq:smallNoisePointSubTime}
\end{equation}
\end{itemize}
\end{minipage}
\end{prop}
In~\eqref{eq:eigValFPSub} and~\eqref{eq:smallNoisePointSubTime}, $\mathcal{O}_{\tilde \beta}(\smallin)$ is the usual asymptotic notation for the small parameter $\smallin$ but with an indication that the remaining terms in the expansions actually depend on the twist factor $\tilde \beta$.

The RP resonances \eqref{eq:eigValFPSub} are represented in Fig.~\ref{fig:artMixingEigVal}-(c) for fixed values of the parameters.
A typical triangular structure is observed, as a result of the {\mk aforementioned} integer linear combination of complex conjugate eigenvalues $\lambda^{\pm} = \delta \pm i \gamma$ of the tangent map $J_{x_*}$.
In the direction of the real axis, these eigenvalues are separated by a gap of $\delta$ given by the real part of the eigenvalues of the tangent map.
Thus, as the control parameter $\delta$ is increased to its critical value,
the decorrelation time $\tau \approx -1 / \delta$ in (\ref{eq:smallNoisePointSubTime}) increases,
 indicative of the weaker stability of the {\mk steady state} of the deterministic system.
Moreover, the eigenvalue $\lambda_{ln}, n > l$ is associated
with an {\mk eigenfunction that is approximated by} the product of a polynomial of degree $n + l$
and the $(n-l)$th harmonic function $\exp{i(n-l)\theta}$.
Thus, {\mk eigenfunctions} associated with eigenvalues further {\mk away} from the real axis (resp.~imaginary axis)
{\mk exhibit a} higher degree of nonlinearity in the radial (resp.~azimuthal) direction, as measured by their number of {\mk sign changes}.
As an example, the eigenfunction $\psi_{01}$ associated
with the eigenvalue $\lambda_{01} \approx \delta + i \gamma$
closest to the imaginary axis is represented in Fig.~\ref{fig:artMixingEigVal}-(e).
Its phase $\arg \psi_{01} = \theta$ is represented by filled contours,
while its amplitude $\psi_{01} e^{-i\arg \psi_{01}} = r$ is represented by dashed contour lines.
The amplitude and phase of the leading secondary eigenfunction is thus the components of the stochastic process in polar coordinates.
This is not surprising, since the {\mk eigenfunctions} are approximated by those
of the (linear) Ornstein-Uhlenbeck process with drift given by the tangent map $J_{x_*}$, {\mk as explained above}.
\begin{rmk}
  In the expansion~\eqref{eq:eigValFPSub}, we do not control for changes in the weights in front of the $\mathcal{O}(\epsilon^2)$ for different eigenvalues.
  Thus, high-order terms may have a larger impact for some eigenvalues than for other, a phenomenon that we describe in the numerical results of Section~\ref{sec:numericalBetaZero}.
\end{rmk}

\subsection{Above the bifurcation ($\delta > 0$)}
\label{sec:ExpansionSup}

After the deterministic Hopf bifurcation, two limit sets coexist,
the unstable {\mk steady state} $x_*$ at the origin and the stable limit cycle $\Gamma$ of radius $R$.
An example of stochastic trajectory is represented in blue in Fig.~\ref{fig:artMixingEigVal}-(b) for $\delta = 1.5$, $\kappa = 1, \gamma = 4, \beta = 0.5$ and $\epsilon = 0.4$ on top of the corresponding stationary density given by (\ref{eq:statDist}), while the {\mk orbit} $\Gamma$ is represented by the dashed line.
{\mk Here small-noise expansions are also illuminating to obtain approximation formulas when applied separately about the unstable {\mk steady state} and the limit cycle.}

\subsubsection{{\mk Small-noise expansions about} the unstable {\mk steady state} $x_*$}
{\mk Repeating similar arguments than in the case $\delta < 0$, the RP resonances associated with the unstable {\mk steady state} are here given for $\delta > 0$, by} 
\begin{align}
  \lambda_{ln} = -(l + n + 2) \delta - i (l - n) \gamma
  + \mathcal{O}_{\tilde \beta}\lr{\lr{\smallin}^2}, \quad l, n \in \mathbb{N}.
  \label{eq:eigValFPSup}
\end{align}
These eigenvalues are represented for fixed values of the parameters as blue triangles in Fig.~\ref{fig:artMixingEigVal}-(d).
A triangular array of eigenvalues is found, as for $\delta < 0$ in panel (c) {\mk of the same figure}.
However, the real part of these eigenvalues satisfies $\Re(\lambda_{ln}) \le -2 \delta.$ {\mk The latter bound actually characterizes the rate of expansion of volumes near the unstable {\mk steady state} $x_*$ (and away from the limit cycle $\Gamma$).
The latter decreases with increasing $\delta$, i.e.~as the instability of $x_*$ increases.}

\subsubsection{{\mk Small-noise expansions about} the limit cycle $\Gamma$}\label{sec:aboutLimit}
On the other hand, another family of eigenvalues associated with the limit cycle can be {\mk identified}.
In order to study small-noise perturbations of the system away from the limit cycle,
we work in Appendix~\ref{sec:proof_eigen_above} from the adimensional version of the Kolmogorov equation~\eqref{eq:BKEIsochron} associated with the radial $r$ and asymptotic phase $\phi$, variables.
{\mk Compared to the original Kolmogorov equation~\eqref{eq:HopfBKEPolar} written in polar coordinates, the Kolmogorov equation~\eqref{eq:BKEIsochron} --- formulated in Sect.~\ref{Sec_phase_BKE} with the help of isochrons --- helps us separate the drift term into two contributions, one in the $r$-coordinate alone, and the other in the $\phi$-coordinate.}
In the unperturbed case, this separation of variables shows that the isochrons can be identified with isolines of phase of the eigenfunctions associated with purely imaginary eigenvalues.
As a result, Fourier averages related to these eigenfunctions have been proposed to estimate the isochrons~\cite{Mauroy2012,Mauroy2013}.
The following proposition, proved in Appendix~\ref{sec:proof_eigen_above}, shows for the SHE~\eqref{eq:HopfSDEPolar} that, when the noise is asymptotically small, the isochrons still coincide with the isoline of phase of the eigenfunctions.

\begin{prop}\label{prop:eigen_above}
{\mk For $\delta > 0$ and $\smallin << 1$ the approximation of the leading eigenvalues and eigenfunctions associated with the limit cycle $\Gamma$ of the SHE~\eqref{eq:HopfSDECart} are given by:}

\begin{minipage}{\linewidth}
\begin{itemize}
\item {\mk Eigenvalues associated with} the stable limit cycle:
\begin{equation}
	\lambda_{ln} =
	\begin{cases}
		-\frac{n^2 \epsilon^2 (1 + \tilde \beta^2)}{2 R^2} + i n \omega_f + \mathcal{O}\lr{\smalleq {}^3},
		\quad &l = 0, \quad n \in \mathbb{Z} \\
		- 2 l \delta + i n \omega_f + \mathcal{O}{\smalleq}, \quad &l \ne 0.
	\end{cases}
	\label{eq:eigValOrbit}
\end{equation}
\item {\mk Eigenfunctions associated with} the stable limit cycle:
\begin{equation}
	\psi_{ln} \approx \left(2^k k!\right)^{-\frac{1}{2}} \enskip
	e^{i n \left(\theta - \tilde \beta \log{\frac{r}{R}} \right)} \enskip
	H_l\left( \frac{\sqrt{2 \delta}}{\epsilon} \left(r - R\right) \right), \enskip \quad l = 0.
	\label{eq:eigvecHarmonic}
\end{equation}
\item Adjoint Eigenfunctions of the stable limit cycle:
\begin{equation}
	\psi^*_{ln} \approx \left(2^k k!\right)^{-\frac{1}{2}} \enskip
	e^{i n \left(\theta + \tilde \beta \log{\frac{r}{R}} \right)} \enskip
	H_l\left( \frac{\sqrt{2 \delta}}{\epsilon} \left(r - R\right) \right) \enskip
	\rho_{x_*}(r), \enskip \quad l = 0.
	 \label{eq:eigvecHarmonicAdjoint}
\end{equation}
\item Decorrelation time:
\begin{equation}
	\tau = \frac{2 R^2}{\epsilon^2 (1 + \tilde \beta^2)} + \mathcal{O}\lr{1}.
	\label{eq:smallNoisePointSupTime}
\end{equation}
\end{itemize}
\end{minipage}
\end{prop}

\begin{figure}
	\centering
	\begin{subfigure}{0.48\textwidth}
		\includegraphics[width=.85\textwidth,height=.85\textwidth]{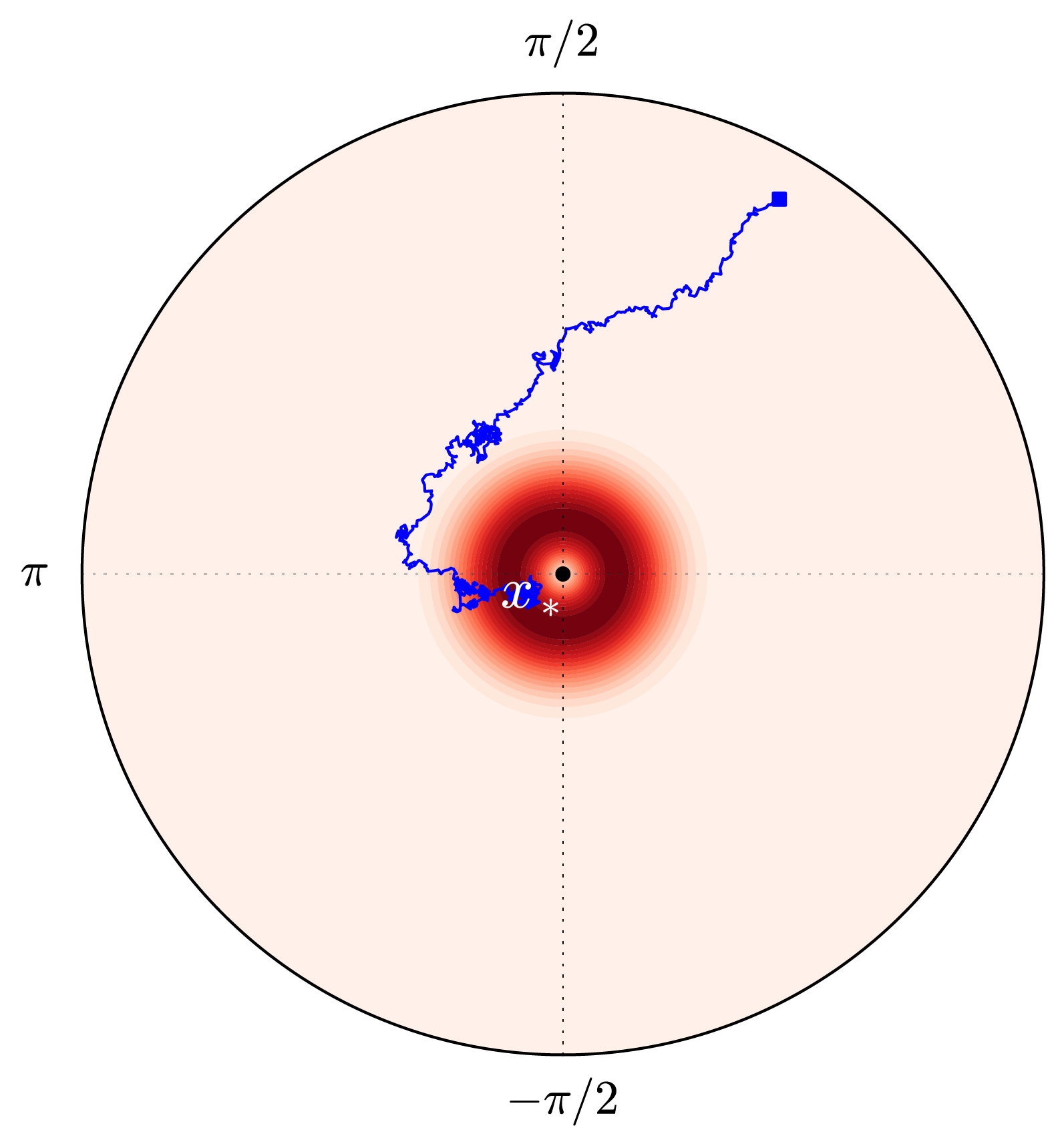}
		\caption{}
	\end{subfigure}
	\begin{subfigure}{0.48\textwidth}
		\includegraphics[width=.85\textwidth,height=.85\textwidth]{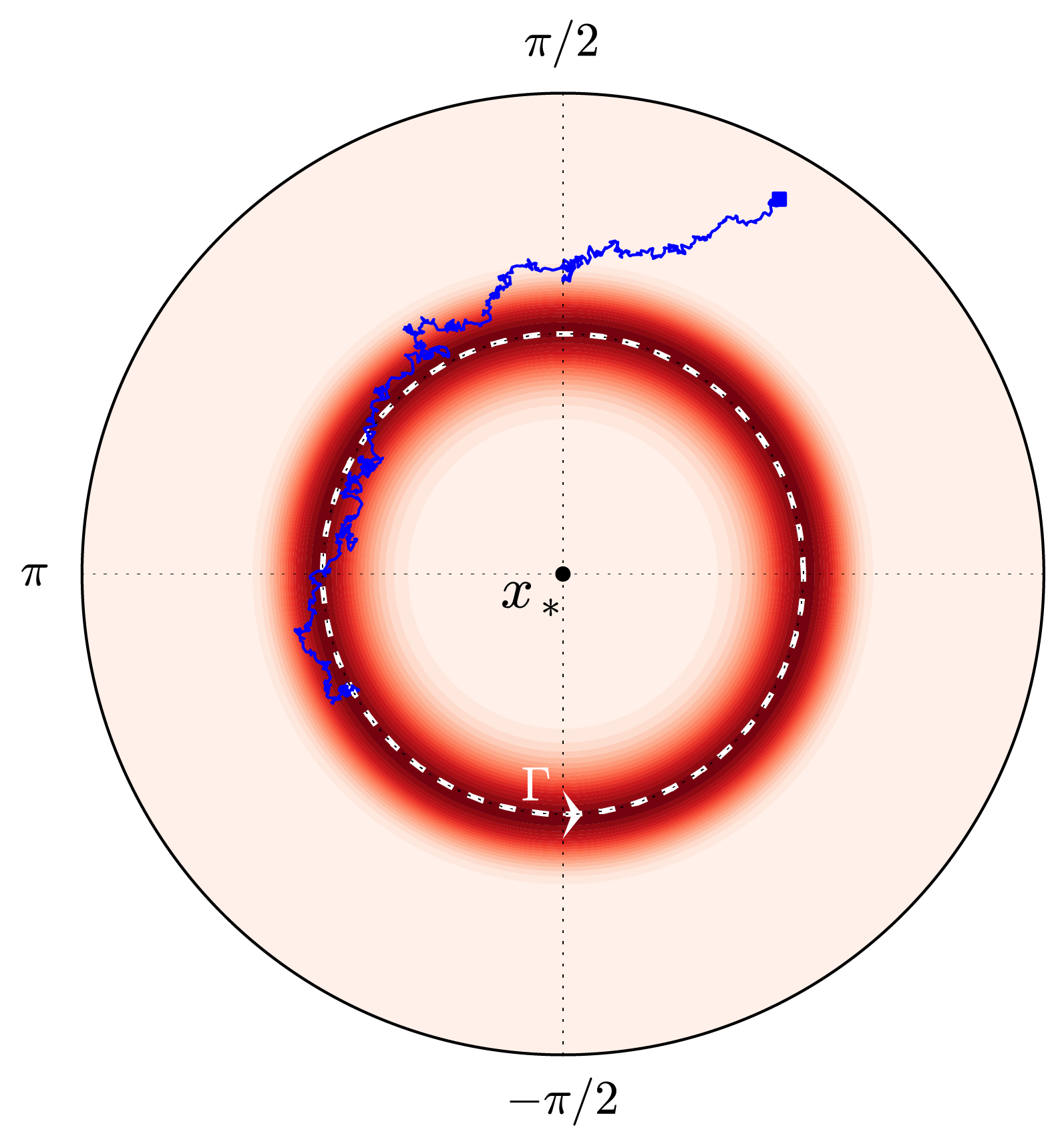}
		\caption{}
	\end{subfigure} \\
	\begin{subfigure}{0.48\textwidth}
		\includegraphics[width=.85\textwidth,height=.85\textwidth]{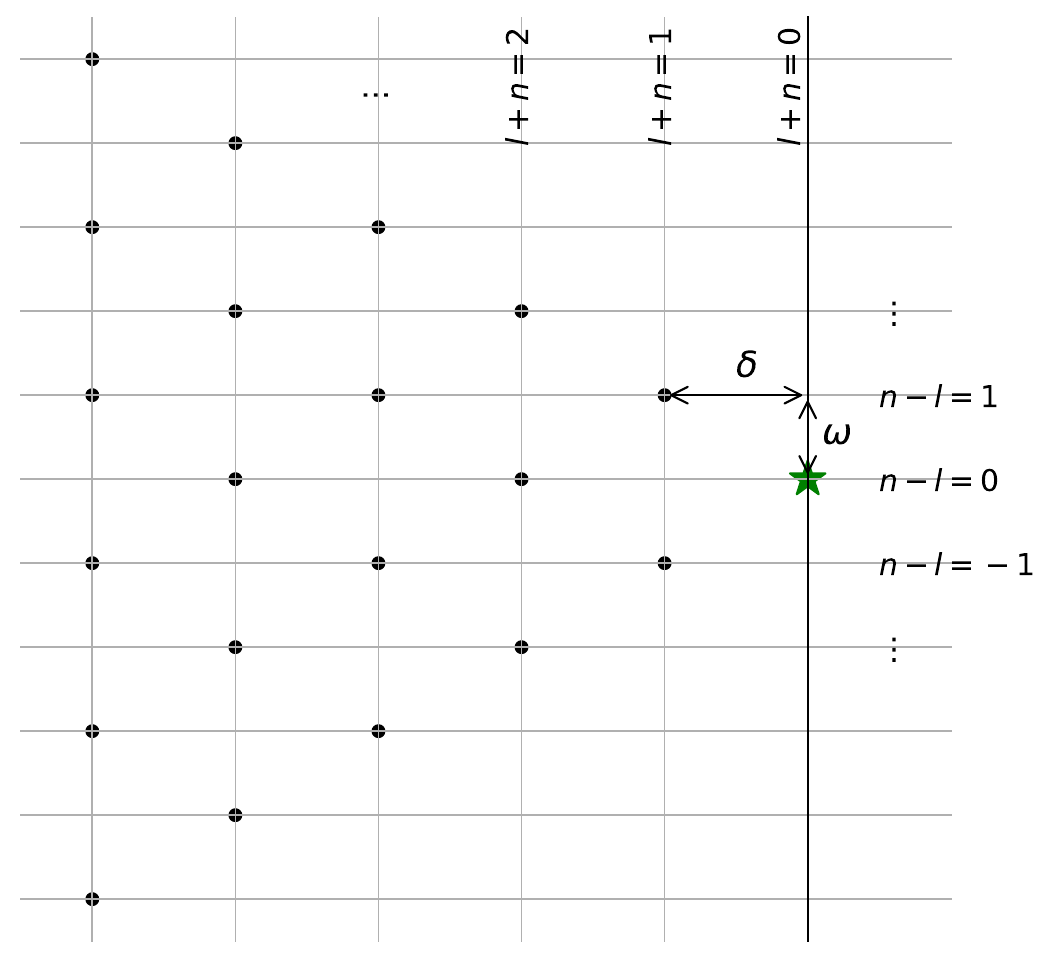}
		\caption{}
	\end{subfigure}
	\begin{subfigure}{0.48\textwidth}
		\includegraphics[width=.85\textwidth,height=.85\textwidth]{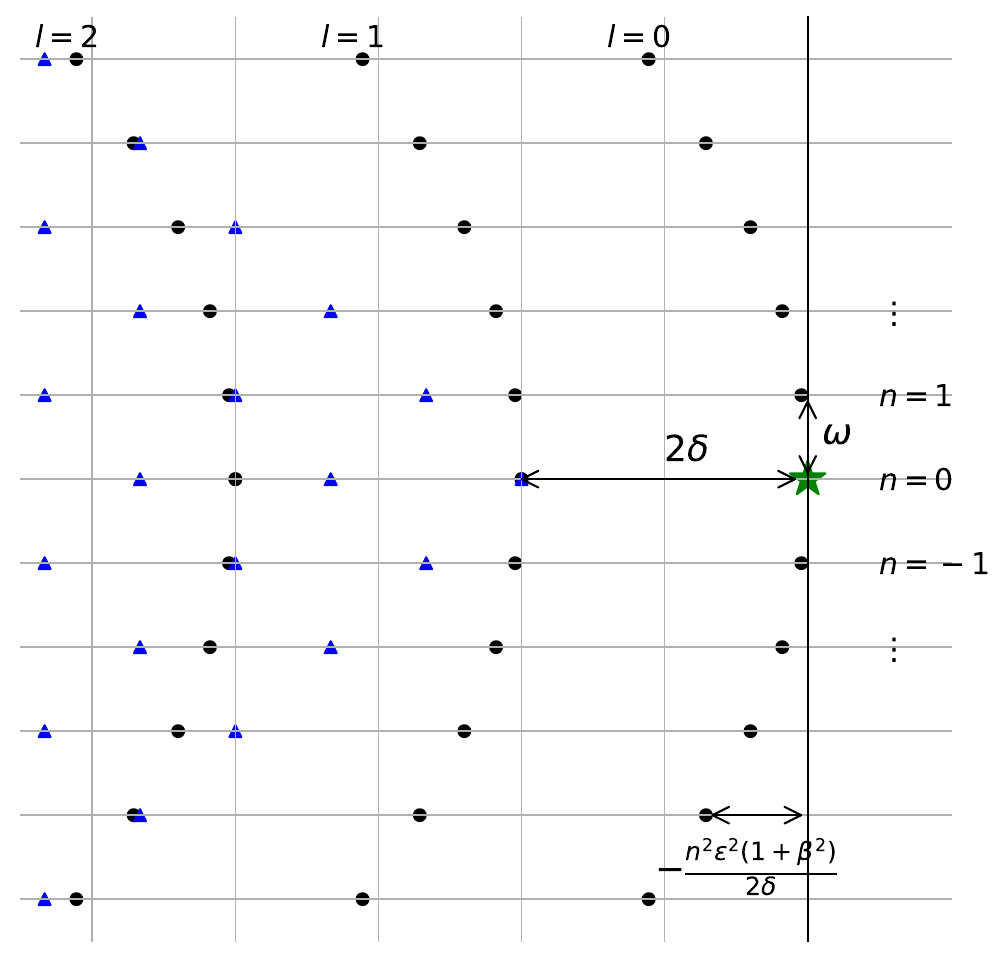}
		\caption{}
	\end{subfigure} \\
	\begin{subfigure}{0.48\textwidth}
		\includegraphics[width=.85\textwidth,height=.8\textwidth]{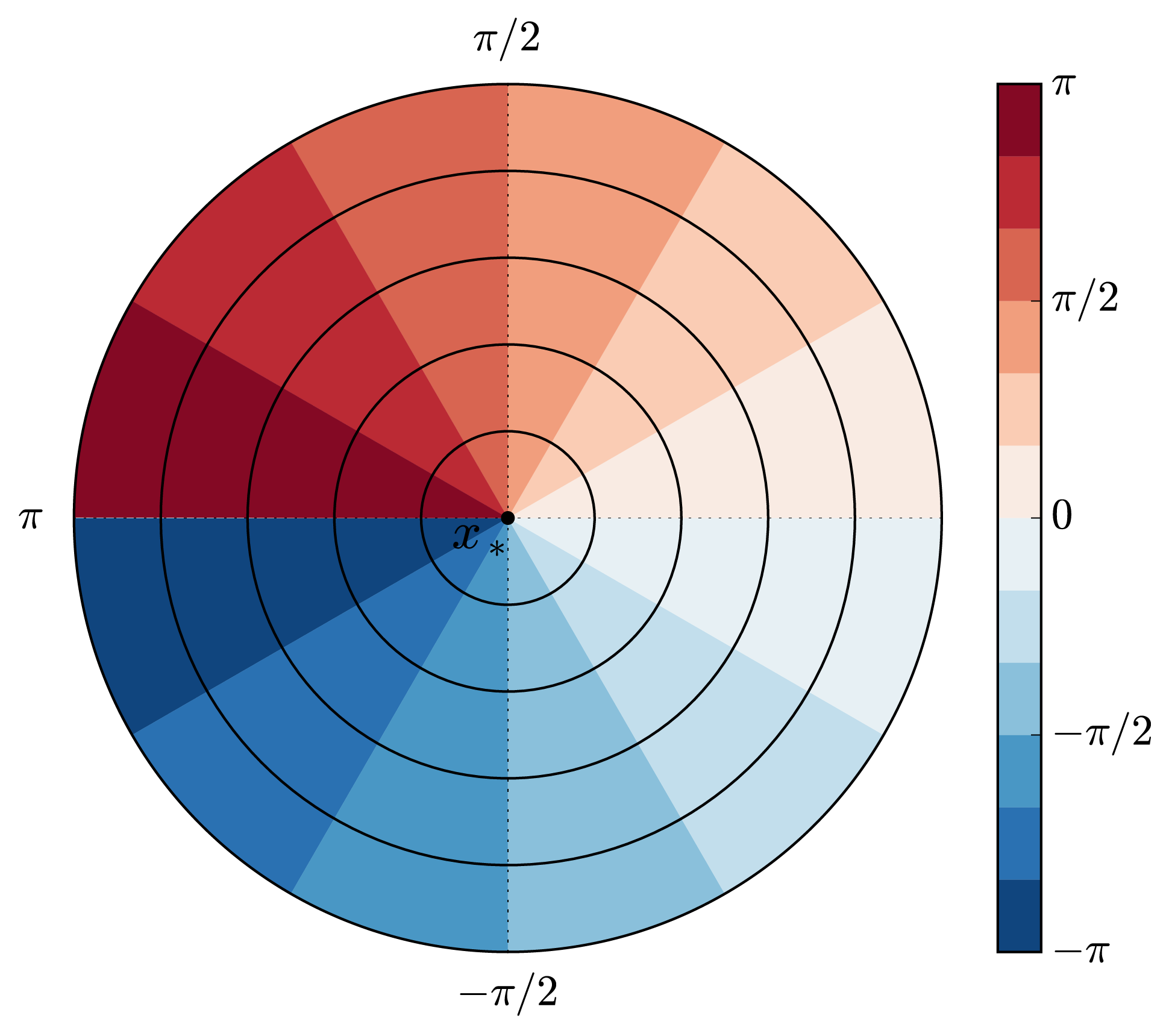}
		\caption{}
	\end{subfigure}
	\begin{subfigure}{0.48\textwidth}
		\includegraphics[width=.85\textwidth,height=.8\textwidth]{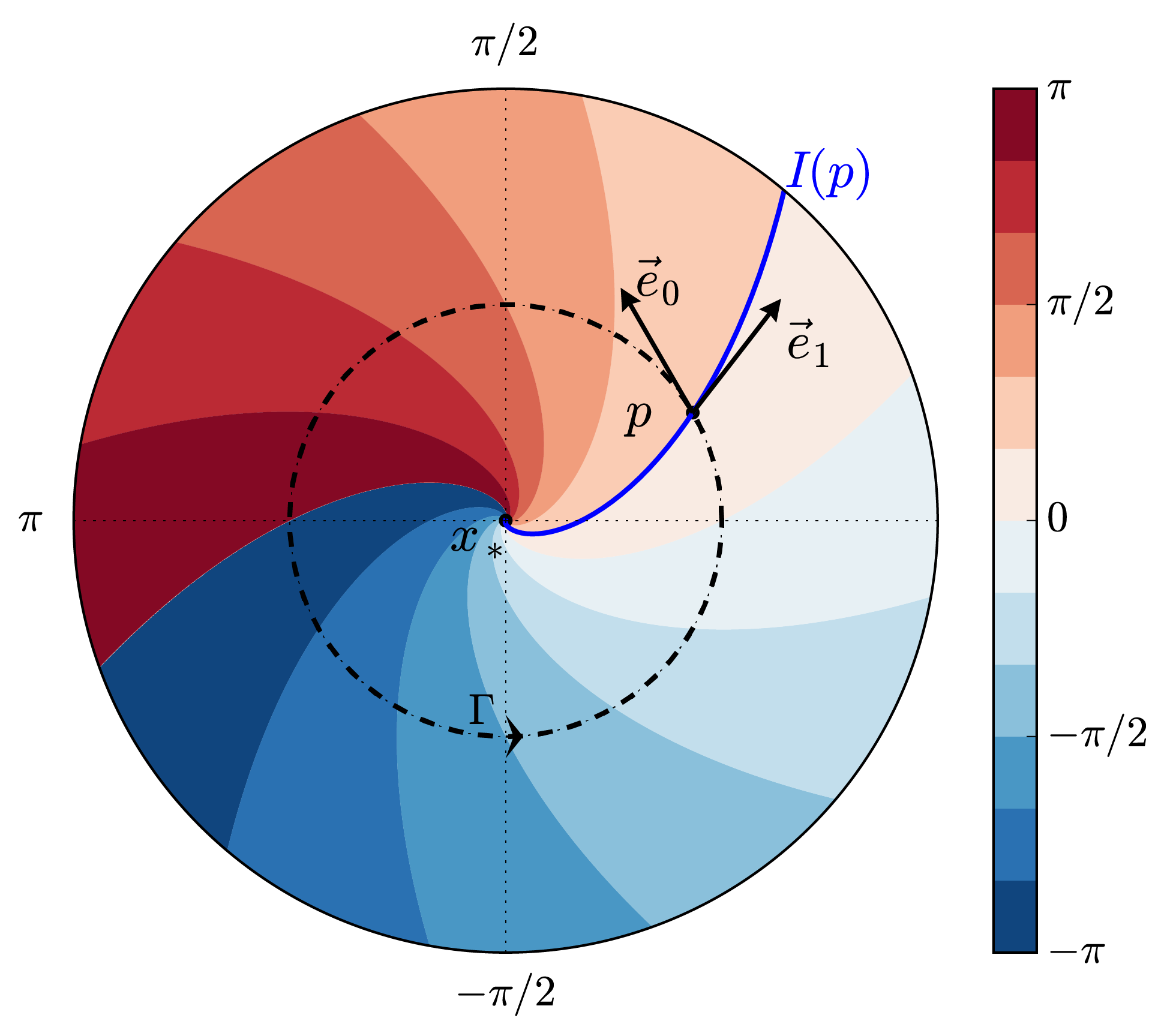}
		\caption{}
	\end{subfigure} \\
	\caption{
          Schematic of the approximated RP spectrum of the SHE~\eqref{eq:HopfSDECart}, for $\kappa = 1, \gamma = 4, \beta = 0.5$ and $\epsilon = 0.4$, and with $\delta = -1 < 0$ (left) or $\delta = 1.5$ (right).
          The top panels (a-b) represent the stationary density~\eqref{eq:statDist} as red filled contours, together with an example of trajectory in blue and the {\mk steady state} $x_*$ at the origin.
          The RP resonances in the complex plane are represented in the central panels (c-d), with their real parts as abscissa and their imaginary parts as ordinates.
	The bottom panels (e-f) represent the second eigenfunction $\psi_{01}$, with its phase as filled contours and its amplitude as thin line contours (for $\delta < 0$).
	For $\delta > 0$, the deterministic limit cycle $\Gamma$ is also represented as a thick dashed line, together with the isochron $I_{x_0}$ of some point $x_0$ on $\Gamma$ as a thick blue line and the eigenvectors $\vec{e}_0$ and $\vec{e}_1$ of the tangent map at this point.}
	\label{fig:artMixingEigVal}
      \end{figure}

To {\mk help interpret} these formulas, the RP resonances for fixed values of the parameters are represented in Fig.~\ref{fig:artMixingEigVal}-(d) together with the eigenfunction $\psi_{01} = \exp{i (\theta - \tilde \beta \log{r / R})}$ associated with the second eigenvalue $\lambda_{01} = - \epsilon^2 (1 + \tilde \beta^2) / (2 R^2) + i n \omega_f$ in Fig.~\ref{fig:artMixingEigVal}-(f).
One can first observe in panel (d) a typical array of parabolas of eigenvalues.
The latter are separated by a spectral gap of $-2\delta$ (see Remark~\ref{Remark_2delta})
given by the characteristic exponent associated with the Floquet vector transverse to the flow and accounting for the stability of the limit cycle $\Gamma$; see also~\ref{sec:Floquet}.
One the other hand, the imaginary part $i n \omega_f$,
for each harmonic, is associated with the neutral dynamics of advection along the limit cycle.
These two contributions jointly coincide with the eigenvalues for the deterministic case found in spaces of distributions by~\cite{Gaspard2001a}.

However, the diffusion along the limit cycle, is responsible for an additional real contribution $-n^2 \epsilon^2 (1 + \tilde \beta^2) / (2 R^2)$,
which is not found in the deterministic case and which is responsible for the parabolic shape of the array of eigenvalues.
As a result, $\lambda_{00} = 0$ (represented as a green star in Fig.~\ref{fig:artMixingEigVal}-(d))
is the only eigenvalue on the imaginary axis.
The presence of noise therefore enforces the system to be mixing, in agreement with the spectral gap result of Section~\ref{sec:discreteSpectrum}.
This ``loss of memory'' is captured by the finiteness
of the decorrelation time $\tau \approx 2 R^2 / (\epsilon^2 (1 + \tilde \beta^2))$,
which decreases as the noise level $\epsilon$ and the curvature $1 / R$ of $\Gamma$ strengthen.

In addition, the phase diffusion becomes stronger with increasing magnitude of the twist factor $\tilde \beta$ as well.
As discussed in Section~\ref{sec:FellerIrreducibility}, a nonvanishing twist factor $\tilde \beta$ allows for a fraction of the noise in the radial direction to be transmitted to the azimuthal direction by the deterministic vector field $F$.
As depicted in panel (f) of Fig.~\ref{fig:artMixingEigVal}, the eigenvector $\vec{e}_2$ of the tangent map to $F$ is tangent to $\Gamma$, while $\vec{e}_1$ is tangent to the isochron.
Thus, when $\beta \ne 0$, the vector $\vec{e}_1$ projects both on the radial and on the azimuthal parts of the stochastic forcing.
Moreover, since $\arg \psi_{01} = \phi = \theta - \tilde \beta \log(r / R)$,
the phase of the second eigenfunction follows the isochrons,
so that the radial dependence of the phase diffusion
results in the characteristic twisting of the eigenfunctions when $\beta \ne 0$.
As a result, the eigenfunctions are not orthogonal when $\beta$ is nonzero and the Kolmogorov operator $\mathcal{K}$ inherits from the nonnormality of the Jacobian $J_\Gamma$.
Finally,~\eqref{eq:EigVecHermite} and~\eqref{eq:eigvecHarmonic} show that the eigenfunctions associated with eigenvalues further from the real axis (imaginary axis) have a higher degree of nonlinearity in the radius (resp. the phase).

To conclude, let us emphasize the difference in structure between the RP spectrum
associated with the stable {\mk steady state} for $\delta < 0$ and the one associated with the
limit cycle for $\delta > 0$.
While the eigenvalues have nonvanishing imaginary parts in both cases (\ref{eq:eigValFPSub})
and (\ref{eq:eigValOrbit}), which must result in peaks in the power spectrum,
the triangular structure for the {\mk steady state} and the parabolic structure
for the limit cycle, as shown in Fig.~\ref{fig:artMixingEigVal}-(c) and Fig.~\ref{fig:artMixingEigVal}-(d),
allow one to discriminate between stochastically forced
linear oscillations and nonlinear oscillations with phase diffusion.
This is also true regarding the eigenfunctions, given by {\mk the formulas} (\ref{eq:complexHermite}) and (\ref{eq:eigvecHarmonic}),
which in the case of the {\mk steady state} (and to zeroth order) are
the product of different polynomials by harmonics with a different sensitivity
to the twist factor $\tilde \beta = \beta / \kappa$.
These effects will be illustrated in the applications of the third part of this contribution~\cite{tantet_ruellepollicott_2019},
with a discussion of  their use to characterize the nature
of the dynamics of complex oscillatory systems.
The investigation of the RP spectrum at the bifurcation does not follow the reasoning above.
Instead, we numerically investigate mixing at the bifurcation in the following Section~\ref{sec:numericalHopf}.

\section{Mixing at the bifurcation point: Numerical results}\label{sec:numericalHopf}

Close to the bifurcation point, the small noise-expansions of the previous Section~\ref{sec:smallNoiseExpansion} are no longer valid since the linear term in $\delta$ vanishes and the rescaling of time in terms of this parameter is no longer possible.
We thus perform a complementary numerical analysis of the Kolmogorov equation to study the RP spectrum at the critical point and test the range of validity of the analytical formulas of the previous Section~\ref{sec:smallNoiseExpansion}.

\subsection{A different scaling}
Let us first note that, even though the limit cycle does not exist, the asymptotic phase $\phi$ for any point different from the origin can still be defined up to a constant as $\phi = \theta - \tilde \beta \log{r}$ and such that the derivative \eqref{eq:dphidr} exists and the Kolmogorov equation~\eqref{eq:BKEIsochron} in  $(r, \phi)$ coordinates holds.
Second, contrary to the deterministic case, a new temporal scale can be defined as $(\epsilon \sqrt{\kappa})^{-1}$ when $\epsilon > 0$.
A corresponding spatial scale may then be defined as $\epsilon^{1/2}/\kappa^{1/4}$.
This time scale thus depends on the coefficient $\kappa$ of the cubic term of the radial vector field in~\eqref{eq:HopfSDEPolar} rather than on the coefficient $\delta$ of the linear term used in Section~\ref{sec:smallNoiseExpansion} for $\delta \ne 0$, and the spatial scale is now proportional to $\sqrt{\epsilon}$ rather than to $\epsilon$.
We thus use the following change of variable to adimensionalize the SHE~\eqref{eq:HopfSDECart},
\begin{align*}
	r' = r \kappa^{1/4}/ \epsilon^{1/2}, \quad \phi' = \phi + \omega_f t \quad s = \epsilon \sqrt{\kappa} t.
\end{align*}
Indeed, the Kolmogorov equation~\eqref{eq:BKEIsochron} with $u'(r', \phi') = u(r, \phi)$ then becomes
\begin{align*}
	\partial_s u'
	= (-r'^3 + \frac{1}{2r'}) \partial_{r'} u'
	+ \frac{1}{2} \partial^2_{r'r'} u' - \frac{\tilde \beta}{r'} \partial^2_{r'\phi'} u' + \frac{1 + \tilde \beta^2}{2r'^2} \partial^2_{\phi'\phi'} u'.
\end{align*}
Interestingly, even though the nonlinear coefficients hinder the full resolution of the associated eigenproblem,
this equation shows no dependance on the noise level $\epsilon$.
This is allowed by the absence of the $r$ term in the drift when $\delta$ is zero.
As a consequence, each eigenvalue must have a real part proportional to $\epsilon \sqrt{\kappa}$, i.e.
\begin{align}
	\Re(\lambda_k) \sim \epsilon \sqrt{\kappa}, \label{eq:rescaleCritical}
\end{align}
and the decorrelation time is proportional to the inverse of $\epsilon\sqrt{\kappa}$,
\begin{align*}
	\tau \sim \frac{1}{\epsilon\sqrt{\kappa}}.
\end{align*}
This simple result is rich in conclusions,
as it shows that the more intense the noise level $\epsilon$,
the larger the spectral gap between the eigenvalues.
Thus, the noise has a stabilizing effect on the statistics, compared to the deterministic case,
which can be understood from its smoothing effect analyzed in Section~\ref{sec:StocAnaHopf}.

\subsection{Parameter dependence close to bifurcation: Numerical results}
To learn more about the RP spectrum of the SHE~\eqref{eq:HopfSDECart} for $\delta \approx 0$,
we proceed to a numerical approximation of the {\mk Kolmogorov operator $\mathcal{K}$ associated with} the Kolmogorov equation~\eqref{eq:HopfBKECart}.
{\mkr Due to its two-dimensional character}, this numerical problem is directly tractable, {\mkr and the RP resonances are estimated from discretization of the Kolmogorov operator; see \cite[Remark 1-(iii)]{Chekroun_al_RP2}.} 
{\mkr In that respect,} the standard finite-difference scheme
proposed by \cite{Chang1970} is chosen for the adjoint $\mathcal{K}^*$ in the Fokker-Planck equation,
since it satisfies the conservation of probabilities and of positivity and is straightforward to implement.
The numerical approximation of $\mathcal{K}$ is then simply given by the transpose of that of $\mathcal{K}^*$.
Here, we impose no-flux boundary conditions for convenience (instead of vanishing at infinity),
but with a sufficiently large domain to avoid boundary effects.
The square $[-5\hat{L}, 5\hat{L}]^2$ is discretized into 200-by-200 boxes,
where $\hat{L}$ is an approximation of the standard deviation of the $x$ and $y$ coordinates.
The spectrum of the finite-difference approximation of the Kolmogorov operator $\mathcal{K}$ is then calculated numerically using the implicitly restarted Arnoldi iterative algorithm implemented in ARPACK \cite{Lehoucq1997}.
The domain and resolution of the grid have been chosen for the approximation of at least the second eigenvalue to converge (tests for particular cases suggest that a resolution of about 300-by-300 boxes would also allow for the third or forth eigenvalues to converge, but at the price of a significant increase in the computing time).
Different experiments for varying $\delta$, $\beta$ and $\epsilon$ will be analyzed, while $\gamma$ and $\kappa$ are kept fixed to 1 (i.e. $\tilde \beta = \beta$).
\subsubsection{Crossing the bifurcation point, with a zero twist factor $\tilde \beta$}
\label{sec:numericalBetaZero}
\begin{figure}
     	\captionsetup[subfigure]{aboveskip=-3pt,belowskip=-3pt}
  	\centering
	\begin{subfigure}{0.42\textwidth}
		\includegraphics[width=.95\textwidth,height=.8\textwidth]{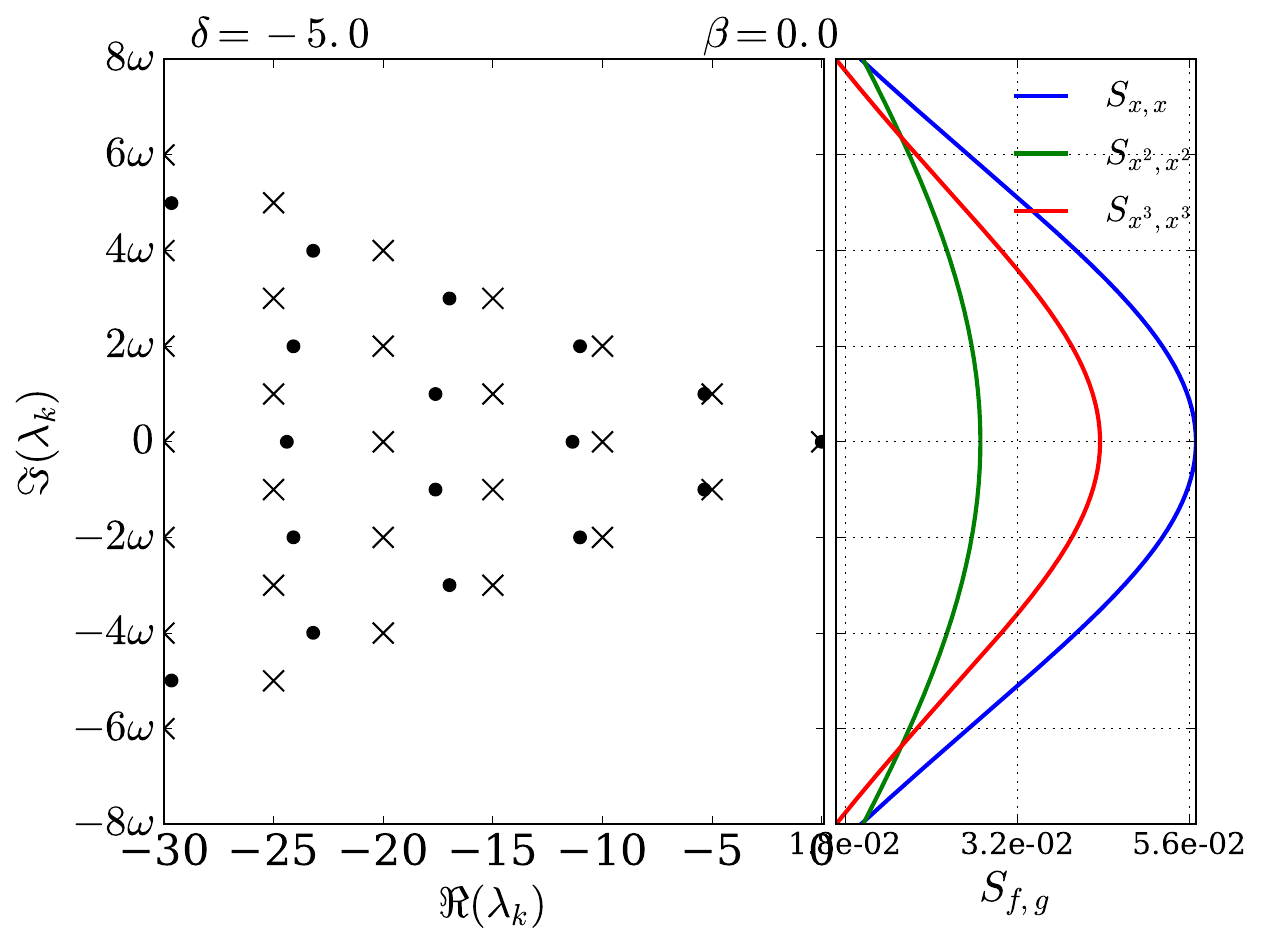}
		\caption{}
	\end{subfigure}
	\begin{subfigure}{0.42\textwidth}
		\includegraphics[width=.95\textwidth,height=.8\textwidth]{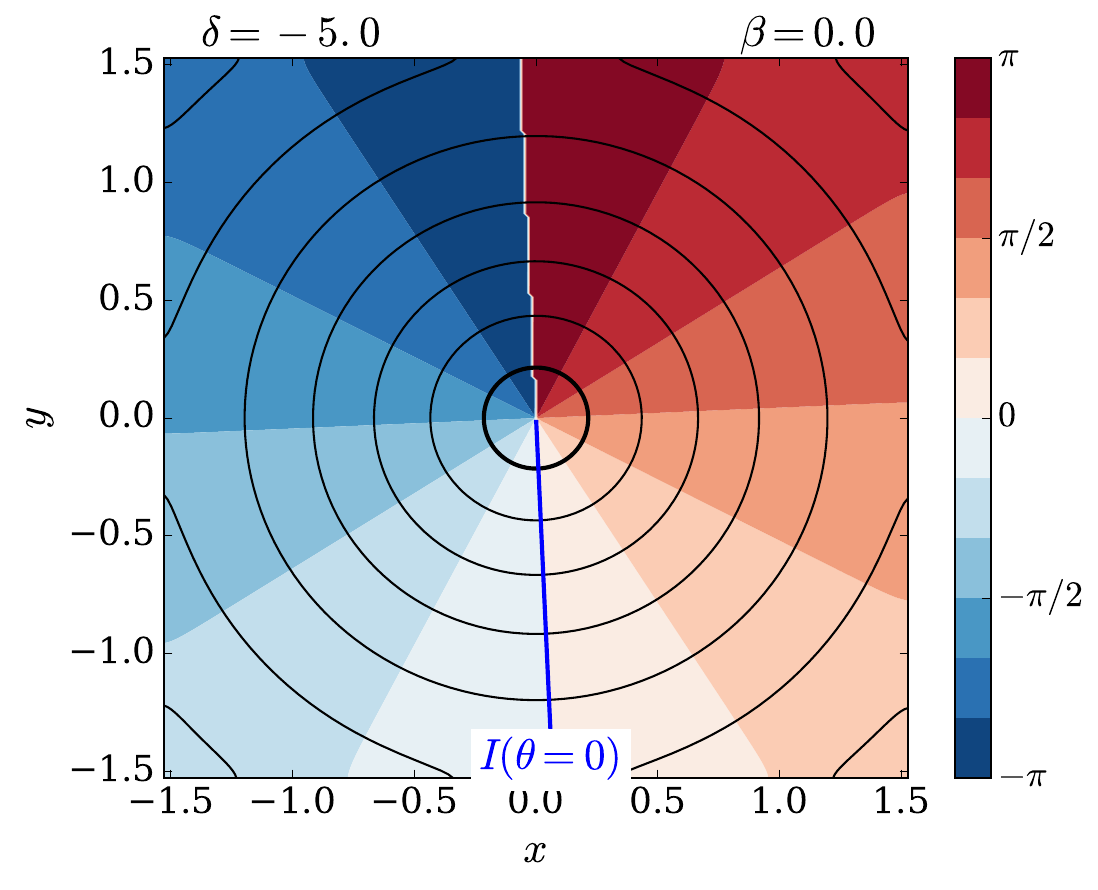}
		\caption{}
	\end{subfigure}\\
	\begin{subfigure}{0.42\textwidth}
		\includegraphics[width=.95\textwidth,height=.78\textwidth]{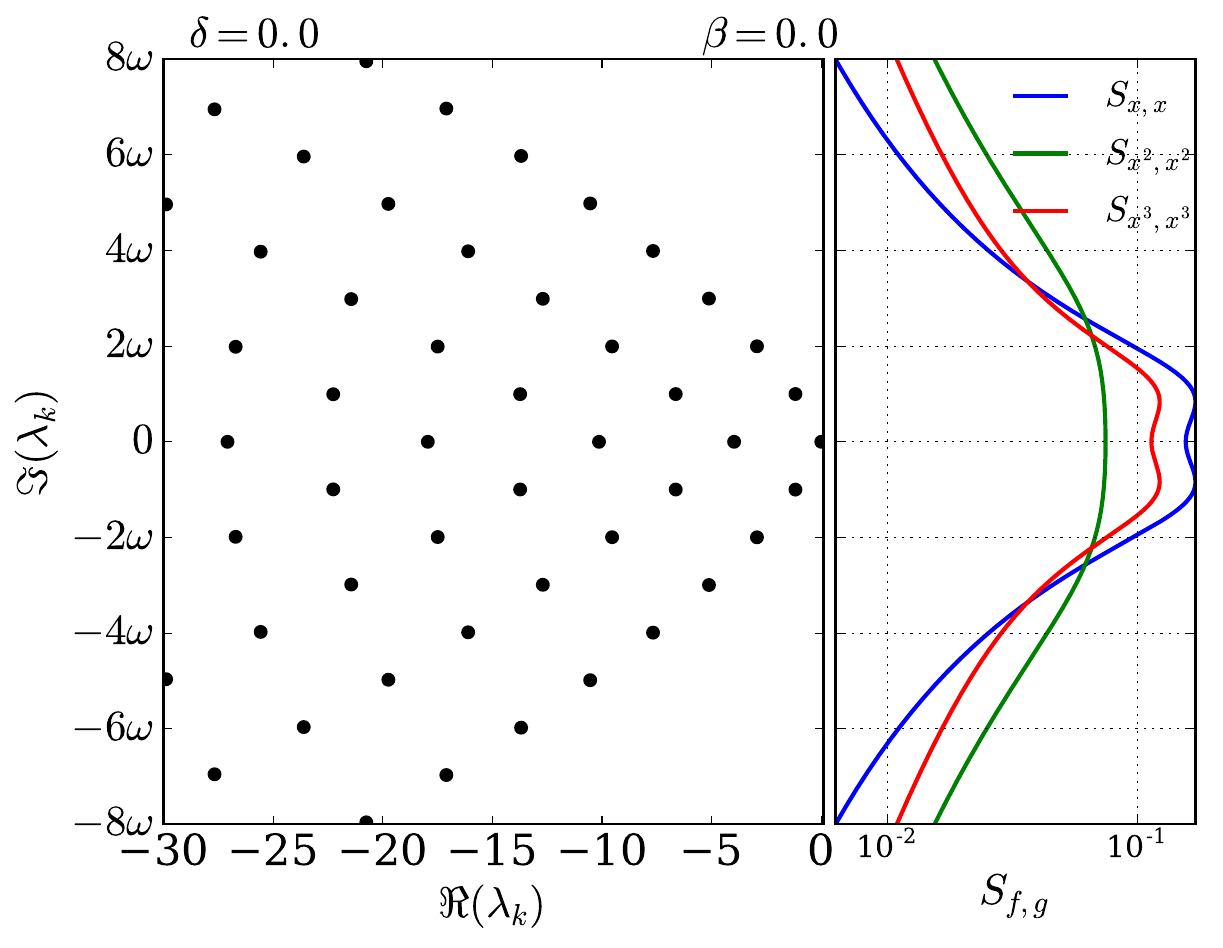}
		\caption{}
	\end{subfigure}
	\begin{subfigure}{0.42\textwidth}
		\includegraphics[width=.95\textwidth,height=.78\textwidth]{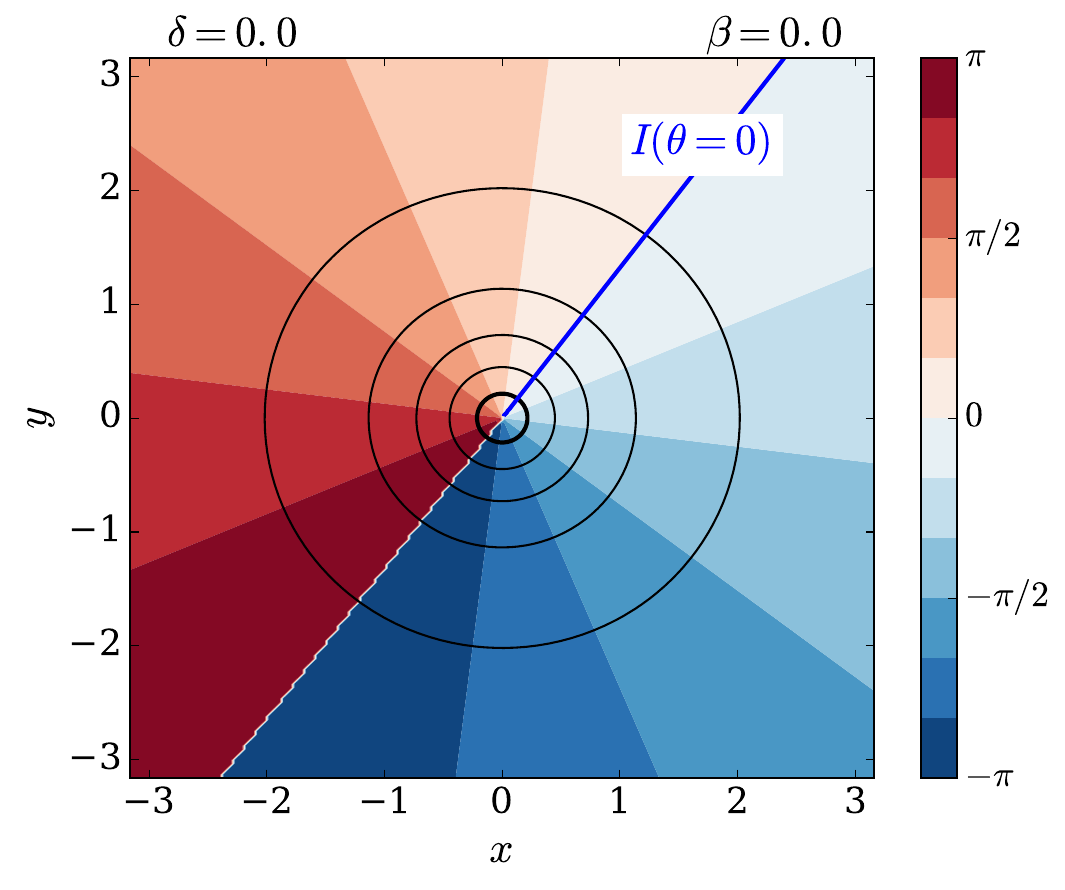}
		\caption{}
	\end{subfigure}\\
	\begin{subfigure}{0.42\textwidth}
		\includegraphics[width=.95\textwidth,height=.78\textwidth]{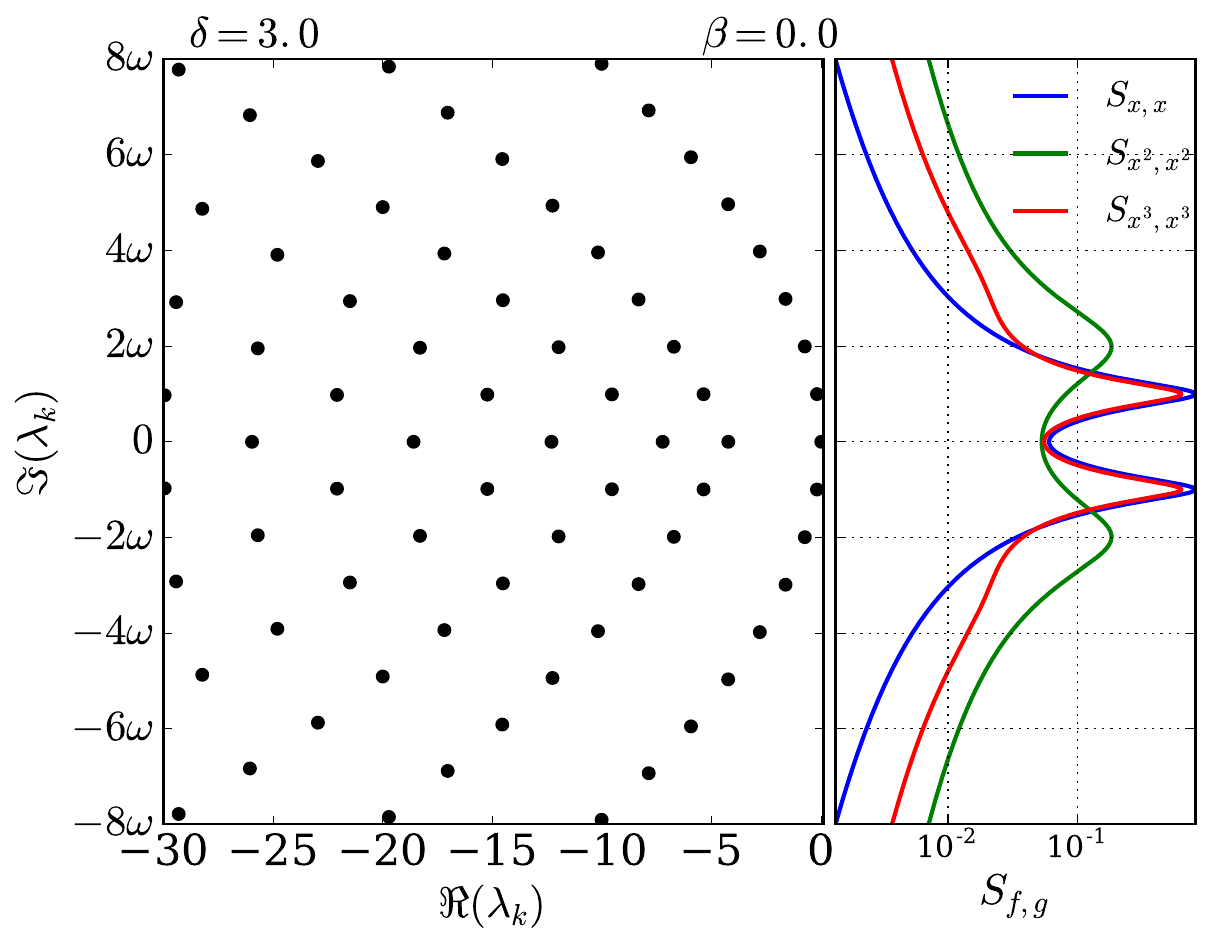}
		\caption{}
	\end{subfigure}
	\begin{subfigure}{0.42\textwidth}
		\includegraphics[width=.95\textwidth,height=.78\textwidth]{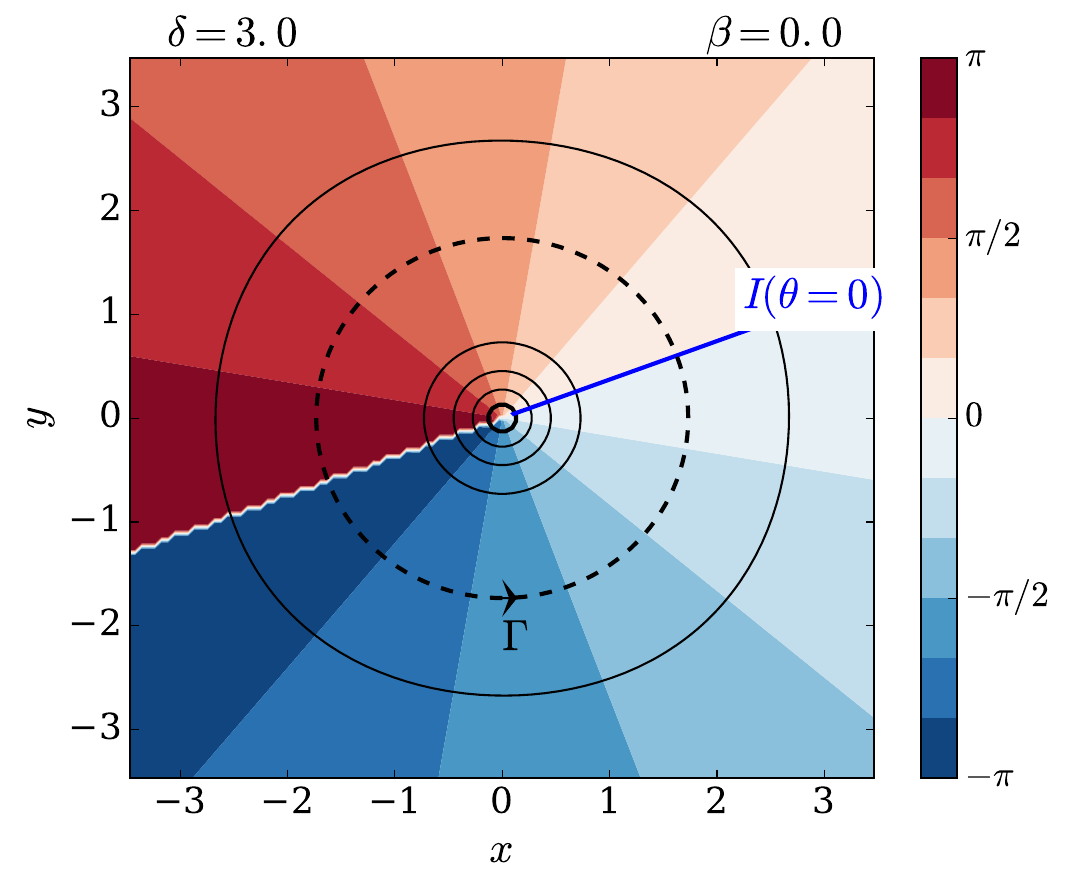}
		\caption{}
	\end{subfigure}\\
	\begin{subfigure}{0.42\textwidth}
		\includegraphics[width=.95\textwidth,height=.78\textwidth]{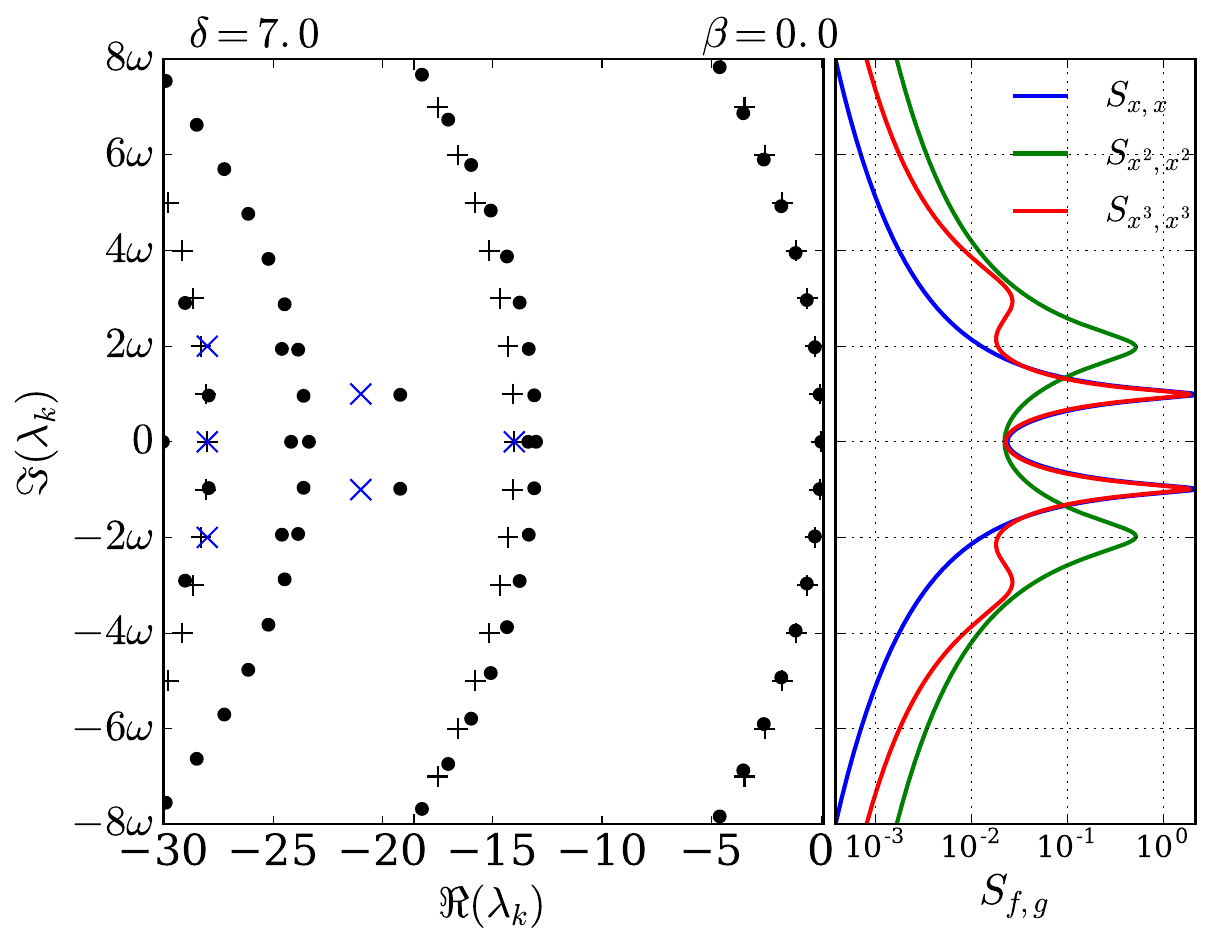}
		\caption{}
	\end{subfigure}
	\begin{subfigure}{0.42\textwidth}
		\includegraphics[width=.95\textwidth,height=.85\textwidth]{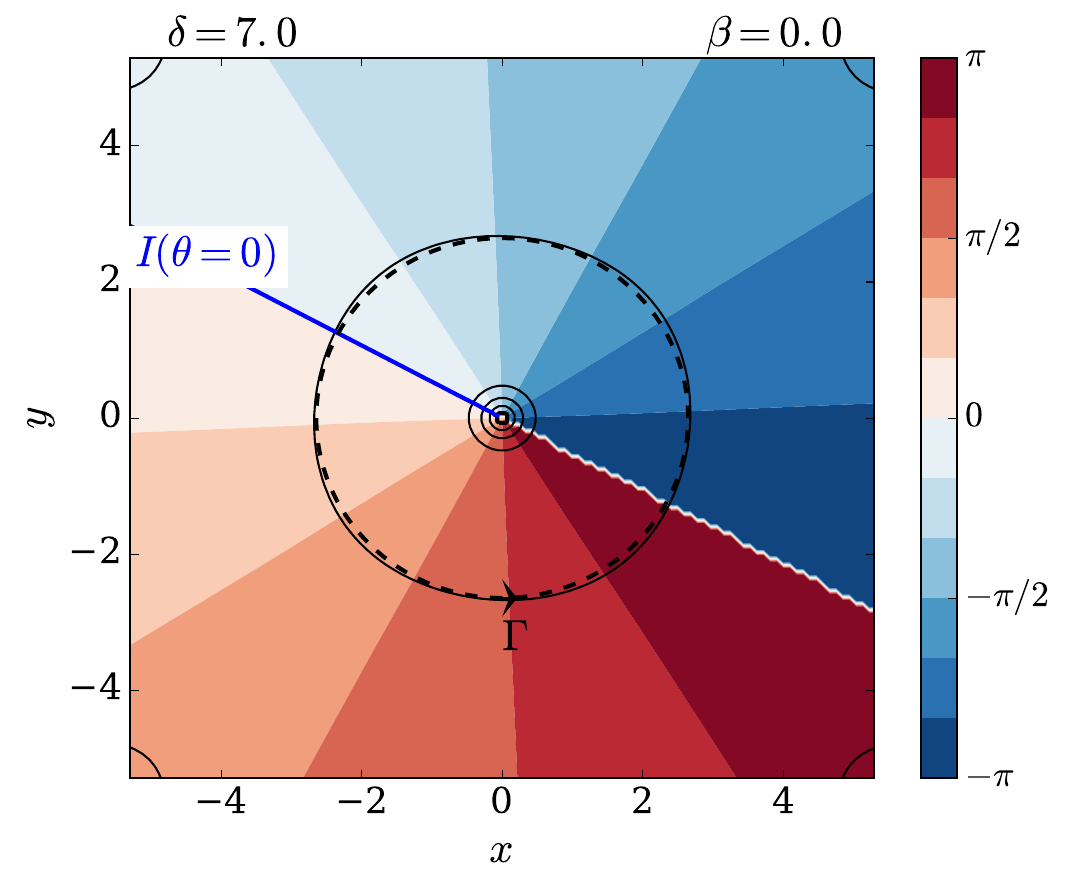}
		\caption{}
	\end{subfigure}
	\caption{
	{\bf Left}: Numerical approximation of the leading eigenvalues (black dots)
	of the Kolmogorov operator $\mathcal{K}$ for $\tilde \beta = 0$, $\gamma = \kappa = 1$ and
	(a) $\delta = -5$, (c) $\delta = 0$, (e) $\delta = 3$ and (g) $\delta = 7$.
	In addition, the small noise prediction (\ref{eq:eigValFPSub}),
	for the RP resonances of the stable fixed point,
	is also represented as black crosses in panel (a).
	In (g), the small noise predictions (\ref{eq:eigValFPSup},~\ref{eq:eigValOrbit}),
	for the eigenvalues of the unstable fixed point and of the limit cycle
	are also represented as blue crosses and black pluses, respectively.
	On the same panels, to the right, the power spectra between the three monomials $x$, $x^2$ and $x^3$
	of the $x$ coordinate are also represented as blue, green and red lines, respectively (end of caption on next page).
        {\bf Right}: Eigenfunction associated with the second eigenvalue with positive imaginary part.
	The phase of the eigenfunction is represented by filled contours 
	and its amplitude by contour lines  $(\bold{0.001}, 0.002, ...)$.}
	\label{fig:numFPeps1}
\end{figure}
We start by analyzing the numerical results for a fixed value of the noise level $\epsilon = 1$
and a vanishing twist factor $\tilde \beta = 0$, but different values of the control parameter $\delta$. 
In Fig.~\ref{fig:numFPeps1}, the leading eigenvalues
of the finite-difference approximation of the Kolmogorov operator
$\mathcal{K}$ are represented as black dots on the
left panels for (a) $\delta = -5$, (c) $\delta = 0$, (e) $\delta = 3$ and (g) $\delta = 7$.
In addition, the small noise prediction (\ref{eq:eigValFPSub}), for the RP resonances of the stable fixed point, is also represented
as black crosses in panel Fig.~\ref{fig:numFPeps1}-(a). In Fig.~\ref{fig:numFPeps1}-(g), the small noise predictions~\eqref{eq:eigValFPSup},~\eqref{eq:eigValOrbit},
for the eigenvalues of the unstable fixed point and of the limit cycle
are also represented as blue crosses and black pluses, respectively.
On the same panels, to the right, the power spectra between the three monomials $x$, $x^2$ and $x^3$ of the $x = r \cos \theta$ coordinate are also represented as blue, green and red lines, respectively.
According to the order of the harmonics in the small-noise expansions~\eqref{eq:complexHermite} and~\eqref{eq:eigvecHarmonic} for the eigenfunctions and adjoint eigenfunctions, the observable $x$ is expected to project mainly on the  eigenfunctions of the first complex pair of eigenvalues, $x^2$ on the  eigenfunctions of the second pair and $x^3$ on the eigenfunctions of both the first and the third pair.
These power spectra are calculated from the numerical approximations of the eigenvalues, eigenfunctions and adjoint eigenfunctions (i.e.~the eigenvectors of the transpose of the finite-difference approximation of $\mathcal{K}$)
according to the spectral decomposition~\eqref{eq:spectralPower}.
Finally, on the right panels, the corresponding eigenvector associated
with the second eigenvalue with positive imaginary part represented
\footnote{Recall that the eigenfunction associated with the first eigenvalue 
is constant \cite[Definition 1.(i)]{Chekroun_al_RP2}, while the eigenfunction of the adjoint corresponds to the invariant measure.}.
The phase of the eigenvectors is represented by filled contours 
and their amplitude by contour lines  $(\bold{0.001}, 0.002, ...)$.

For a small value of $\delta$, panel (a) of Fig.~\ref{fig:numFPeps1}, a triangular structure of eigenvalues is found and,
because of the large gap between the eigenvalues and the imaginary axis,
the power spectra are broad, with no distinct resonance.
The leading eigenvalues are in quantitative agreement
with the small-noise expansion~\eqref{eq:eigValFPSub} around the stable fixed point
represented in Fig.~\ref{fig:artMixingEigVal}-(c).
The corresponding second eigenfunction in panel (b) of Fig.~\ref{fig:numFPeps1} also agrees with the expansion $\psi_{01}$ of
\eqref{eq:complexHermite} represented in Fig.~\ref{fig:artMixingEigVal}-(e).
On the other hand, the secondary columns of eigenvalues are farther from the imaginary axis
than the small-noise expansions.
Since the numerical results have converged, this must be due to higher-order terms in the
expansions which are not taken into account and which can depend on the noise level $\epsilon$ and be responsible for more mixing.
This points at the fact that, in the expansion~\eqref{eq:eigValFPSub}, we do not control the weight of the higher-order terms in $\epsilon^2$ as we switch from one eigenvalue to the next.
One should thus take this effect into account when the noise level is strong with respect to the contraction measured by $\delta$.
This is particularly important when considering eigenvalues farther from the imaginary axis.
Indeed, the latter typically exhibit more complex nodal properties, as is the case in the small-noise expansion \eqref{eq:complexHermite} and in general for multi-dimensional Ornstein-Uhlenbeck processes for which the eigenfunctions are polynomials of increasing degree \cite{Metafune2002a}, and are thus more difficult to approximate \cite[see e.g.]{Varga1971}.

As the control parameter $\delta$ is increased (from panel (a) to (c) in Fig.~\ref{fig:numFPeps1})
the eigenvalues get closer to the imaginary axis,
as expected from the weaker stability of the limit cycle and
as predicted by the expansion (\ref{eq:eigValFPSub}) for the stable fixed point.
One can also see from the larger gaps between the contour lines in Fig.~\ref{fig:numFPeps1}-(d) compared {\mk to those of Fig.~\ref{fig:numFPeps1}-(b)}
that the amplitude of the second eigenvector flattens, in agreement with \eqref{eq:complexHermite}.
Because of the approach of the first complex pair of eigenvalues to the imaginary axis,
in agreement with the spectral decomposition (\ref{eq:spectralPower})
and the eigenfunction expansions (\ref{eq:complexHermite},~\ref{eq:eigvecHarmonic}),
broad peaks begin to appear in the power spectra of the observables $x$ and $x^3$
at angular frequencies given by the imaginary part of the eigenvalues.
On the other hand, the second pair is still too far for the observable $x^2$ to resonate.

As $\delta$ is further increased (panels (c-d) to (g-h) of Fig.~\ref{fig:numFPeps1}) and the bifurcation point is crossed,
a rather smooth transition from the small-noise expansions for $\delta > 0$
and then $\delta < 0$ occurs,
in which the first line of eigenvalues gets closer and closer to the imaginary axis.
As a result, strong resonant behavior occurs for all three observables,
as can be seen from the sharpening of the spectral peaks at the position of the first three harmonics.
The peaks remain finite, however, since, in agreement with the small-noise expansion (\ref{eq:eigValOrbit}),
a spectral gap persists between the eigenvalues and the imaginary axis, due to the noise.
Finally, for $\delta = 7$ inn panel Fig.~\ref{fig:numFPeps1}-(g), one finds the superposition of a family of parabolas
and of a triangular family of eigenvalues, in very good agreement
with the small-noise expansions (\ref{eq:eigValOrbit}) and (\ref{eq:eigValFPSup})
for the limit cycle and for the unstable fixed point, respectively,
while the corresponding eigenvector on panel Fig.~\ref{fig:numFPeps1}-(h) has an almost uniform amplitude, in agreement with (\ref{eq:eigvecHarmonic}), except at the origin (c.f. Fig.~\ref{fig:artMixingEigVal}-(d, f)).

In agreement with the results of Section~\ref{sec:StocAnaHopf},
the spectrum remains discrete during the transition, as opposed to the deterministic case (c.f.~\cite{Gaspard2001a}).
On the other hand, precisely how the transition occurs
could not be predicted analytically from the geometric properties of the deterministic flow.
In particular, eigenvalues farther away from the real axis tend to approach the imaginary axis at a faster rate than the others, resulting in a curving of the triangle array of eigenvalues, while the second eigenvector continues to flatten away from the origin.
Eventually (from panel Fig.~\ref{fig:numFPeps1}-(e) to Fig.~\ref{fig:numFPeps1}-(g)), parabolas of eigenvalues detach one after the other,
while other eigenvalues persist as a triangular family.

So far, these numerical experiments have mostly allowed to test the validity of
the small-noise expansions of Section~\ref{sec:smallNoiseExpansion}
when the twist factor $\tilde \beta$ is vanishing and to reveal unpredicted phenomena close to the bifurcation point.
Next, the role of $\tilde \beta$ is investigated and a more detailed
numerical analysis of the change of the RP spectrum close to the bifurcation point is given.

\subsubsection{Crossing the bifurcation point, with a nonzero twist factor}
\label{sec:numericalBetaPos}
\begin{figure}
	\centering
	\begin{subfigure}{0.42\textwidth}
		\includegraphics[width=\textwidth]{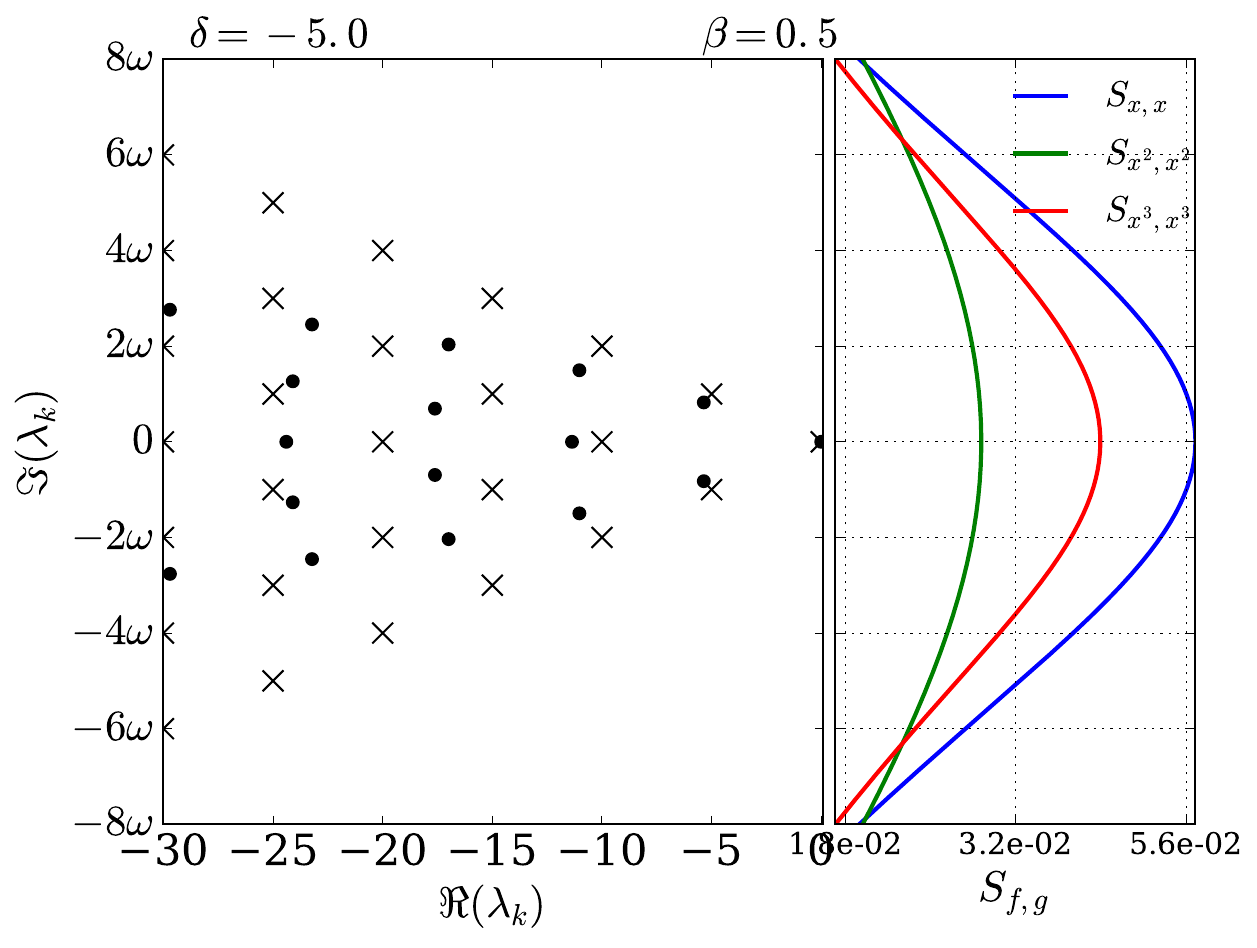}
		\caption{}
	\end{subfigure}
	\begin{subfigure}{0.42\textwidth}
		\includegraphics[width=\textwidth]{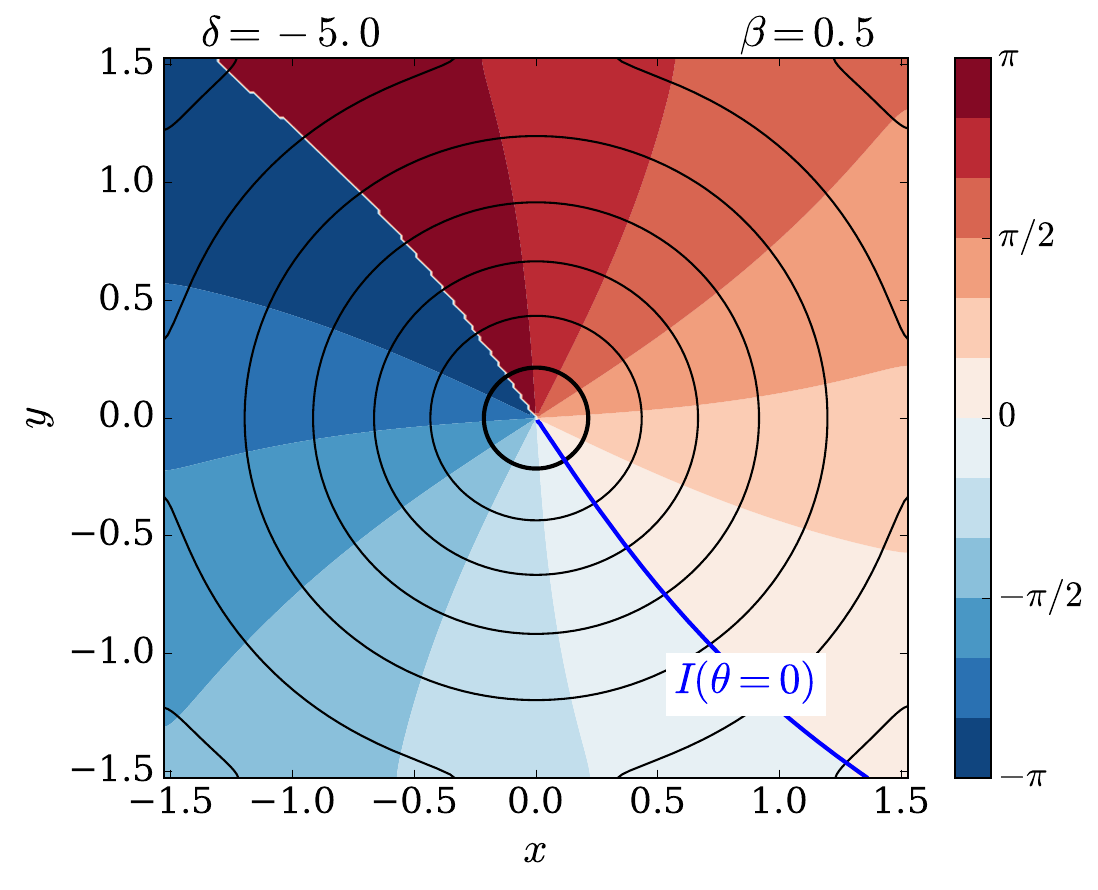}
		\caption{}
	\end{subfigure}\\
	\begin{subfigure}{0.42\textwidth}
		\includegraphics[width=\textwidth]{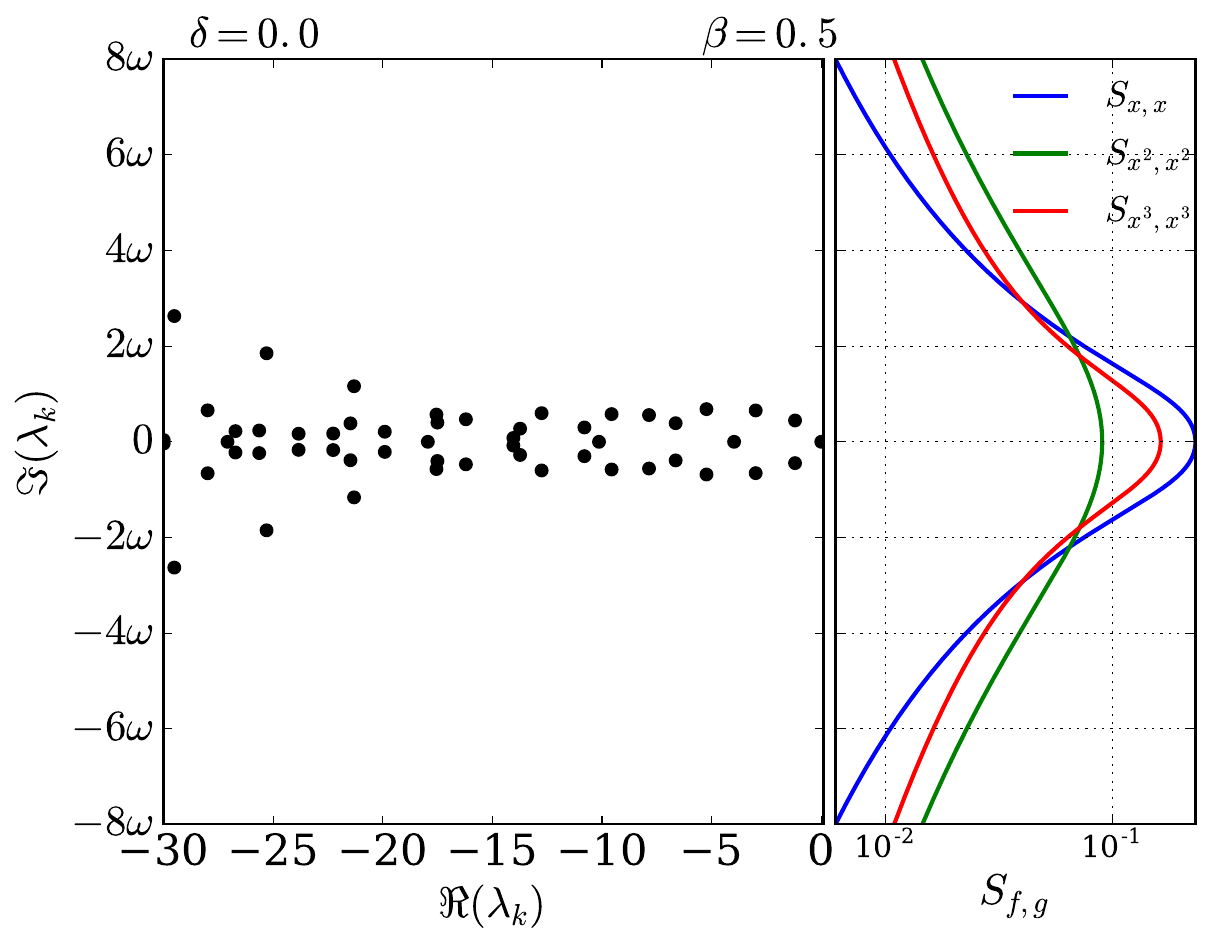}
		\caption{}
	\end{subfigure}
	\begin{subfigure}{0.42\textwidth}
		\includegraphics[width=\textwidth]{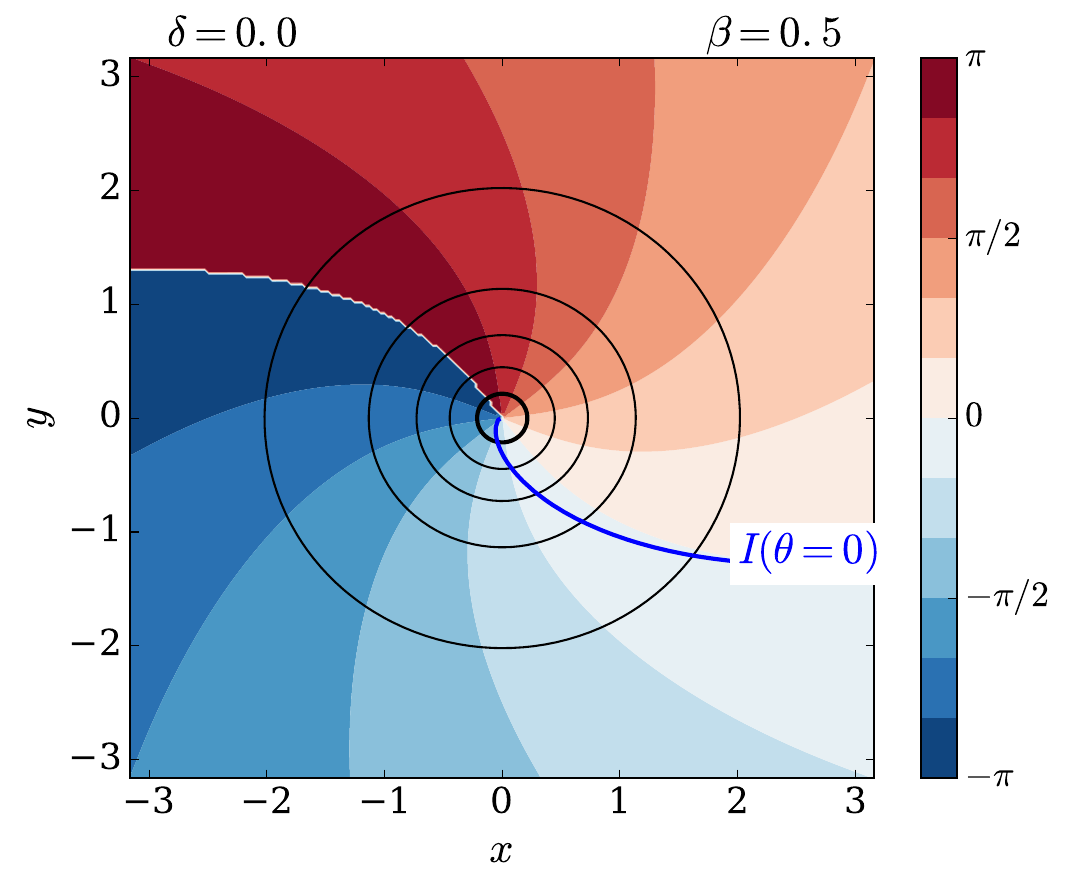}
		\caption{}
	\end{subfigure}\\
	\begin{subfigure}{0.42\textwidth}
		\includegraphics[width=\textwidth]{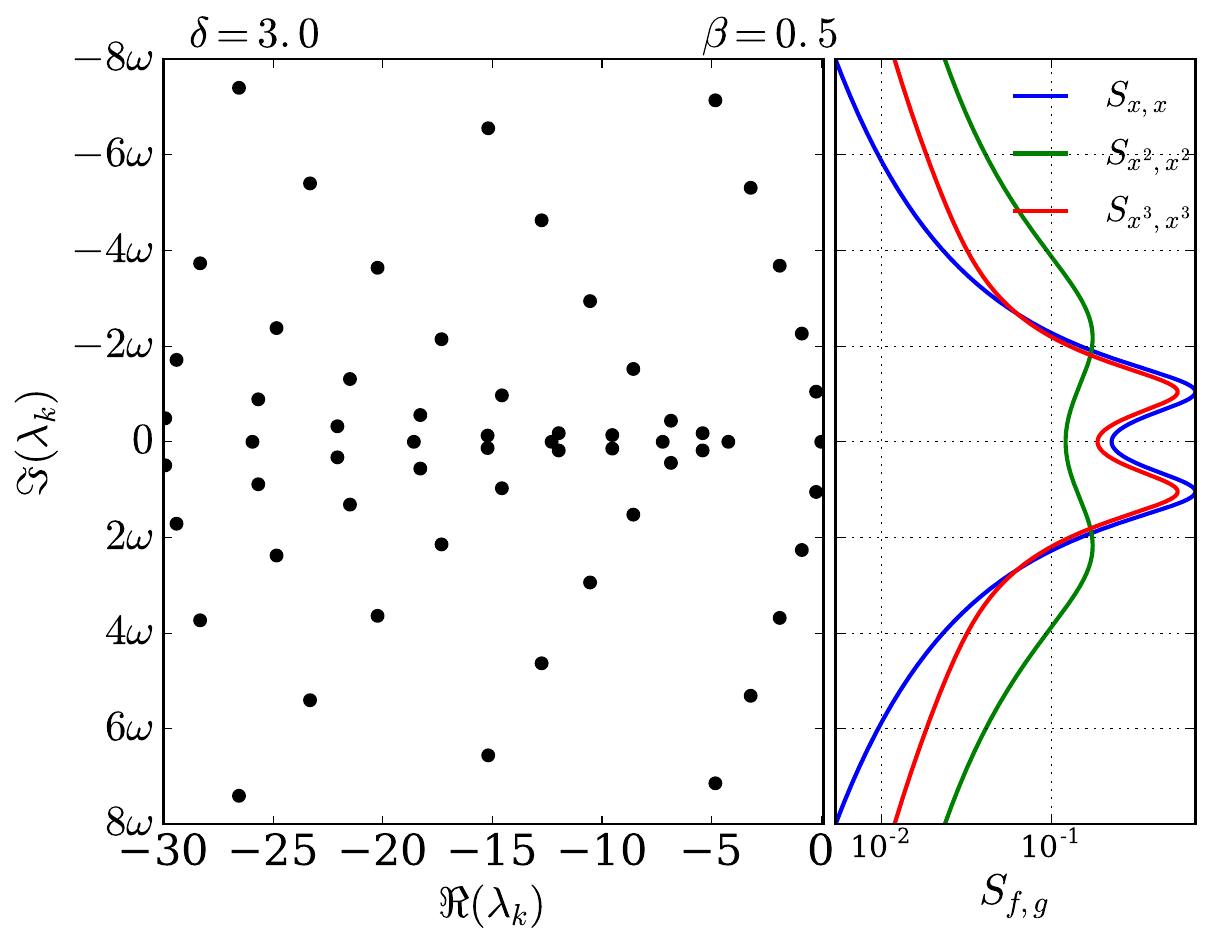}
		\caption{}
	\end{subfigure}
	\begin{subfigure}{0.42\textwidth}
		\includegraphics[width=\textwidth]{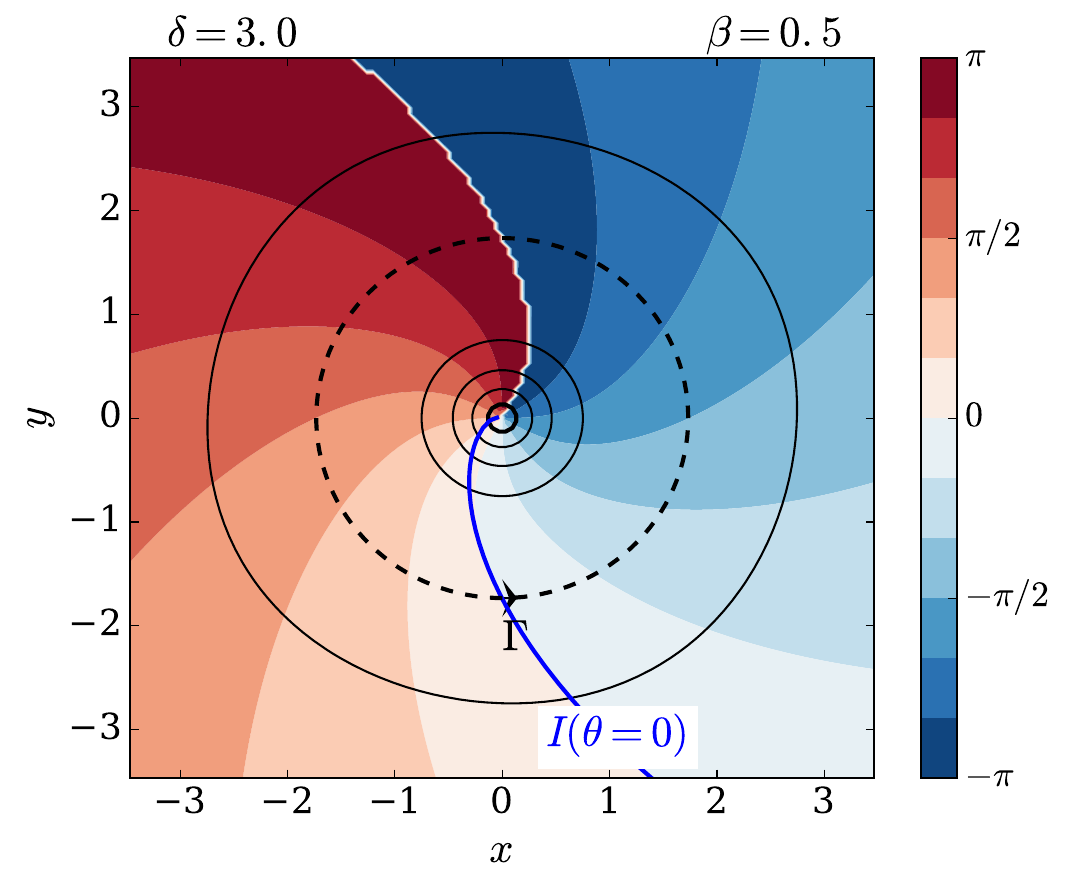}
		\caption{}
	\end{subfigure}\\
	\begin{subfigure}{0.42\textwidth}
		\includegraphics[width=\textwidth]{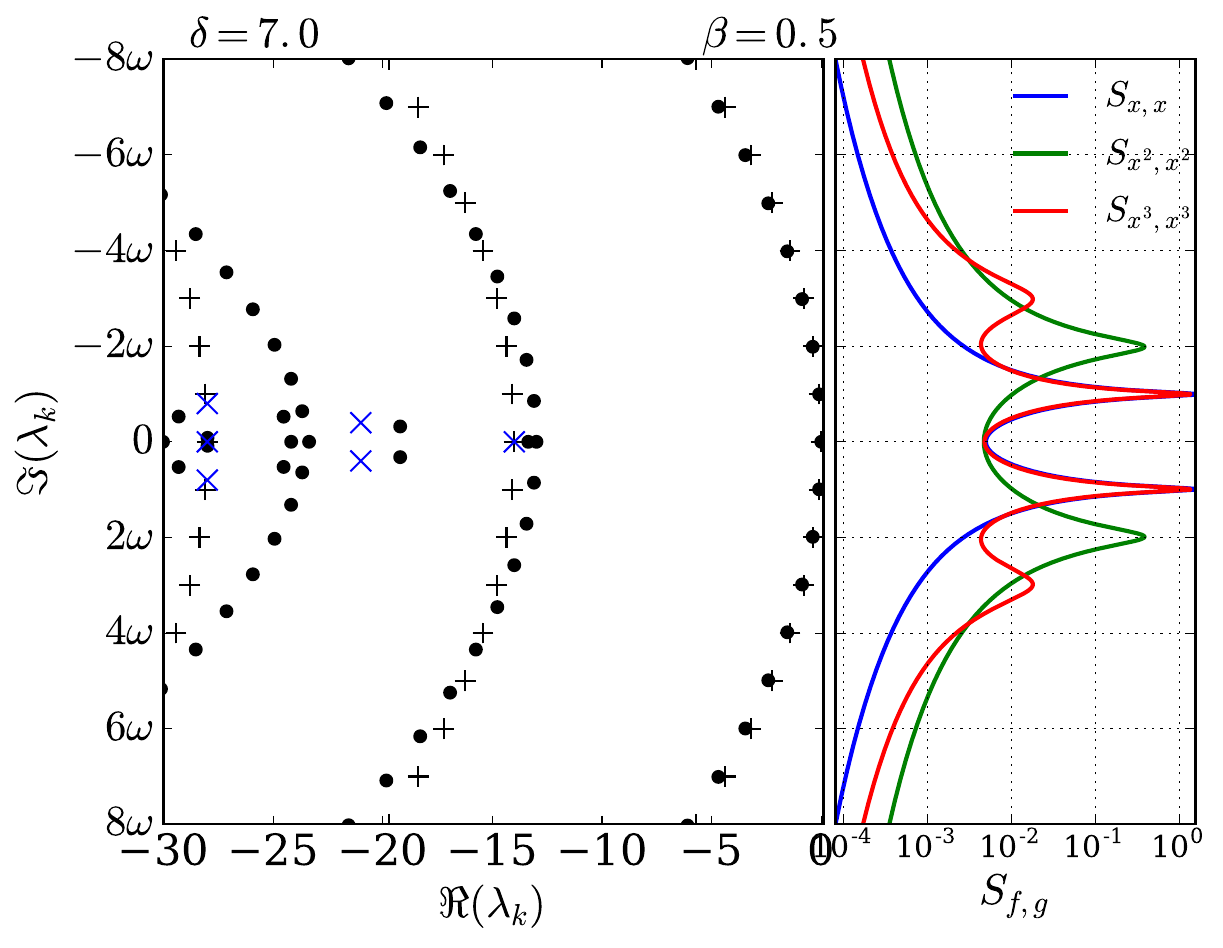}
		\caption{}
	\end{subfigure}
	\begin{subfigure}{0.42\textwidth}
		\includegraphics[width=\textwidth]{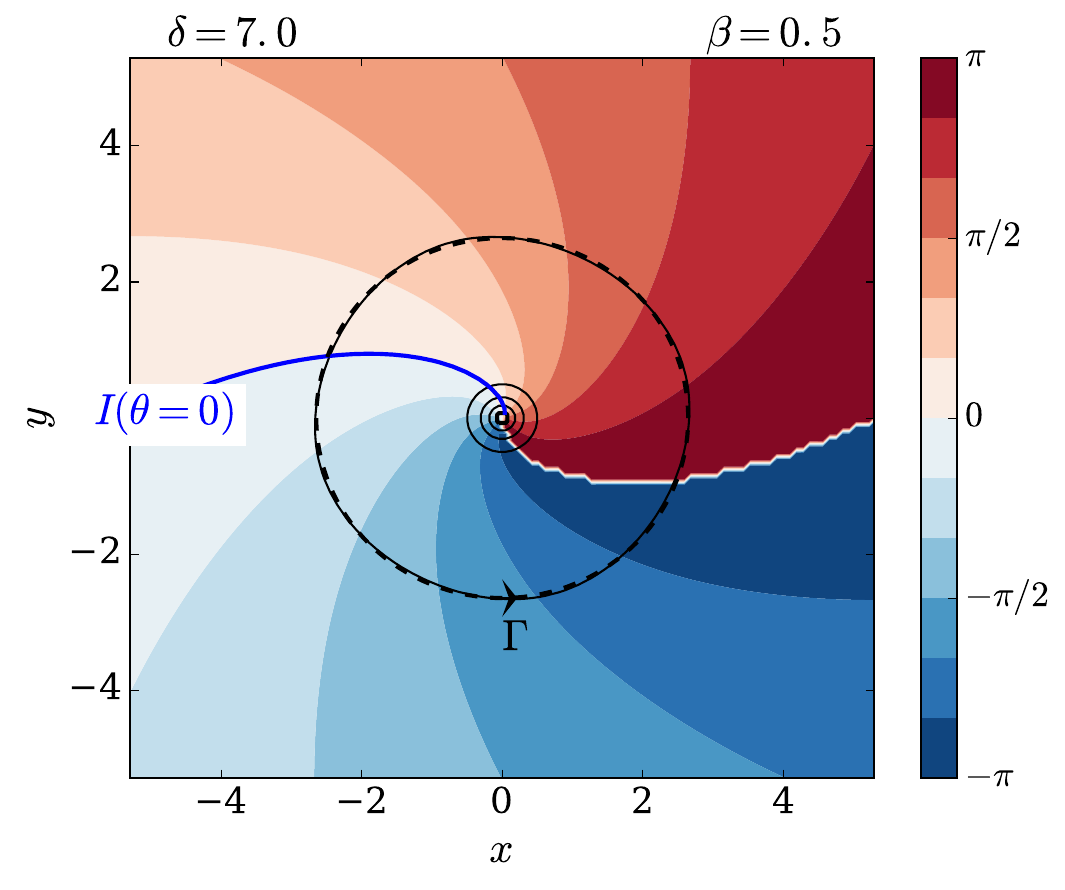}
		\caption{}
	\end{subfigure}
	\caption{Same as Fig.~\ref{fig:numFPeps1} but with $\tilde \beta = 0.5$.}
	\label{fig:numFPeps1beta}
\end{figure}
To learn more about the change in the spectrum when the twist factor $\tilde \beta$ is nonzero, the same set of numerical experiments as in the previous subsection~\ref{sec:numericalBetaZero} is performed, but with $\tilde \beta = 0.5 > 0$.
The results are reported in Fig.~\ref{fig:numFPeps1beta} in the same way as in Fig.~\ref{fig:numFPeps1}.
Below the bifurcation point, the small-noise expansions \eqref{eq:eigValFPSub} and \eqref{eq:complexHermite} do not depend on $\tilde \beta$, so that panels (a) and (b) of Fig.~\ref{fig:numFPeps1} and~\ref{fig:numFPeps1beta}
should be identical.
{\mkr As closer inspection shows} this is not exactly the case, so that the noise level is strong enough to excite higher-order terms in $\epsilon$ which depend on $\tilde \beta$, in agreement with the $\mathcal{O}_{\tilde \beta}$ in the expansions of Proposition~\ref{prop:eigen_below} and~\ref{prop:eigen_above}.
As a result, the imaginary parts of the eigenvalues are smaller, due the decrease of the frequency of the fundamental and its harmonics induced by the twist factor $\tilde \beta$.
In addition, the isolines of phase of the second eigenvector (panel (b) of Fig.~\ref{fig:numFPeps1beta}) are slightly tilted.
One {\mkr discerns} on panels (c) and (d) of Fig.~\ref{fig:numFPeps1beta} that both effects become more prominent closer to the bifurcation point,
i.e.~the eigenvalues are even closer to the real axis and the isolines of phase even more tilted.
In particular the fact that the eigenvalues get closer to the real axis, and even cross it,
results in a dramatic change in the power spectra where the resonances are much more centred,
so that no spectral peak is visible away from $0$ in Fig.~\ref{fig:numFPeps1beta}-(c). 

On the other hand, one {\mkr distinguishes} on panels (g) and (h) of Fig.~\ref{fig:numFPeps1beta} that the small-noise expansions \eqref{eq:eigValFPSup}, \eqref{eq:eigValOrbit} and \eqref{eq:eigvecHarmonic}
are in very good agreement with the numerical results far above the bifurcation point.
In particular, the increase of the spectral gap associated with the
increase of the phase diffusion due to the nonzero twist factor $\tilde \beta$
as well as the tilt of the isolines of phase of the second eigenvector with the isochrons are captured.
To summarize, the twist factor $\tilde \beta$ is responsible for increasing the mixing, changing the position of the harmonics and twisting the eigenvectors.

\subsubsection{Parameter dependence close to bifurcation}
\begin{figure}
	\centering
	\begin{subfigure}{0.48\textwidth}
		\includegraphics[width=\textwidth]{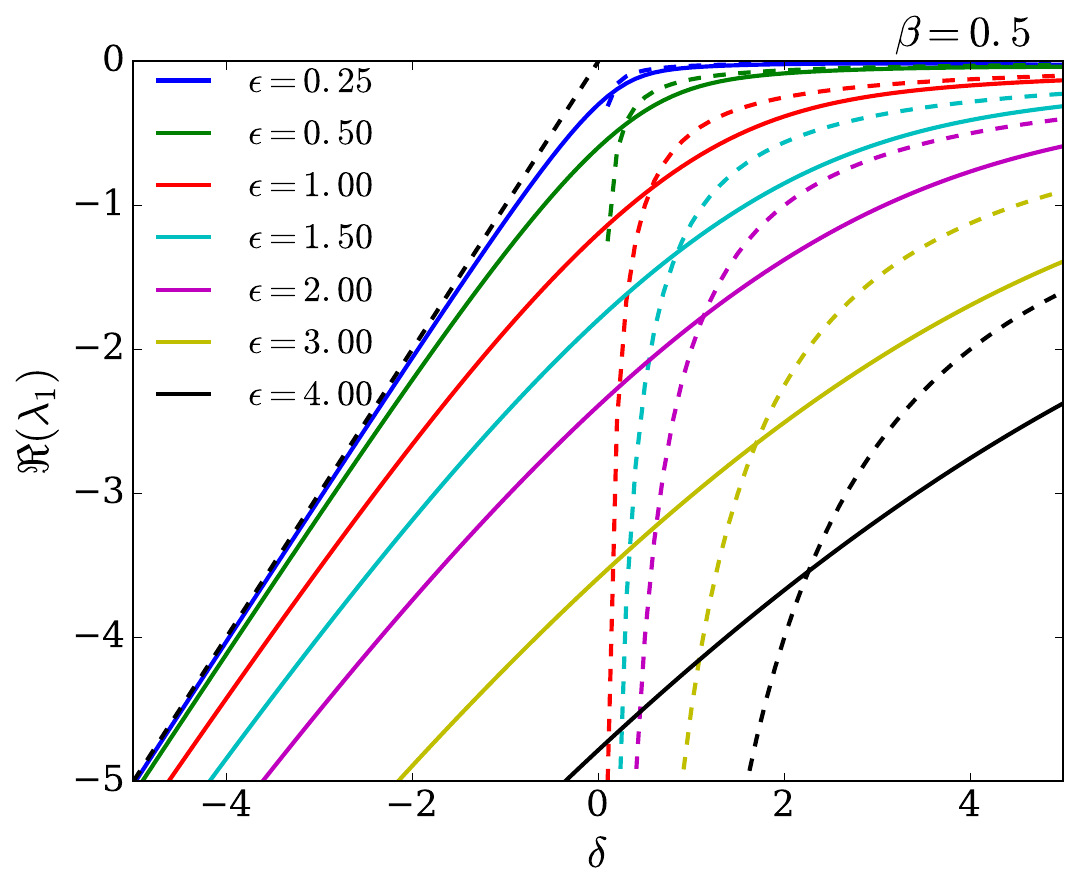}
		\caption{}
	\end{subfigure}
	\begin{subfigure}{0.48\textwidth}
		\includegraphics[width=\textwidth]{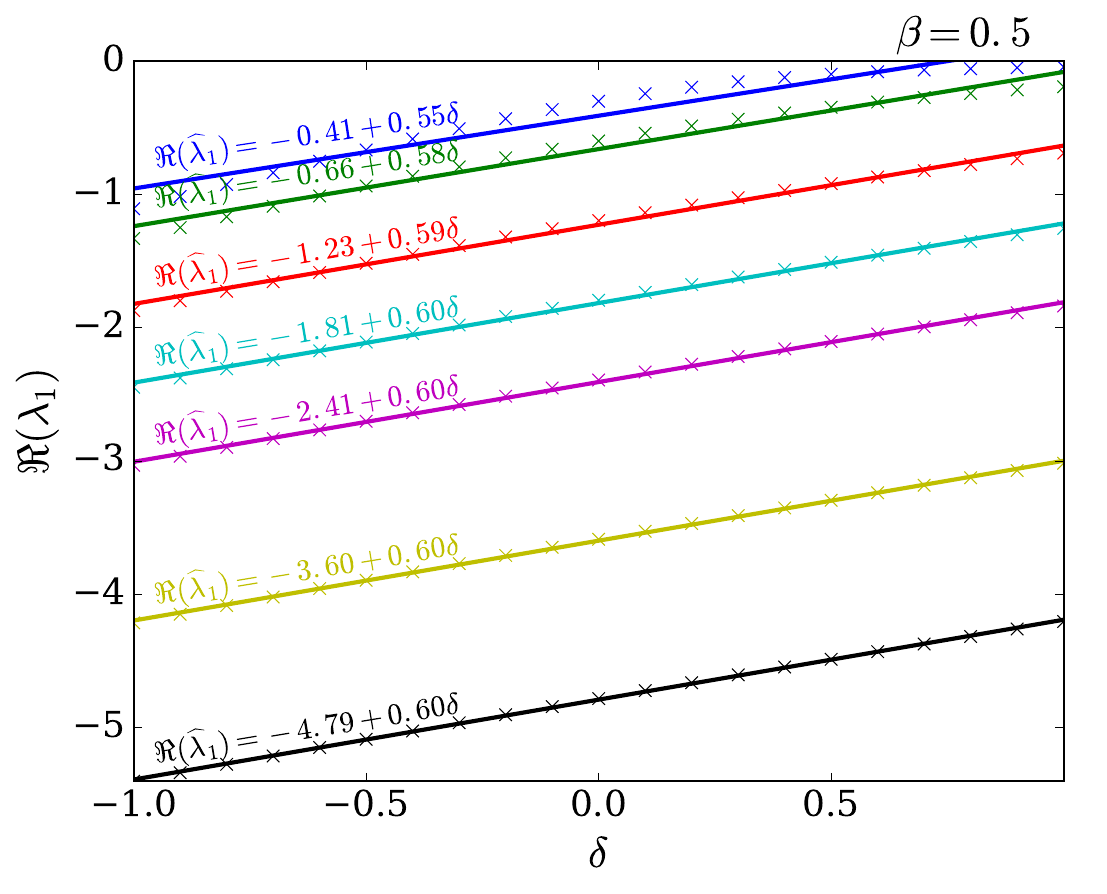}
		\caption{}
	\end{subfigure}\\
	\begin{subfigure}{0.48\textwidth}
		\includegraphics[width=\textwidth]{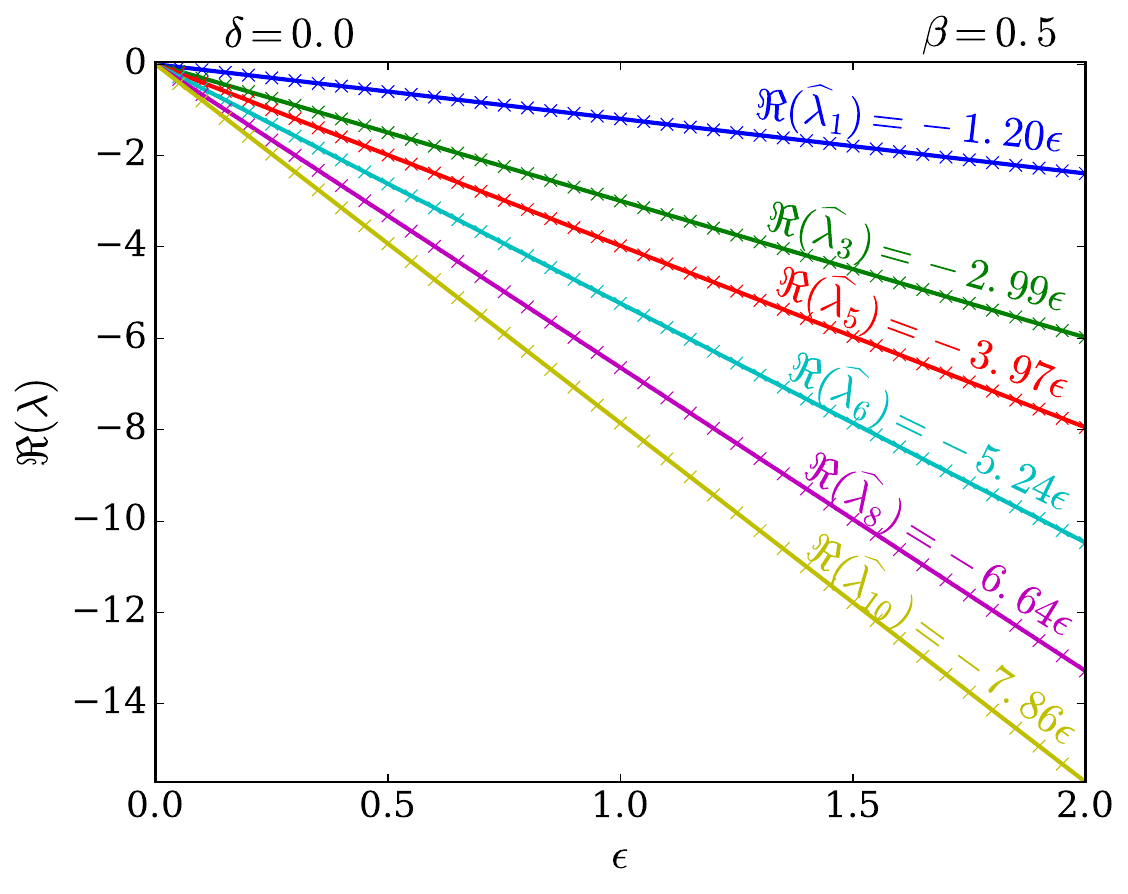}
		\caption{}
	\end{subfigure}
	\caption{{\bf Top left}: Real part of the approximated second eigenvalue $\lambda_1$ versus $\delta$ for $\tilde \beta = 0.5$ (thick lines) and with $\gamma = \kappa = 1$.
	Different colors correspond to different values of the noise level $\epsilon$ (see the legend).
	For $\delta < 0$, the curve $\Re(\lambda_1) = \delta$ corresponding to the small-noise expansion
	(\ref{eq:eigValFPSub}) is plotted as a dashed black line.
	For $\delta > 0$, the curves $\Re(\lambda_1) = -\epsilon^2 (1 + \tilde \beta^2) / (2 R^2)$
	corresponding to the small-noise expansion
	(\ref{eq:eigValOrbit}) are plotted as dashed lines in the color corresponding that
	of the numerical approximation for a given $\epsilon$.\\
	{\bf Top right}: Zoom to $\delta$ in the interval $[-1, 1]$. The numerical approximations are now represented as
	crosses in the same color as on the right together with a least-square fit of the line $y = a + b \delta$.\\
	{\bf Bottom}: Real part of the approximated leading eigenvalues versus 
	he noise level $\epsilon$ for $\delta = 0$ (crosses).
	The lines represent least square fits $y = a + b \epsilon$.}
	\label{fig:w2VSmu}
\end{figure}
In order to better understand the parameter dependence of the RP spectrum close to bifurcation,
we focus now on the real part of the second eigenvalue $\lambda_1$.
Its numerical approximation is represented in Fig.~\ref{fig:w2VSmu} for varying $\delta$ and $\epsilon$ with fixed $\tilde \beta = 0.5$.
On the left panel, each line corresponds to the numerical approximation of $\Re(\lambda_1)$
for different values of the noise level $\epsilon$ (color code in the legend).
In addition, the dashed black line $\Re(\lambda_1) = \delta$
corresponds to the small-noise expansion (\ref{eq:eigValFPSub}) for $\delta < 0$
and the colored dashed lines $\Re(\lambda_1) = -\epsilon^2 (1 + \tilde \beta^2) / (2 R^2)$ 
correspond to the small-noise expansion (\ref{eq:eigValOrbit}) for $\delta > 0$
and different values of $\epsilon$.
As expected, for smaller values of $\epsilon$ and larger absolute values of $\delta$,
the numerical approximations converge to the small-noise expansions.
On the other hand, strong deviations occur when the noise level is increased
or {\mk when the system is placed} closer to the bifurcation point.
There, the eigenvalue transits smoothly from the small-noise expansions for
$\delta < 0$ to $\delta > 0$.
Interestingly, this change occurs more slowly when $\epsilon$ is large,
so that the noise has a stabilizing effect on the dependence of the eigenvalue of $\delta$.

On the right panel of Fig.~\ref{fig:w2VSmu}, a zoom to $\delta \in [-1, 1]$
allows for a more detailed analysis of the changes in the second eigenvalue.
There, the numerical approximations of $\Re(\lambda_1)$ are represented
by crosses in the same colors as the left panel for the same values of $\epsilon$.
On top of them is plotted their least-square fit of the line $y = a + b \delta$.
Interestingly, {\mkr the linear regressions performs} very well
for a range of {\mkr $\delta$'s values close to} $0$, the latter increasing with $\epsilon$.
Even more surprising, the slope of the linear regressions does not seem
to depend on the noise level $\epsilon$.
In other words, the dependence of the minimum decay rate of correlations $\Re(\lambda_1)$
on the control parameter $\delta$ around $0$ is close to linear,
on a range which increases with the noise level $\epsilon$ but with a coefficient which does not depend on $\epsilon$.

To learn more about the role of the noise for $\delta = 0$,
the approximation of the real part of the leading eigenvalues versus $\epsilon$
are represented on the bottom panel of Fig.~\ref{fig:w2VSmu} by crosses.
Least square fits $y = a + b \epsilon$ are also represented by lines.
In agreement with the {\mk scaling relationship} \eqref{eq:rescaleCritical},
all real parts depend linearly on $\epsilon$.
Yet, it is interesting to see that the slope of the lines is steeper
for higher-rank eigenvalues, farther from the imaginary axis.
In other words, eigenvalues farther from the imaginary axis
are more sensitive to the noise, so that, as the noise level is increased,
they move away from the imaginary axis at a faster rate.

\begin{figure}
	\centering
	\begin{subfigure}{0.48\textwidth}
		\includegraphics[width=\textwidth]{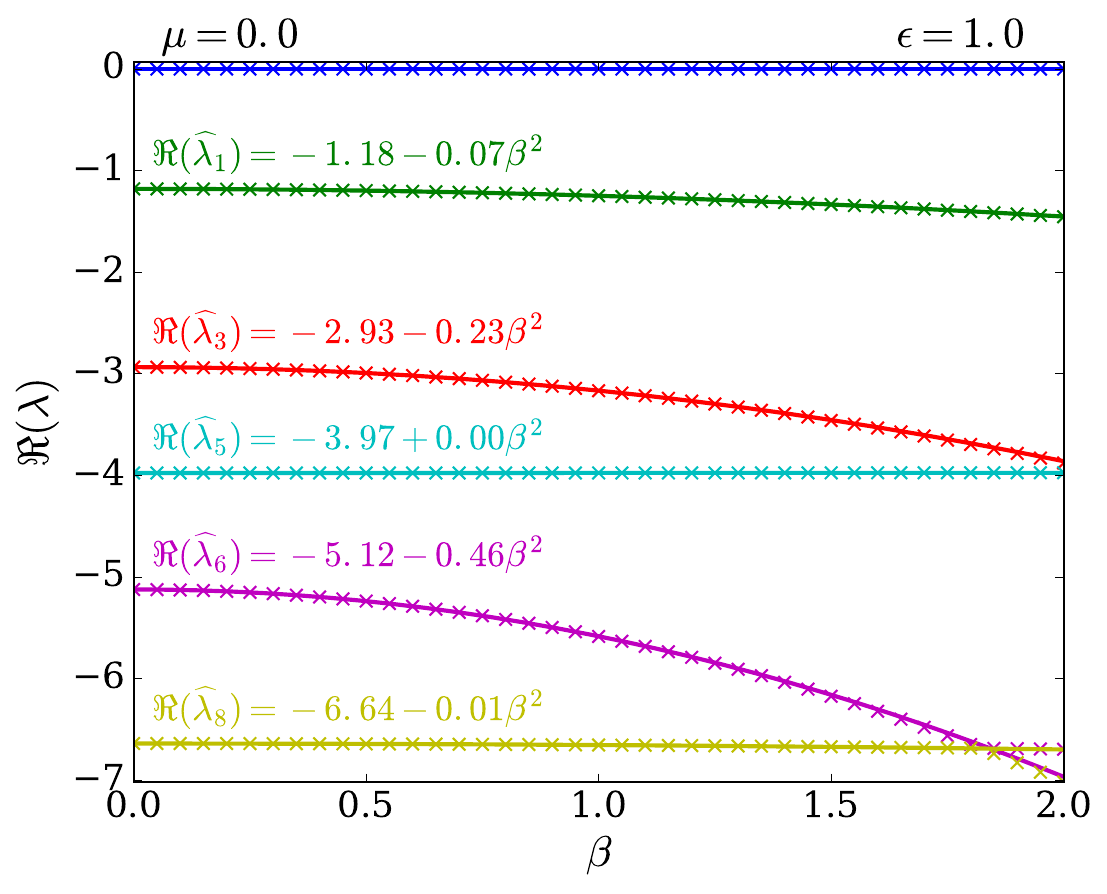}
		\caption{}
	\end{subfigure}
	\begin{subfigure}{0.48\textwidth}
		\includegraphics[width=\textwidth]{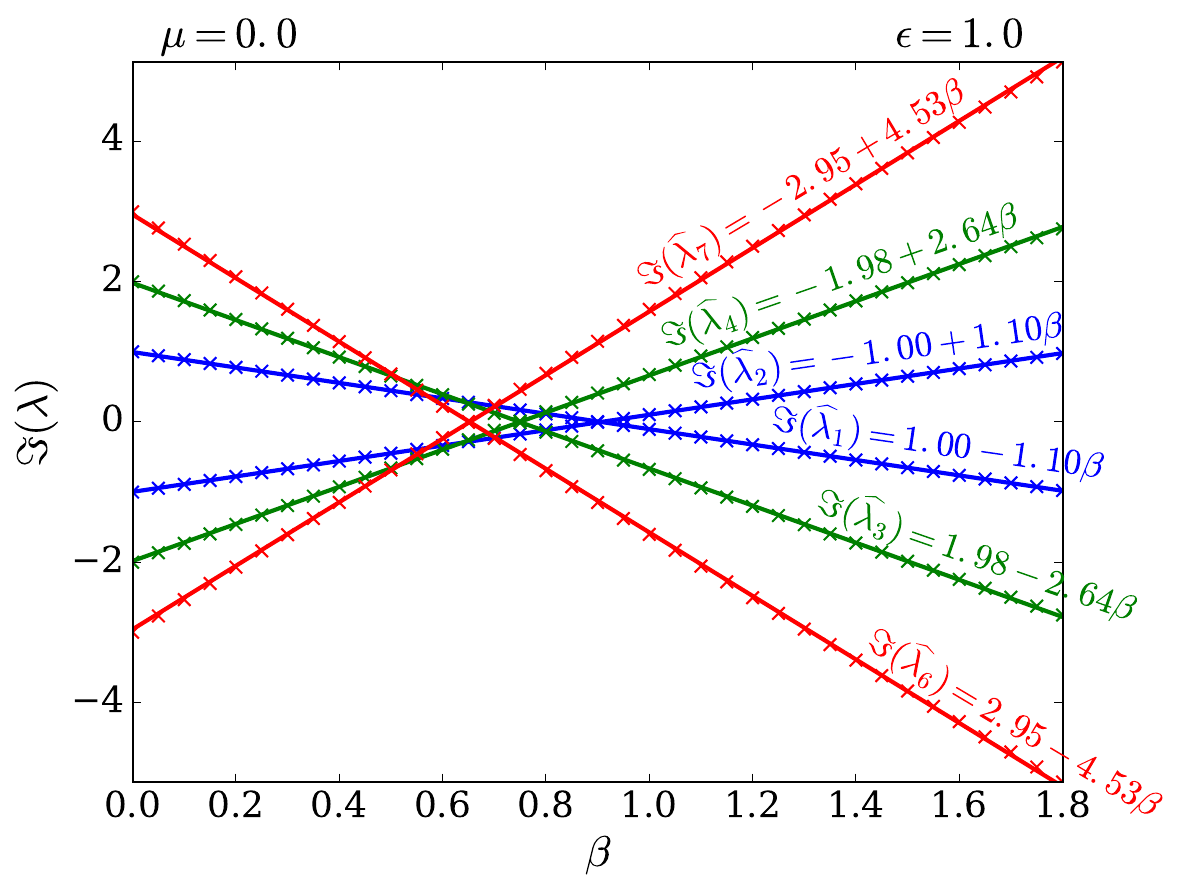}
		\caption{}
	\end{subfigure}
	\caption{Left: Real part of the finite-difference approximation of the second RP resonance $\lambda_1$ versus $\delta$ (thick lines),
	for $\epsilon = 0.25$ (blue), $\epsilon = 0.5$ (green), $\epsilon = 1$ (red), $\epsilon = 1.5$ (cyan), $\epsilon = 2$ (magenta).
	For $\delta > 0$, the small-noise curves $\Re(\lambda_1) = -\epsilon^2/(2\delta)$
	are plotted as dashed lines in the corresponding color.
	For $\delta < 0$, the small noise curve $\Re(\lambda_1) = \delta$
	is plotted as a black dashed line.\\
	Right: Real part of the finite-difference approximation of the second RP resonance
	$\lambda_1$ versus $\epsilon$,
	for $\delta = 0$ (plus) and $\delta = 5$ (cross). For $\delta = 0$,
	the least-square linear regression with coefficient $-1.18$
	is also represented as a dashed line.
	For $\delta = 0$, the curve $\Re(\lambda_1) = -\epsilon^2/(2\delta)$
	is also represented as dotted dashed line.}
	\label{fig:w2VSBeta}
\end{figure}
Finally, we investigate the dependence of the RP spectrum on $\tilde \beta$ at the bifurcation point.
In Fig.~\ref{fig:w2VSBeta} is represented by crosses the evolution of the real parts (left panel)
and imaginary parts (right panel) of the approximated leading eigenvalues,
for $\delta = 0$ and $\epsilon = 1$.
On top of the real parts on the left are also represented the least square fits $y = a + b \tilde \beta^2$.
Their perfect match reveals the quadratic dependence of the real parts of the leading eigenvalues on $\tilde \beta$, as is also the case for the small-noise expansions (\ref{eq:eigValOrbit}) for $\delta > 0$.
This dependence is, however not uniform, as the real part of some eigenvalues are more sensitive to $\tilde \beta$ than others.
Second, the perfect match of the least square fits $y = a + b \epsilon$ on the imaginary parts (right panel) reveals their linear dependence on $\tilde \beta$.
Moreover, this dependence is much stronger for eigenvalues initially farther from the real axis when $\tilde \beta = 0$.
This result is in agreement with the strong folding of the high harmonics
from one side of the real axis to the other {\mk in Fig.~\ref{fig:numFPeps1beta}-(c) and Fig.~\ref{fig:numFPeps1beta}-(d)}.

\section{Summary and Discussion}
\label{sec:Conclusion}

We examined the stochastic Hopf bifurcation from the perspective of the
Markov semigroup and the spectrum of its generator. {\mk The latter---the RP spectrum---provides a characterization
of the dynamics, permitting among other things a decomposition of the correlation functions and related power spectra; {\mkr see \cite[Corollary 1 and Sec.~2.3]{Chekroun_al_RP2}.}
Contrary to the traditional characterization of a bifurcation in terms of crossing of the imaginary axis by the eigenvalues of the linearized problem, the RP resonances have real parts that remain negative as one crosses the criticality. Bifurcations are instead characterized by a change in the {\mkr geometric patterns formed by} the RP spectrum {\mkr in the left half complex plane}, and {\mkr in particular the} decay of correlations.
The RP spectrum allows {\mkr for a} unifying framework of stochastic analysis and Fokker-Planck equations {\mkr relevant and useful for} the study of stochastic bifurcations.
For instance, the stochastic analysis techniques reviewed in~\cite{Chekroun_al_RP2} {\mkr and related decomposition formulas of correlations and power spectra}, are applied here to the Hopf bifurcation case and small-noise expansions are derived for the corresponding eigenvalues and eigenfunctions in terms of an adimensional small parameter involving the noise level and the coefficients controlling the stability of the deterministic solutions.}

As in the deterministic case, weak values of the parameter $\delta$ controlling the distance to the deterministic bifurcation point are associated with the slowing down of the decay of correlations, as given by the gap between the leading RP resonances and the imaginary axis; see Section~\ref{sec:smallNoiseExpansion}.
However, the application, in Section~\ref{sec:StocAnaHopf}, of the theory of Lyapunov functions and ultimate bounds {\mkr (see~\cite[Theorems 5 \& 6]{Chekroun_al_RP2})} allows us to show that {\mkr noise implies the existence of a spectral gap that stays away from zero,} even at and above the bifurcation point,  {\mkr leading to correlations that} always decay exponentially.
In particular, while above the bifurcation point the neutral deterministic dynamics on the limit cycle
is associated with purely imaginary eigenvalues and is thus non-mixing, phase diffusion due to the noise ensures mixing along this limit cycle as well as the existence of a stationary density.
This phase diffusion results in the characteristic parabolic structure  of the RP resonances above the bifurcation point~\eqref{eq:eigValOrbit}, as opposed to the {\mk triangular} one below this point (see~\eqref{eq:eigValFPSub}), in the small-noise case considered in Sec.~\ref{sec:smallNoiseExpansion}.

{\mk To understand this phase diffusion, we use the concept of isochrons provided by the set of points that 
share the same asymptotic phase, on the limit cycle.  This allows us to derive an SDE for the phase evolution in which two contributions appear: one coming directly from the azimuthal direction, and one resulting from the interaction of the deterministic vector field with the radial noise; see Eq.~\eqref{eq:SDEIsochron}. This interaction is quantified thanks to the H\"ormander condition which in terms of isochrons can be summarized as (see Theorem~\ref{thm:isoHypo} and Fig.~\ref{fig:isochronlie}):}
\begin{quote}
	\emph{For phase diffusion to occur, it is necessary that at least one component of the noise acts transverse to the isochrons.}
\end{quote}
In particular, even if the stochastic vector field is tangent to the radial direction,
phase diffusion can occur as long as the isochrons are transverse to this direction.
In the case of the stochastic Hopf bifurcation considered here,
it was shown in Section~\ref{sec:normalForm} that the tilt of the isochrons
is controlled by the twist factor $\tilde \beta = \beta / \kappa$
measuring the dependence of the frequency of the oscillations on their amplitude.
This effect was directly visible from the isolines of phase of the RP eigenfunctions
obtained in Section~\ref{sec:smallNoiseExpansion}, in the small-noise case,
as well as in the phase diffusion coefficient
$ \epsilon^2 (1 + (\beta/\kappa)^2) / R^2$ entering the real parts
of the small-noise expansion about the limit cycle of the RP resonances.
These formulas for the resonances are particularly useful,
as they allow one to quantitatively relate the phenomenon of phase diffusion to
the broadening of the peaks in the power spectrum.

While, in the general case of a hyperbolic limit cycle in the presence of noise,
the tilt of the isochrons with respect to the stochastic forcing
can in principle be measured, no single parameter playing the role of twist factor
can in general be singled out.
In addition, the calculation, even numerical,
of the global isochrons of a high-dimensional system is a difficult task. However, in the small-noise case, only local information on the isochrons
about the limit cycle is necessary.
Indeed, it was shown by~\cite{Gaspard2002a} that
the effect of the interaction of the noise with the deterministic vector field
integrated over one period of the limit cycle is measured by the coefficient
\begin{align}
	\Phi = - \frac{\epsilon^2 \omega_f^2}{T}
	\frac{\left< C(T) \vec{f}^R_2, \vec{f}^R_2 \right>}{\left<\vec{e}^R_2, \vec{f}^R_2 \right>},
	\label{eq:diffusionCorrelation}
\end{align}
where $\epsilon$ is the noise level and $\vec{e}^R_2$ and $\vec{f}^R_2$
are respectively the right and left eigenvectors of the matrix $R$
of the Floquet representation~\eqref{eq:defR} of the fundamental matrix $M(t)$
associated with the eigenvalue $0$ (care should be taken to normalize $\vec{e}^R_2$ to the same magnitude of the vector field for~\eqref{eq:diffusionCorrelation} to be valid) ; see Section~\ref{sec:isochrons}.
The matrix $C(t)$ is given by
\begin{align}
	C(t) &= \int_0^t M(t) M(-s) D_\Gamma(s) (M(t) M(-s))^* ds
	\label{eq:covarianceOU}
\end{align}
and corresponds to the \emph{correlation matrix}~\cite{DaPrato1995,Geissert2008a}
of a periodic Ornstein-Uhlenbeck process with a drift given by the Jacobian matrix $A(t)$ (see Appendix~\eqref{eq:variational})
and with a diffusion matrix $D_\Gamma(t)$, both evaluated along the limit cycle.
A simple calculation, given in Appendix~\ref{sec:diffusionCoefficient}, shows that,
in the case of the stochastic Hopf bifurcation considered here, the coefficient $\Phi$ correctly coincides with the diffusion coefficient entering the real parts of the small-noise expansion about the limit cycle of the RP resonances.
Thus, in the small-noise case, the phase diffusion coefficient (\ref{eq:diffusionCorrelation}) is readily accessible from the local properties of the deterministic system about the limit cycle.

From the difference in structure of the RP spectrum below and above the bifurcation point {\mk identified} in Section~\ref{sec:smallNoiseExpansion},
one could hope to distinguish the case of a single stable stationary point perturbed by noise below the bifurcation, from the case of a perturbed limit cycle above the bifurcation.
However, close to the bifurcation point and for a high level of noise, the small-noise expansions of the eigenvalues are no longer in agreement with  the numerical approximations of Section~\ref{sec:numericalHopf}.
The stochastic dynamics can no longer be understood in terms
of small perturbations of the deterministic dynamics.
The numerical approximations, however, give evidence
that new constraints emerge at the bifurcation point.
These results can be summarized as follows:

\fbox{
\begin{minipage}{0.8\textwidth}
\begin{itemize}
	\item The numerical approximations are in good agreement with the small-noise
	expansions of Section~\ref{sec:smallNoiseExpansion}
	far from the bifurcation point and for a small noise-level,
	even though eigenvalues farther from the imaginary axis tend to be more sensitive to the truncation.
	\item As the bifurcation point is crossed,
	a transition occurs from a triangular structure of eigenvalues to a parabolic one
	and the eigenvalues remain discrete.
	\item For $\delta \approx 0$, there is a $\delta$-interval over which the real parts of the eigenvalues
	are to a large extent linearly dependent on the bifurcation parameter $\delta$.
	This interval widens with the noise level, but the slope does not depend on it.
	\item For $\delta = 0$, the real parts of the eigenvalues depend linearly on the noise,
	but eigenvalues farther from the imaginary axis are more sensitive to the noise.
	\item The effect of the twist factor $\tilde \beta = \beta / \kappa$ on the RP eigenfunctions and eigenvalues
	is already visible below the bifurcation point.
	\item For $\delta = 0$, the real parts and the imaginary parts of the leading eigenvalues
	depend quadratically and linearly on the twist factor $\tilde \beta$, respectively. 
	\item Eigenvalues that would correspond to higher harmonics for $\tilde \beta = 0$ can evolve rapidly with changes in $\tilde \beta$ and $\delta$
	and need not correspond to integer multiples of a constant fundamental frequency.
	This is particularly noticeable near the bifurcation
	and can have subtle consequences for the power spectrum.
\end{itemize}
\end{minipage}}
\vspace{1ex}

{\mk The} geometric characterization of the phenomenon of phase diffusion
and the formulas for the RP eigenvalues and eigenfunctions
allow one to gain novel insights on the dependence of the regularity of
nonlinear oscillators forced by noise not only on the noise level,
but also on the stability of the underlying limit cycle and on the twist of its isochrons.

{\mk The analysis conducted here on the stochastic Hopf equation points out a rich set of properties that builds intuition for  examination of more complex nonlinear oscillations in presence of noise.} In the third part of this contribution~\cite{tantet_ruellepollicott_2019}, these results are applied to models of a leading mode of climate variability, El Ni\~no-Southern Oscillation, for which understanding the dynamics
behind its aperiodicity remains a challenge.

\begin{acknowledgements}
  The programs used for this analysis are available as an open-source C++ library at \url{https://github.com/atantet/ergoPack/} together with a link to its documentation.

  The authors would like to thank the reviewers for their very useful and constructive comments.
  This work has been partially supported by the European Research Council under the European Union’s Horizon 2020 research and innovation program (grant agreement No. 810370 (MDC)), by the Office of Naval Research (ONR) Multidisciplinary University Research Initiative (MURI) grant N00014-16-1-2073 (MDC), by the National Science Foundation grants OCE-1658357, DMS-1616981 (MDC), AGS-1540518 and AGS-1936810 (JDN), by the LINC project (No. 289447) funded by EC’s Marie-Curie ITN (FP7-PEOPLE-2011-ITN) program (AT and HD) and by the Utrecht University Center for Water, Climate and Ecosystems (AT).
\end{acknowledgements}

\appendix
\section{Floquet theory applied to the Hopf normal form}\label{sec:Floquet}

Floquet theory allows one to characterize the local stability properties
of deterministic flows about a periodic orbit.
These properties are essential to the response of the system to stochastic forcing
studied in Section~\ref{sec:StocAnaHopf}  and to the small-noise expansions
of the RP spectrum obtained in Section~\ref{sec:smallNoiseExpansion}.
We thus review here standard results from the application of Floquet theory
to the normal form~\eqref{eq:normalHopf} of the Hopf bifurcation.

Small deviations $x'(t)$ from {\mk the orbit} $x_\Gamma(t)$ {\mk associated with the limit cycle} $\Gamma$,
satisfy the \emph{variational equation} (\cite{Kuznetsov1998}, Chap.~1.5)
\begin{align}
	\dot{x'}(t) = A(t) x'(t), \quad x'(t) \in {\mk \mathbb{R}^2}, \quad t \in \mathbb{R}.
	\label{eq:variational}
\end{align}
{\mk Here,} $A(t): = (D F)_{x_\Gamma(t)}$ {\mk denotes the Jacobian matrix about the orbit $x_\Gamma(t)$, of the vector field
$F$ associated with the Hopf normal form written in Cartesian coordinates, i.e.~the RHS of Eq.~\eqref{eq:HopfSDECart} for $\epsilon=0$}.
In other words, $A(t)$ {\mk provides} the tangent map of $F$ {\mk along} $x_\Gamma(t)$.
Thus, $A$ is periodic, i.e., $A(t + T) = A(t)$, {\mk for any $t$ in} $\mathbb{R}$.
Let $M(t)$ be a \emph{fundamental solution} (\cite{Hartman1964}, Chap.~IV.1)
of \eqref{eq:variational}, i.e.,
\begin{align}
	\dot{M}(t) = A(t) M(t) \quad \mathrm{and~} \det M(t) \ne 0, \quad t \in \mathbb{R}.
	\label{eq:variationalMatrix}
\end{align}
Then {\mk the Floquet theorem} (e.g.~\cite{Hartman1964}, Theorem~IV.6.1) {\mk ensures that} $M(t)$ {\mk has the following} representation 
\begin{align}
	M(t) = Z(t) e^{t R}, \quad \mathrm{where~} Z(t + T) = Z(t), \quad t \in \mathbb{R},
	\label{eq:Floquet}
\end{align}
and $R$ is a constant matrix.
Imposing, without loss of generality, that $M(0) = I$ yields $Z(T) = Z(0) = I$
and $M(T) = e^{T R}$.

While determining the Floquet representation of a fundamental matrix
is in general a difficult task, in the case of the Hopf normal form \eqref{eq:normalHopf},
it can easily be found from the linearization
of the vector field in polar coordinates. {\mk In that respect, we assume furthermore that $\delta$  in Eq.~\eqref{eq:normalHopf} is positive.}
The {\mk orbit $x_\Gamma(t)$  writes  then $(R, \theta_0 + \omega_f t)$},
for some initial phase $\theta_0$. The linearization about $\Gamma$ of the vector field (\ref{eq:HopfField})
in polar coordinates is given by the matrix
\begin{align}
	J_\Gamma(t) =
	\begin{pmatrix}
		-2\delta			& 0 \\
		-2\beta R 	& 0
	\end{pmatrix}
\end{align}
and depends {\mk implicitely} on time only through the evolution of the tangent space on which it acts
with the reference solution $x_\Gamma(t)$, so that the time argument will be dropped in the sequel.

To proceed, let us introduce the Jacobian matrix of the transformation $(x, y) \to (r, \theta)$
and its inverse, respectively given by
\begin{align*}
	J_{\mathrm{polar}}(r, \theta) &= 
	\begin{pmatrix}
		\cos \theta & \sin \theta \\
		- r^{-1} \sin \theta & r^{-1} \cos \theta
	\end{pmatrix}
	 =~ S^{-1}(r) L(-\theta), \quad r > 0 \\
	J_{\mathrm{polar}}^{-1}(r, \theta) &=
	\begin{pmatrix}
		\cos \theta & - r \sin \theta \\
		\sin \theta & r \cos \theta
	\end{pmatrix}
	=~ L(\theta) S(r),
\end{align*}
where we have used the rotation and diagonal matrices
\begin{align*}
	L(\theta) =
	\begin{pmatrix}
		\cos \theta & -\sin \theta \\
		\sin \theta & \cos \theta
	\end{pmatrix}
	\quad \mathrm{and} \quad	
	S(r) = 
	\begin{pmatrix}
		1 & 0 \\
		0 & r \\
	\end{pmatrix}.
\end{align*}
%
%
The matrix $J_\Gamma$ is then related to the matrix $A(t)$
of the tangent map $(DF)_{x_\Gamma(t)}$ in Cartesian coordinates by
\begin{align}
  A(t) = J_{\mathrm{polar}}^{-1}(R, \theta_0 + \omega_f t)
  ~ J_\Gamma ~ J_{\mathrm{polar}}(R, \theta_0)
  + \omega_f L(\frac{\pi}{2}), \quad t \in \mathbb{R}.
\end{align}
That the conversion of $J_\Gamma$ to Cartesian coordinates
coincides with the matrix $A(t)$ of the tangent map
but for the term $\omega_f L(\pi/2)$ is due to the rotation
of the polar frame along the limit cycle $\Gamma$,
which was not taken into account when calculating $J_\Gamma$.

One can then verify that the matrix
\begin{align}
	M(t) =
  J_{\mathrm{polar}}^{-1}(R, \theta_0 + \omega_f t)
  ~ e^{t J_\Gamma}
  ~ J_{\mathrm{polar}}(R, \theta_0), \quad t \in \mathbb{R},
\end{align}
is a solution to (\ref{eq:variationalMatrix}),
for the reference solution $x_\Gamma(t)$ on $\Gamma$.
Since 
\bes
{\mk J_{\mathrm{polar}}^{-1}(R, \theta_0 + \omega_f t)
= L(\omega_f t) ~ J_{\mathrm{polar}}^{-1}(R, \theta_0),}
\ees
it follows that the fundamental matrix $M(t)$ has a Floquet representation
\begin{align}
  &M(t) = Z(t) ~ e^{t R}, \quad t \in \mathbb{R}, \label{eq:defR} \\
  \mathrm{with} \enskip
  Z(t) = L(\omega_f t) \quad
  &\mathrm{and} \enskip
    R = J_{\mathrm{polar}}^{-1}(R, \theta_0)
    ~ J_\Gamma
    ~ J_{\mathrm{polar}}(R, \theta_0). \nonumber
\end{align}
Applying $M(t)$ to a vector $x'$ at time $0$ thus corresponds to
converting this vector to polar coordinates, integrating to a time $t$
according to the generator $J_\Gamma$ and converting back
from polar coordinates at time $t$.
In other words, the polar frame at $x_\Gamma(t) = (R, \theta_0 + \omega_f t)$
constitutes a co-moving frame adapted to the Floquet representation of $M(t)$.

Note next that $J_\Gamma$ can be diagonalized as
\begin{align}
	&J_\Gamma = E ~ \Lambda ~ F^* \\
	\mathrm{with}~ E = 
	\begin{pmatrix}
		1 & 0 \\
		\frac{\tilde \beta}{R} & 1
	\end{pmatrix}, \quad
	&\Lambda = 
	\begin{pmatrix}
		-2 \delta & 0 \\
		0 & 0
	\end{pmatrix} \quad \mathrm{and} \quad
	F^* = E^{-1} = 
	\begin{pmatrix}
		1 & 0 \\
		-\frac{\tilde \beta}{R} & 1
	\end{pmatrix}, \nonumber
\end{align}
where $F^*$ denotes the complex conjugate of the matrix $F$.
Then, from the definition (\ref{eq:defR}) of $R$,
\begin{align}
  &R = E_R ~ \Lambda ~ F_R^* \label{eq:eigR} \\
  \mathrm{with}~
  E_R = J_{\mathrm{polar}}^{-1}(R, \theta_0) ~ E
  \quad &\mathrm{and} \quad
          F^*_R = E_R^{-1} = F^* ~ J_{\mathrm{polar}}(R, \theta_0). \nonumber
\end{align}
Thus, the eigenvalues of $R$ coincide with those of $J_\Gamma$
and its eigenvectors are given by converting those of $J_\Gamma$
from polar coordinates.

The eigenvalues $\alpha_1$ and $\alpha_2$ of $R$
are called the \emph{characteristic exponents}
of $\Gamma$ and the eigenvalues of $e^{TR}$ its
\emph{characteristic multipliers} (\cite{Guckenheimer1983}, Chap.~1.5).
The eigenvector associated with $\alpha_2$ is in the direction of the flow,
so that $e^{T \alpha_2}$ is always unity.
On the other hand, the other eigenvalue $\alpha_1 = -2 \delta$ determines the
stability of the periodic orbit.
It is in fact the eigenvalue of the tangent map $D S_T$ of the Poincar\'e map.

\subsection{Calculation of the phase diffusion coefficient from the correlation matrix}
\label{sec:diffusionCoefficient}

In the case of the stochastic Hopf bifurcation considered here,
the diffusion matrix $D_\Gamma$ in~\eqref{eq:covarianceOU} for any point on $\Gamma$ is given in polar coordinates by

\begin{align}
	D_\Gamma &= 
	\begin{pmatrix}
		1 & 0 \\
		0 & \frac{1}{R^2}
	\end{pmatrix}, \quad t \in \mathbb{R}
\end{align}
and is hence constant in time.
Since $\vec{f}^R_2$ is a left eigenvector of the matrix $M(T) = e^{TR}$
with $R$ given by (\ref{eq:defR}), it follows that

\begin{align}
  \Phi
  = -\epsilon^2 \omega_f^2 \left< D_\Gamma \vec{f}_2, \vec{f}_2 \right>
  = -\epsilon^2 \frac{1 + \tilde \beta^2}{R^2}
  = -\epsilon^2 \frac{1 + \beta^2}{\delta \kappa},
\end{align}
where $\vec{f}_2 = \omega_f^{-1} (-\tilde \beta /  R), 1)$
is the conversion to polar coordinates of the left eigenvector
$\vec{f}^R_2$ of $R$ and, according to (\ref{eq:eigR}), coincides with the
left eigenvector of the polar Jacobian matrix $J_\Gamma$ in (\ref{eq:JacobianCycle}) at initial time.
The factor $\omega_f^{-1}$ in $\vec{f}_2 = (-\tilde \beta / (\omega_f R), 1)$ is due to the normalization of $\vec{e}^R_2$ to the magnitude of the vector field $F$ on $\Gamma$, which is essential for \eqref{eq:diffusionCorrelation} to hold.

\section{Proofs of the stochastic analysis results of Section~\ref{sec:StocAnaHopf}}

\subsection{Proof of Theorem~\ref{thm:isoHypo}: isochrons and H\"ormander condition}\label{sec:proofIso}

For two arbitrary smooth vector fields $V$ and $W$, recall that the Lie bracket $[V, W]$
coincides with the Lie derivative $\mathcal{L}_V W$ of $W$ along $V$.
The Lie derivative can be defined in terms of pullback of a {\mk vector field} by a diffeomorphism.
The pullback, or Lie transport, $(S_t^{V*} W)(q)$ at a point $q$
of a vector field $W$ by the flow $S_t^{V}$ generated by $V$
can be defined as the vector at $q$ tangent to the image by $S_{-t}^{V}$
of any curve to which $W(S_t^{V} q)$ is tangent.
The Lie derivative at a point $q$ is then defined in terms of {\mk pullback of a vector field}, by
\begin{align}\label{Eq_Lie_flow}
	\mathcal{L}_V W = \at{\frac{\d}{\d t}}{0} S_t^{V*} W.
\end{align}
{\mk This expression says that $\mathcal{L}_V W$} measures the rate of change of $W$ due to the Lie transport~\cite[Chap.~3-4]{Fecko2006}.
Note that the Lie derivative is well defined because both the vector field at some point
and its pullback at the same point live in the tangent space to the manifold at this point.
The following is derived from the fact that the isochrons are permuted by the flow $S_t$ generated by $V_0$ {\mk (Proposition~\ref{prop:isochron}-(ii))}:
if a vector field $V_i$
is tangent to an isochron $W_{ss}(S_t p)$ at some point $S_t q$,
i.e.~if $V_i(S_t q) \in T W_{ss}(S_t p)$, where $T W_{ss}(p)$ denotes the tangent space to $W_{ss}(p)$,
then its pullback to {\mk a point $q$ in} $ U_\Gamma$ by $S_t$ is necessarily
tangent to the isochron $W_{ss}(p)$, i.e~$V_i(q) \in T W_{ss}(p)$.
Thus, as a linear combination of vectors in the tangent space $T W_{ss}(p)$,
the Lie derivative $(\mathcal{L}_{V_0} V_i)(q) = [V_0, V_i](q)$
is also in $T W_{ss}(p)$.
The same argument holds for the Lie derivative $\mathcal{L}_{V_i} V_j$
between two vector fields tangent to the isochrons everywhere in $U_\Gamma$,
with the difference that the vector fields are Lie transported along the same isochron, in this case.
Lastly, any iteration of Lie brackets between the family $\{V_i, 0 \le i \le m\}$,
where $V_0$ is the vector field of the deterministic system with a hyperbolic limit cycle
and the $\{V_i, 0 < i \le m\}$ are vector fields tangent to the isochrons of the limit cycle,
yields the same outcome.
It follows that
\begin{align*}
	\cup_{k \ge 1} \Span\{V(q): V \in \mathcal{V}_k\} = TW_{ss}(p), \quad \text{for any} \enskip q \in U_\Gamma,
\end{align*}
where $W_{ss}(p)$ is the isochron passing through $q$.

\subsection{Proof of Proposition~\ref{prop:spectral_gap}: spectral gap}\label{sec:proofGap}

{\mkrr Proposition~\ref{prop:spectral_gap} can be obtained as application of \cite[Theorem~6]{Chekroun_al_RP2} which provides conditions ensuring existence of a spectral gap and 
exponential decay of correlations.
Since, as shown in Section~\ref{sec:FellerIrreducibility}, the Markov semigroup $(P_t)_{t\geq 0}$ associated with the SHE~\eqref{eq:HopfSDECart} is irreducible and strong Feller in 
it is thus sufficient to check the ultimate bound condition of~\cite[Theorem~6]{Chekroun_al_RP2} to conlude, which we do hereafter.}

%

{\mk More specifically, denoting by $X_t^x$ the stochastic process solving the SHE~\eqref{eq:HopfSDEPolar} and emanating from 
$x = (r, \theta)$}, we show that there exists $k, c, d > 0$ such that
\begin{align}
	\mathbb{E} |X_t^x|^2 = \mathbb{E}[r_t^2] < k r^2 e^{-c t} + d, \enskip t \ge 0, \enskip r \ge 0,
	\label{eq:ultimateBound}
\end{align}
for any value of the control parameters $\delta$ in $\mathbb{R}$, $\beta$ in $\mathbb{R}$, $\kappa > 0$, and $\epsilon > 0$.

{\mk As evolution of the observable $\varphi(r,\theta)=r^2$ by the Markov semigroup $P_t$, note that the function $t \to \mathbb{E}[r_t^2]$ solves the Kolmogorov equation~\eqref{eq:HopfBKEPolar}, which leads here to the differential equation 
\begin{align}
	\frac{d}{dt} \mathbb{E}[r_t^2]
	&= 2 \epsilon^2 + 2 \left(\delta \mathbb{E}[r_t^2] - \kappa \mathbb{E}[r_t^4]\right).
	\label{eq:2ndMoment}
\end{align}
}
%
{\mk To derive a bound \eqref{eq:ultimateBound}, we} bound the right-hand side
of the ODE~\eqref{eq:2ndMoment} in $\mathbb{E}[r_t^2]$ and to apply comparison results of Gronwall-Bellman-Bihari type; {\mk see e.g.~\cite{Bainov1992}.}

For $\delta < 0$, below the bifurcation, the estimate
\begin{align*}
	\frac{d}{dt} \mathbb{E}[r_t^2]
	&\le 2 \epsilon^2 + 2 \delta \mathbb{E}[r_t^2],
\end{align*}
holds, {\mk since $\mathbb{E}[r_t^{4}] > 0$}.
It follows from the {\mk standard} Gronwall inequality for linear {\mk differential inequalities}
(e.g.~\cite[Chap.~1, Lemma~1.1]{Bainov1992}) that 
%
\begin{align}
	\mathbb{E}[r_t^2]
	\le r^2 e^{2\delta t} + \frac{\epsilon^2}{\delta} (1 - e^{2\delta t})
	\le r^2 e^{2\delta t} + \frac{\epsilon^2}{\delta}, \quad t \ge 0.
	\label{eq:boundSub}
\end{align}
Thus, one can choose $k = 1$, $c = -2\delta$ and $d > -\epsilon^2 / \delta$,
for the ultimate bound (\ref{eq:ultimateBound}) to be satisfied.

Next, for $\delta \ge 0$, above the bifurcation,~\eqref{eq:2ndMoment} is equivalent to
\begin{align*}
	\frac{d}{dt} \mathbb{E}[r_t^2]
	&= 2 \epsilon^2 - 2 \mathbb{E}\left[r_t^2 (\kappa r_t^{2} - \delta)\right],
\end{align*}
and it follows, by applying \emph{Jensen's inequality} (e.g.~\cite{Kallenberg2002}, Lemma~2.5), that
\begin{align}
	\frac{d}{dt} \mathbb{E}[r_t^2]
	&\le 2 \epsilon^2 - 2 \mathbb{E}[r_t^2] (\kappa \mathbb{E}[r_t^{2}] - \delta), \quad t \ge 0.
	\label{eq:2ndMomentSup}
\end{align}

{\mk A classical comparison theorem on differential inequalities} 
\cite[Chap.~2, Theorem~6.3]{Bainov1992} {\mk ensures that} 
the inequality \eqref{eq:2ndMomentSup}
implies {\mk boundedness from above of the $2nd$} moment $\mathbb{E}[r_t^2]$  by 
a maximal solution $y$ of the {\mk scalar} ODE
\begin{align*}
	y'(t) = 2 \epsilon^2 - 2 y(t) (\kappa y(t) - \delta), \quad y(0) = r^2, \quad t \ge 0.
\end{align*}
{\mk By solving this equation},
one finds the maximal solution
\begin{align*}
	y(t) = R_\epsilon(\delta, \kappa)^2 - \frac{w \sqrt{\Delta}}{w - \exp \left(2 \sqrt{\Delta} t \right)}, \quad t \ge 0,
\end{align*}
where $w$ is a constant of integration, $\Delta = R^2 + 4 \epsilon^2 / \kappa$ and $R_\epsilon(\delta, \kappa)^2 = (R + \sqrt{\Delta}) / 2$ is the {\mk equilibrium} to which $y(t)$ converges as $t$ goes to infinity.
For the initial condition $y(0) = r^2$, one finds 
\bes
w = (r^2 - R_\epsilon(\delta, \kappa)^2) (r^2 - R_\epsilon(\delta, \kappa)^2 + \sqrt{\Delta})^{-1},
\ees

Let us look for exponential bounds on $y(t)$.
First,
\begin{align*}
	r \le R_\epsilon(\delta, \kappa) \enskip \Rightarrow \enskip w \le 0
	\enskip \Rightarrow \enskip y(t) \le R_\epsilon(\delta, \kappa)^2, \quad \text{for} \enskip t \ge 0,
\end{align*}
while
\begin{align*}
	r \ge R_\epsilon(\delta, \kappa) \enskip \Rightarrow \enskip 0 \le w \le 1
	\enskip \Rightarrow \enskip y(t) \le R_\epsilon(\delta, \kappa)^2 + (r^2 - R_\epsilon(\delta, \kappa)^2) \exp \left(-2 \sqrt{\Delta} t \right),
	\quad \text{for} \enskip t \ge 0.
\end{align*}
We have thus shown that the second moment $\mathbb{E}[r_t^2]$
satisfies the {\mk inequality}
\begin{align*}
	\mathbb{E}[r_t^2] \le y(t) \le R_\epsilon(\delta, \kappa)^2 + r^2 \exp \left(-2 \sqrt{\Delta} t \right),
	\quad r \ge 0, \quad t \ge 0.
\end{align*}
Thus, for $\delta \ge 0$ and $\epsilon > 0$,
the second moment satisfies the ultimate bound (\ref{eq:ultimateBound})
with $k = 1$, $c = 2 \sqrt{\Delta}$ and $d = R_\epsilon(\delta, \kappa)^2$.
This estimate is valid even at the critical value $0$ of $\delta$,
as long as the noise level $\epsilon$ is nonzero.
In this case, the exponential decay rate $a = 4 \epsilon$ is proportional to the noise level.

\section{Proofs of the small-noise expansions of Section~\ref{sec:smallNoiseExpansion}}\label{sec:proof_eigen}

\subsection{Proof of Proposition~\ref{prop:eigen_below}: expansions for $\delta < 0$ about the stationary point}\label{sec:proof_eigen_below}

We proceed to the small-noise expansion of the Kolmogrov equation corresponding to the SHE~\eqref{eq:SHEScaled} in adimensional Cartesian coordinates, $x' = x / L_\epsilon(\delta)$, $y' = y / L_\epsilon(\delta)$ and $t' = \delta t$,
\begin{align*}
  \partial_{t'} u
  &= \lrs{\lr{-1 - \sigma_\epsilon^2 \lr{x'^2 + y'^2}} x'
    - \lr{\tilde \gamma
    - \tilde \beta \sigma_\epsilon^2 \lr{x'^2 + y'^2}} y'} \partial_{x'} u
    + \frac{1}{2} \partial^2_{x'x'} u\\
  &+ \lrs{\lr{\tilde \gamma - \tilde \beta \sigma_\epsilon^2 \lr{x'^2 + y'^2}} x'
    + \lr{-1 - \sigma_\epsilon^2 \lr{x'^2 + y'^2}} y} \partial_{y'} u
    + \frac{1}{2} \partial^2_{y'y'} u.
\end{align*}
Since the small parameter $\sigma_\epsilon = 1 / r_\epsilon$ appears squared only, we can expand the eigenvalues and eigenfunctions in $\sigma_\epsilon^2$.
To zeroth order, we have
\bea
  \lambda^{(0)} \psi^{(0)}
  &= \mathcal{K}_{x_*}^{(0)} \psi^{(0)}, \label{eq:smallBKESub}\\
  \mathrm{with}\enskip \mathcal{K}_{x_*}^{(0)}
  &= \lr{-x' - \tilde \gamma y'} \partial_{x'}
    + \lr{\tilde \gamma x' - y'} \partial_{y'}
    + \frac{1}{2} \partial_{x'x'}
    + \frac{1}{2} \partial_{y'y'}
\eea
This equation yields to the eigenvalue problem of a two-dimensional nonsymmetric Ornstein-Uhlenbeck process with Kolmogorov operator $\mathcal{K}_{x_*}^{(0)}$.
Its linear drift and diffusion have the following  matrix representation in adimensional Cartesian coordinates $(x', y')$:
\begin{align*}
	J_{x_*} =
	\begin{pmatrix}
		-1 & -\tilde \gamma \\
		\tilde \gamma & -1
	\end{pmatrix},
	\quad \text{and} \quad D = \frac{1}{2} I.
\end{align*}
Here, $J_{x_*}$ {\mk corresponds also to  the tangent map at the origin of the vector field $F$ associated with the Hopf normal form \eqref{eq:HopfSDECart} for $\epsilon = 0$, while  $I$ denotes the $2\times 2$ identity matrix.}
The stationary density of this Ornstein-Uhlenbeck process is given in adimensional polar coordinates $(r', \theta')$ by
\begin{align}
	\rho_{x_*}(r') = \frac{1}{\pi} \enskip r' e^{-r'^2}.\label{eq:statDenSub}
\end{align}
{\mk For the weighted inner-product $\langle \cdot,\cdot \rangle_{\rho_{x_*}}$ with respect to this density, the Kolmogorov operator associated with this Ornstein-Uhlenbeck process is asymmetric.
  This asymmetry comes from the anti-symmetry of the rotation operator
\bes
\Omega = -\tilde \gamma y' \partial_{x'} + \tilde \gamma x' \partial_{y'},
\ees
i.e. $\langle\Omega f, g \rangle_{\rho_{x_*}} = -\langle f, \Omega g\rangle_{\rho_{x_*}}$, while the operator
\bes
-x' \partial_{x'} - y' \partial_{y'} + \frac{1}{2} \partial_{x'x'} + \frac{1}{2} \partial_{y'y'},
\ees 
encapsulating the diffusion and contraction effects, is symmetric.}


The RP spectrum of one-dimensional Ornstein-Uhlenbeck processes is well studied (see e.g.~\cite[Chap.~5]{Risken1989}).
In several dimensions, the more recent work~\cite{Metafune2002a} {\mk shows} that
the spectrum of an Ornstein-Uhlenbeck process is discrete and composed of eigenvalues ---{\mk corresponding here to the set of $\lambda_k^{(0)}$ solving~\eqref{eq:smallBKESub} with the $\psi_k^{(0)}$ in $L^2_{\rho_{x_*}}(\mathbb{R}^2)$} --- are given by integer linear combinations  of the eigenvalues of the drift matrix $J_{x_*}$, i.e.~the complex conjugate {\mk pair} $-1 \pm i \tilde \gamma$, in our case.
In dimensional terms, the eigenvalues of the SHE~\eqref{eq:HopfSDECart} are thus given to first order by the combiations $(l + n) \delta + i (n - l) \gamma$,
with $n, l \in \mathbb{N}$, which coincides with the eigenvalues of the deterministic normal form~\eqref{eq:normalHopf}; c.f.~\cite{Gaspard2002a}.

In addition, it has recently been shown by~\cite{Chen2014} that the solutions to~\eqref{eq:smallBKESub} are given by products of Laguerre polynomials with harmonic functions.
In adimensional polar coordinates $(r', \theta')$ this yields in our case,
\begin{align*}
  \psi_{ln}^{(0)}(r', \theta')
  &=
    \begin{cases}
      e^{i(n-l)\theta'} \enskip \sqrt{\frac{l!}{n!}} \lr{r'}^{n-l}
      L_l^{n-l}\lr{-r'^2}, \quad &n \ge l \\
      e^{i(l-n)\theta'} \enskip \sqrt{\frac{n!}{l!}} \lr{r'}^{l-n}
      L_n^{l-n}\lr{-r'^2}, \quad &n < l,
    \end{cases}
\end{align*}
or in dimensional polar coordinates $(r, \theta)$,
\begin{align*}
  \psi_{ln}^{(0)}(r, \theta)
  &=
    \begin{cases}
      e^{i(n-l)\theta} \enskip
      \sqrt{\frac{l!}{n!}}
      \lr{\sqrt{-\frac{\delta}{\epsilon^2}} r}^{n-l}
      L_l^{n-l}\lr{-\frac{\delta r^2}{\epsilon^2}}, \quad &n \ge l \\
      e^{i(l-n)\theta} \enskip
      \sqrt{\frac{n!}{l!}}
      \lr{\sqrt{-\frac{\delta}{\epsilon^2}}r}^{l-n}
      L_n^{l-n}\lr{-\frac{\delta r^2}{\epsilon^2}}, \quad &n < l.
    \end{cases}
\end{align*}

From the orthogonality of the Laguerre polynomials~\cite[p.~84]{Lebedev1972} and of the harmonic functions, it follows that the appropriately normalised eigenfunctions form a complete orthonormal family of {\mk $L^2_{\rho_{x_*}}(\mathbb{R}^2)$}.
{\mk The product of these eigenfunctions with the density $\rho_{x_*}$ thus yield the eigenfunctions of the Fokker-Planck equation dual to the Kolmogorov equation~\eqref{eq:smallBKESub}.}

To first order in $\sigma_\epsilon^2$,
\begin{align*}
  \lambda^{(0)} \psi^{(1)} + \lambda^{(1)} \psi^{(0)}
  &= \lrs{-\lr{x'^2 + y'^2} x' + \tilde \beta \lr{x'^2 + y'^2} y'}
     \partial_{x'} \psi^{(0)}
    + \lrs{-\tilde \beta \lr{x'^2 + y'^2} x' -\lr{x'^2 + y'^2} y'}
    \partial_{y'} \psi^{(0)} \\
  &+ \lr{-x' - \tilde \gamma y'} \partial_{x'} \psi^{(1)}
    + \lr{\tilde \gamma x' - y'} \partial_{y'}  \psi^{(1)}
    + \frac{1}{2} \partial_{x'x'} \psi^{(1)}
    + \frac{1}{2} \partial_{y'y'} \psi^{(1)}
\end{align*}
Thus the magnitude of this term depends on the twist factor $\tilde \beta = \beta / \kappa$.
For this reason, we use the asymptotic notation $\mathcal{O}_{\tilde \beta}((\smallin)^2)$ to represent it.

\subsection{Proof of Proposition~\ref{prop:eigen_above}: expansions for $\delta > 0$ about the limit cycle $\Gamma$}\label{sec:proof_eigen_above}

We are here interested in the finding the leading eigenvalues and eigenfunctions originating from the ruins of the deterministic limit cycle $\Gamma$ when $\sigma_\epsilon = \frac{1}{r_\epsilon}$ is small.
We thus proceed to an additional change of variables from the adimensional coordinates $(r', \phi')$ to a frame centered on $\Gamma$ and rotating at the angular frequency $\tilde \omega_f = \tilde \gamma - \tilde \beta$ of the adimensional deterministic dynamics on $\Gamma$,
\begin{align*}
  \hat{r} &= r' - r_\epsilon = r' - \sigma_\epsilon^{-1}\\
  \hat{\phi} &= \phi' + \tilde \omega_f t'.
\end{align*}
The \eqref{eq:SHEScaled} then reads in $(\hat{r}, \hat{\phi})$ coordinates,
\bea
\label{eq:SHEScaledCycle}
\d \hat{r}
&= \lr{\hat{r} + \sigma_\epsilon^{-1}}
\lr{1 - \lr{\sigma_\epsilon \hat{r} + 1}^2
  + \frac{\sigma_\epsilon^2}{2\lr{\sigma_\epsilon \hat{r} + 1}^2}} \d t' + \d W_r \\
\mathrm{or~} \d \hat{\phi}
&= - \tilde \beta \frac{\sigma_\epsilon}{\lr{\sigma_\epsilon \hat{r} + 1}} \d W_r
+ \frac{\sigma_\epsilon}{\lr{\sigma_\epsilon \hat{r} + 1}}\d W_\theta,
\eea
and the corresponding Kolmogorov equation, with $\hat{u}(\hat{r}, \hat{\phi}) = u(r, \phi)$, is,
\begin{align*}
  \partial_{t'} \hat{u}
  &= \lr{\hat{r} + \sigma_\epsilon^{-1}}
    \lr{1 - \lr{\sigma_\epsilon \hat{r} + 1}^2
    + \frac{\sigma_\epsilon^2}{2\lr{\sigma_\epsilon \hat{r} + 1}^2}}
    \partial_{\hat{r}} \hat{u} \\
  &+ \frac{1}{2} \partial_{\hat{r} \hat{r}} \hat{u}
    - \tilde \beta \frac{\sigma_\epsilon}{\lr{\sigma_\epsilon \hat{r} + 1}} \partial_{\hat{r} \hat{\phi}} \hat{u}
    + \frac{\sigma_\epsilon^2 (1 + \tilde \beta^2)}{2 \lr{\sigma_\epsilon \hat{r} + 1}^2} \partial_{\hat{\phi} \hat{\phi}} \hat{u}.
\end{align*}

In this case, we have no choice but to expand the eigenvalues and eigenfunctions in $\sigma_\epsilon$.
We have that
\begin{align*}
  \frac{\sigma_\epsilon}{\sigma_\epsilon \hat{r} + 1}
  &= \sigma_\epsilon - \sigma_\epsilon^2 \hat{r} + \sigma_\epsilon^3 \hat{r}^2
    + \mathcal{O}\lr{\sigma_\epsilon^4} \\
  \frac{\sigma_\epsilon^2}{2 \lr{\sigma_\epsilon \hat{r} + 1}^2}
  &= \frac{\sigma_\epsilon^2}{2} - \sigma_\epsilon^3 r
    + \mathcal{O}\lr{\sigma_\epsilon^4}, \\
\end{align*}
and the radial component of the drift expands as
\begin{align*}
  - 2 \hat{r}
  + \sigma_\epsilon\lr{\frac{1}{2} - 3 \hat{r}^2}
  - \sigma_\epsilon^2 \lr{\frac{\hat{r}}{2} + \hat{r}^3}
  + \mathcal{O}\lr{\sigma_\epsilon^3}.
\end{align*}
The terms of order $1 / \sigma_\epsilon$, which are associated with the deterministic solution on the limit cycle, vanish.

The eigenvalue equation yields to zeroth order,
\begin{align}
  \lambda^{(0)} \psi^{(0)}
  &= -2 \hat{r} \partial_{\hat{r}} \psi^{(0)}
    + \frac{1}{2} \partial^2_{\hat{r} \hat{r}} \psi^{(0)}
    \label{eq:HermiteRadial}
\end{align}
This Hermite equation in {\mk the $\hat{r}$-coordinate} corresponds to the eigenvalue problem of a one-dimensional {\mk stable} Ornstein-Uhlenbeck process (see e.g.~\cite{Pavliotis2014}) with {\mk damping coefficient given by the dimensional Floquet exponent $-2 \delta$ associated with the dimensional Floquet vector $\vec{e}_1 = (1, \tilde \beta / R)$ transverse to $\Gamma$; see Section~\ref{sec:normalForm} and Appendix~\ref{sec:Floquet}.}
The stationary density for this one-dimensional Ornstein-Uhlenbeck process is given in adimensional polar coordinates $(r', \theta')$ by
\begin{align}
	\rho_{\Gamma}(r) = \frac{1}{2\pi}\sqrt{\frac{2}{\pi \epsilon^2}} e^{-2 (r' - r_\epsilon)^2}. \label{eq:statDenSup}
\end{align}
The solutions to the eigenproblem (\ref{eq:HermiteRadial})
for any $\lambda_l^{(0)} = -2 l,$  $l$ in $\mathbb{N}$, are given by the rescaled Hermite polynomials~\cite[Chap.~5.5]{Risken1989}
\begin{align}
  \psi_{l}^{(0)}(\hat{r}, \hat{\phi})
  &= \eta(\hat{\phi}) \enskip H_l(\sqrt{2} \hat{r}), \label{eq:EigVecHermite}
\end{align}
where $H_l$ is the $l^{th}$ Hermite polynomial~\cite[p.~60]{Lebedev1972}
and $\eta$ is some function of $\hat{\phi}$ only.

The function $\eta$ in~\eqref{eq:EigVecHermite} {\mkr is} determined for $l = 0$ by solving for the higher-order equations.
In general, the first and second-order terms of the expansion yield,
\begin{align*}
  \mathcal{O}(\sigma_\epsilon)
  &: \quad \lambda^{(0)} \psi^{(1)} + \lambda^{(1)} \psi^{(0)}
    = \lr{\frac{1}{2} - 3 \hat{r}^2} \partial_{\hat{r}} \psi^{(0)}
    - \tilde \beta \partial^2_{\hat{r} \hat{\phi}} \psi^{(0)}
    -2 \hat{r} \partial_{\hat{r}} \psi^{(1)}
    + \frac{1}{2} \partial^2_{\hat{r} \hat{r}} \psi^{(1)}\\
  \mathcal{O}(\sigma_\epsilon^2)
  &: \quad \lambda^{(0)} \psi^{(2)} + \lambda^{(1)} \psi^{(1)}
    + \lambda^{(2)} \psi^{(0)}
    = -\lr{\frac{\hat{r}}{2} + \hat{r}^3} \partial_{\hat{r}} \psi^{(0)}
    - \tilde \beta \hat{r} \partial^2_{\hat{r}\hat{\phi}} \psi^{(0)}
    + \frac{1 + \tilde \beta^2}{2} \partial_{\hat{\phi} \hat{\phi}}  \psi^{(0)}\\
  & \quad \quad \quad \quad \quad \quad \quad \quad \quad \quad \quad \quad
     \quad \quad \quad \quad
    + \lr{\frac{1}{2} - 3 \hat{r}^2} \partial_{\hat{r}} \psi^{(1)}
    - \tilde \beta \partial^2_{\hat{r} \hat{\phi}} \psi^{(1)}
    -2 \hat{r} \partial_{\hat{r}} \psi^{(2)}
    + \frac{1}{2} \partial^2_{\hat{r} \hat{r}} \psi^{(2)}
\end{align*}
The special case $\psi^{(0)} = \psi_{l}^{(0)}$ with $l = 0$ is such that $\partial_{\hat{r}} \psi^{(0)} = 0$.
Thus, for $\psi^{(1)} = \psi^{(2)} = 0$, to first and second order,
\bea
  \mathcal{O}\lr{\sigma_\epsilon}
  &: \quad \lambda^{(1)} \psi^{(0)} = 0 \\
  \mathcal{O}\lr{\sigma_\epsilon^2}
  &: \quad \lambda^{(2)} \eta
    = \frac{1 + \tilde \beta^2}{2} \partial_{\hat{\phi} \hat{\phi}} \eta. \label{eq:diffusionCircle}
\eea
The first equation in~\eqref{eq:diffusionCircle} implies that $\lambda^{(1)} = 0$,
while the second equation corresponds to the eigenproblem for pure diffusion on the circle with diffusion coefficient $(1 + \tilde \beta^2) / 2$.
Its solutions for $\lambda_n^{(2)} = -n^2 (1 + \tilde \beta^2) / 2, n \in \mathbb{Z}$ are given by the harmonics $\eta_{\pm n} = \exp{(\pm i n \hat{\phi})}$, such that $\psi_{l, \pm n}^{(0)} = H_l(\sqrt{2} \hat{r}) \exp{(\pm in\hat{\phi})}$.

Unfolding the change of variables, $\hat{r} = \sqrt{\delta} (r - R) / \epsilon, \hat{\phi} = \theta - \tilde \beta \log (r / R) + \omega_f t$ and $t' = \delta t$, yields the small-noise expansion from Proposition~\ref{prop:eigen_above} of the eigenvalues and eigenfunctions of the SHE~\eqref{eq:HopfSDEPolar} for $\delta > 0$ and $\sigma_\epsilon$ small.
A term $\exp{(i \omega_f t)}$ appears in front of the eigenfunctions that can be canceled out since multiples of eigenfunctions are also eigenfunctions.
To find the adjoint eigenfunctions $\psi^{(0)*}_{ln}$,
orthonormal to the eigenfunctions $\psi^{(0)}_{ln}, l \in \mathbb{N}, n \in \mathbb{Z}$, one {\mkr uses} the orthogonality of the Hermite polynomials~\cite[p.~65]{Lebedev1972}.
Finally, note that higher-order terms in the expansion depend on $\tilde \beta$.





\bibliographystyle{amsalpha}
\bibliography{biblio_AT}


\end{document}